\documentclass[twocolumn,trackchanges,amsmath]{aastex702}

\begin{document}

\title{A Coherent Search for New Galactic Magnetars using Fermi/GBM and follow-up Swift/BAT: 7 Candidates and a Large Burst Catalog}

\author[0009-0004-8898-248X]{Ariel Perera}
\affiliation{Department of Particle Physics \& Astrophysics, Weizmann Institute of Science, Rehovot 76100, Israel}
\email[show]{ariel.perera@weizmann.ac.il}

\author[0000-0001-5162-9501]{Barak Zackay}
\affiliation{Department of Particle Physics \& Astrophysics, Weizmann Institute of Science, Rehovot 76100, Israel}
\email{barak.zackay@weizmann.ac.il}

\author[0000-0002-1661-2138]{Tejaswi Venumadhav}
\affiliation{Department of Physics, University of California at Santa Barbara, Santa Barbara, CA 93106, USA}
\affiliation{International Centre for Theoretical Sciences, Tata Institute of Fundamental Research, Bangalore 560089, India}
\email{teja@ucsb.edu}

\begin{abstract}

    Galactic magnetars are a rare class of neutron stars with extreme magnetic fields, with only about 30 known sources. Their short hard X-ray and soft gamma-ray bursts provide an important channel for discovering magnetar activity. Wide-field instruments such as the Fermi Gamma-ray Burst Monitor (GBM) are well suited to detecting these bursts, but their broad localization uncertainties make it difficult to associate individual bursts with specific sources. Coded-aperture telescopes, in contrast, provide arcminute-scale localizations but have narrower fields of view.
    We search for previously unidentified magnetars and construct a magnetar burst catalog by combining a broad sample of short, soft triggers from our coherent GBM detection pipeline over 2013-2025 with targeted searches of archival Swift/BAT images. We develop a statistical association framework that uses BAT detections to localize GBM bursts to arcminute precision, together with a spatio-temporal clustering framework that associates poorly localized bursts with common sources. Together, these methods identify 1720 magnetar bursts, of which 1170 are confidently associated with 11 known magnetars. The remaining 550 bursts could not be assigned to specific sources. We identify seven candidate new magnetars through three complementary discovery channels.
    The catalog, composed of the associated bursts, extends to the lowest fluences observed by GBM. We also provide spectral and temporal analyses of the catalog. The new candidate sources provide targets for X-ray follow-up that could reveal previously unidentified Galactic magnetars. More broadly, this work demonstrates how combining instruments with complementary capabilities can maximize the scientific return of high-energy transient observations.

\end{abstract}


\section{Introduction}
Magnetars are a rare class of neutron stars whose radiative activity is powered primarily by their magnetic fields \citep[e.g.,][]{duncan_formation_1992,mereghetti_magnetars_2015,kaspi_magnetars_2017}. 

Only $\sim30$ magnetars and magnetar candidates are currently known in the Milky Way and the Magellanic Clouds \citep{olausen_mcgill_2014}. They are typically slowly rotating with a period of 2--12~s and possess inferred dipole magnetic fields of $\sim10^{13}$--$10^{15}$~G, although magnetar-like activity has also been observed from lower magnetic field neutron stars \citep[e.g.,][]{rea_low-magnetic-field_2010,wang_multiwavelength_2020,blumer_reactivation_2021}.

Their most characteristic transient behavior is the emission of short bursts of hard X-rays and soft gamma rays, typically lasting from a few milliseconds to one second and having spectra softer than those of cosmological short gamma-ray bursts \citep[e.g.,][]{turolla_magnetars_2015,collazzi_five_2015,perera_expanding_2026}. Magnetars also exhibit persistent X-ray emission, longer-lived outbursts, and, more rarely, giant flares reaching energies of $\sim10^{44}$-$10^{46}$~erg \citep[e.g.,][]{coti_zelati_systematic_2018,palmer_giant_2005,turolla_magnetars_2015}.

Because magnetars can remain faint or inactive for long intervals, transient bursts provide an important channel for discovering previously unknown sources. Wide-field hard-X-ray and gamma-ray instruments can identify the onset of activity, while subsequent observations with focusing X-ray telescopes provide precise localization and measurements of pulsations and spin-down. This strategy has led to the discovery of several magnetars with \textit{Swift}, where bursts detected with BAT triggered rapid XRT follow-up and source identification \citep{esposito_very_2020,palmer_swift_2021,page_swift_2020,gogus_discovery_2020,younes_nicer_2020}.

Fermi/GBM \citep{meegan_fermi_2009} is particularly well suited to searches for magnetar bursts because of its nearly all-sky coverage, long observing baseline, and high sensitivity over the hard-X-ray and soft-gamma-ray energy range. Searches of both triggered and continuous GBM data have therefore produced large samples of magnetar bursts, including events below the onboard triggering threshold \citep{perera_expanding_2026, godwin_17_2026}. Identifying the astrophysical sources of these bursts, however, remains considerably more difficult. GBM source positions are inferred statistically from the relative count rates measured by its detectors and typically have localization uncertainties of several degrees for bright bursts, with substantially larger uncertainties for faint events \citep{connaughton_localization_2015,burgess_awakening_2018,perera_expanding_2026}. In addition, GBM localizations are affected by systematic errors. Although the magnitude and detailed origin of these systematics remain under study, characteristic localization uncertainties of several degrees, and in some cases up to approximately $15^\circ$, have been reported \citep{berlato_improved_2019,goldstein_evaluation_2020, lopez_evaluation_2024,perera_expanding_2026}. This is particularly problematic for magnetars, which are concentrated toward the Galactic plane and can be separated from one another by angular distances much smaller than a typical GBM localization uncertainty. For example, the angular separations between Swift J1818.0$-$1607 and AX J1818.8$-$1559, SGR 1806$-$20 and SGR 1808$-$20, and SGR 1833$-$0832 and Swift J1834.9$-$0846 are approximately $0.23^\circ$, $0.26^\circ$, and $0.37^\circ$, respectively \citep{kaspi_magnetars_2017}. Precise localizations from other instruments can provide the missing information needed to associate GBM activity with known sources or to identify previously unknown ones.

In this work, we combine a sensitive coherent search of continuous Fermi/GBM data with new localization and clustering methods to search for previously unidentified Galactic magnetars. We first construct a broad sample of short, soft magnetar-like transients using the detection and classification framework of \citet{perera_new_2025,perera_expanding_2026}. We then develop a targeted \textit{Swift}/BAT imaging search to identify and localize counterparts to GBM triggers, together with a spatio-temporal clustering framework that uses precisely localized bursts as anchors for otherwise poorly localized GBM activity. The large burst sample obtained through this analysis is also used to construct a catalog of magnetar bursts.

The clustering analysis also provides an independent discovery channel: coherent concentrations of GBM bursts in time and sky position can identify candidate magnetar sources even when no usable BAT image is available. These complementary approaches therefore enable source identification both through precise BAT localizations and through statistically significant clustered GBM activity alone. Using this framework, we identify 7 new magnetar candidate sources and present an accompanying catalog of magnetar-like bursts detected with improved sensitivity.

Several magnetar burst catalogs have recently been published using complementary instruments and analysis strategies. These include an \textit{INTEGRAL}/IBIS catalog \citep{pacholski_integral_2025}, based on a narrower-field instrument with an energy range overlapping that of GBM; a NICER catalog \citep{chu_nicer_2026}, which probes the complementary soft X-ray band of $0.5$--$10$~keV; and an independent GBM magnetar burst catalog \citep{godwin_17_2026}. The different instrumental sensitivities, fields of view, burst-detection procedures, and source-association strategies make these catalogs complementary. Over periods of overlapping coverage, our coherent burst search and source-association framework recovers more events - 1170 compared to 735 from \citep{godwin_17_2026} - associated with known magnetars, including both bright bursts and a substantial population of very faint events, thereby providing an expanded sample for studying magnetar burst populations.

This paper is organized as follows. In Sec.~\ref{sec:pipeline_overview}, we summarize the coherent GBM burst detection and parameter estimation pipeline. In Sec.~\ref{sec:data_selection}, we define the magnetar-like trigger sample used in this analysis. In Sec.~\ref{sec:association_with_sources_methods}, we describe the targeted BAT imaging search and the statistical and spatio-temporal clustering methods used to associate GBM bursts with astrophysical sources. In Sec.~\ref{sec:assoc_bursts_to_clusters}, we present the resulting source associations, while in Sec.~\ref{sec:identifying_unknown_sources}, we investigate candidate previously unidentified magnetars identified through BAT localizations and clustered GBM activity. In Sec.~\ref{sec:burst_catalog}, we describe the construction of the burst catalog and examine its temporal, spectral, and fluence distributions. Finally, in Sec.~\ref{sec:summary}, we summarize our results and discuss prospects for future searches. Supporting methodological details and additional results are provided in the Appendices.


\section{Burst Detection and Parameter Estimation Pipeline} \label{sec:pipeline_overview}

\subsection{Overview}
    The magnetar bursts analyzed in this work were identified using the
    coherent Poisson matched-filter pipeline introduced by
    \citet{perera_new_2025} and subsequently applied to the full
    2013--2025 Fermi/GBM data set by \citet{perera_expanding_2026}.
    The latter work also introduced the parameter-estimation,
    classification, and empirical significance framework used to
    characterize the detected events. That search revealed a large
    population of previously unidentified magnetar bursts. Here we briefly
    summarize the aspects of the detection and parameter-estimation
    framework relevant to the present analysis.
    
    \subsection{Detection} \label{sec: detection}
    
    The search was performed on the continuously available Fermi/GBM
    time-tagged event (TTE) data from 2013 through the end of 2025. All
    14 GBM detectors and all available energy channels were analyzed
    coherently. The detection statistic is based on a likelihood-ratio test
    for Poisson-distributed photon counts and is constructed to be
    near-optimal in the low-count regime \citep{perera_new_2025}.

    The search uses a stochastically placed bank of approximately 500
    templates spanning sky position and spectral shape. The bank was
    constructed such that this relatively small number of templates
    provides high coverage of the much larger underlying parameter space.
    To search over burst duration, the pipeline additionally uses 22
    boxcar templates with logarithmically spaced widths from 3 ms to $\sim 6$~s; the spacing was chosen such that the maximum SNR loss due to mismatch between adjacent
    durations is approximately $7\%$. For each spectral-spatial template
    and burst duration, the detection statistic is
    \begin{equation}
    \label{eq:statistic}
    \mathcal{S}_t(\theta) =
    \frac{
    \sum_n (d_{t,n}-b_{t,n}) \star_t
    \log\left[1+\frac{A T_{t,n}(\theta)}{b_{t,n}}\right]
    }{
    \sqrt{
    \sum_{n,t} b_{t,n} \star_t
    \log^2\left[1+\frac{A T_{t,n}(\theta)}{b_{t,n}}\right]
    }
    },
    \end{equation}
    where $d_{t,n}$ is the number of observed photon counts in time bin
    $t$ and detector--energy channel $n$, $b_{t,n}$ is the estimated
    background, and $T_{t,n}(\theta)$ is the expected signal template.
    The template parameters $\theta$ describe the source sky position and
    the Band-function spectral parameters \citep{band_batse_1993}. The
    amplitude $A$ is fixed to the template-dependent detection limit
    determined from simulations, and $\star_t$ denotes cross-correlation
    along the time axis.
    
    The resulting matched-filter time series is locally renormalized using
    the background spectral drift correction introduced in
    \citet{perera_new_2025}, which reduces the effect of slowly varying
    background errors. Candidate triggers are subsequently subjected
    to a series of statistical vetoes designed to reject single-detector
    events, timing glitches, particle-precipitation events, and Earth
    occultation steps.
    
    Because different temporal templates correspond to different effective
    search volumes and therefore different trials factors, the detection
    statistic is empirically calibrated as a function of duration and
    reported in units of a standard normal variate. The trigger is assigned
    the duration associated with the highest calibrated SNR at the
    detection stage.

    In addition to the detection statistic, the pipeline estimates the
    signal amplitude for each trigger using the best-fitting template.
    While the amplitude $A$ appearing in Eq.~\ref{eq:statistic} is fixed
    to the template-dependent detection limit when constructing the
    matched-filter statistic, the physical amplitude of a detected burst
    is subsequently estimated from the data. We denote this estimate by
    $\hat{A}$. This quantity provides the normalization of the
    best-fitting spectral template and is used later in the parameter estimation, the calculation of the burst fluence and the $T_{90}$ duration.
    
    \subsection{Parameter estimation} \label{sec: pe}
    For each surviving trigger, we infer the source sky position and
    spectral parameters using the Poisson likelihood underlying the
    detection statistic. We write the model parameters as
    $\theta=(\Omega,\xi)$, where $\Omega$ denotes sky position and $\xi$
    the Band-function spectral parameters. For a uniform prior, the
    posterior is proportional to
    \begin{align}
    &P(\theta|d,\mathcal{H}_1)
    \propto
    \frac{p(d|\theta,\mathcal{H}_1)}
         {p(d|\mathcal{H}_0)}
    \nonumber\\
    &\propto
    \exp\left\{
    \sum_n
    \left[
    d_n
    \log\left(
    1+\frac{A T_n(\Omega,\xi)}{b_n}
    \right)
    -
    A T_n(\Omega,\xi)
    \right]
    \right\},
    \label{eq:posterior}
    \end{align}
    where $\mathcal{H}_1$ and $\mathcal{H}_0$ denote the signal and
    noise hypotheses, respectively. A derivation from the Poisson
    likelihood is given in Appendix E of \citet{perera_new_2025}. The
    count-space template is
    \begin{equation}
    T_n(\Omega,\xi)
    =
    \int R_n(\Omega,E)\,N(E,\xi)\,dE ,
    \end{equation}
    where $R_n(\Omega,E)$ is the direction- and energy-dependent detector
    response and $N(E,\xi)$ is the incident photon spectrum, modeled here
    with a Band function.
    
    We sample the posterior using Markov Chain Monte Carlo (MCMC), with
    the maximum-likelihood parameters obtained during the detection stage
    used to initialize the sampler. To make parameter estimation feasible
    for the large number of triggers in the archival search, detector
    responses are precomputed on a $2^\circ$ sky grid and the nearest
    response is used at each MCMC sample. The Band-function priors are
    chosen to be sufficiently broad to accommodate both non-GRB
    transients and weakly constrained low-SNR events
    \citep{perera_expanding_2026}. After sampling, we refine the burst duration by scanning a dense logarithmic grid of boxcar widths and adopt the width that maximizes the likelihood as the refined peak-flux duration.

    All the posterior samples used for this parameter estimation, including the sky samples are available in a Zenodo repository: \href{https://doi.org/10.5281/zenodo.22884088} {10.5281/zenodo.22884088} \citep{perera_catalog_2026}, and the parameter estimation code is available online in the \href{https://github.com/PeAriel/grpype}{GitHub repository}\footnote{\url{https://github.com/PeAriel/grpype}} \texttt{grpype} \citep{ariel_perera_pearielgrpype_2026}.
    
    \subsection{Trigger classification and astrophysical probability}
    
    The detected population contains several physically distinct classes of gamma-ray transients, including GRBs, soft gamma repeaters (SGRs), terrestrial gamma-ray flashes (TGFs), and solar flares (SFs). We therefore apply the classification scheme introduced in \citet{perera_expanding_2026}, which uses the inferred spectral, temporal, and directional properties of each trigger, including $E_{\rm peak}$, duration, sky position, and Bayes factors for solar and terrestrial origins.
    
    The magnetar population considered in this work is dominated by the
    \texttt{SGR2} and \texttt{SGR3} classes. \texttt{SGR2} consist of triggers with $10 < E_{\rm peak} < 20~{\rm keV}$ and maximum-likelihood Galactic latitude $|b|<30^\circ$ while \texttt{SGR3} consists of triggers with $20 < E_{\rm peak} < 50~{\rm keV}$, maximum-likelihood Galactic latitude $|b|<20^\circ$, and duration $\Delta<0.85~{\rm s}$.
    These cuts isolate the prominent population of soft events
    concentrated near the Galactic plane identified in
    \citet{perera_expanding_2026}.
    
    To quantify the probability that a trigger corresponds to a real
    transient rather than a statistical fluctuation, we apply the complete
    detection pipeline to a time-slid realization of the GBM data. The
    relative arrival times in different detectors are shifted so that
    true coherent signals are destroyed while the statistical and
    instrumental properties of the individual detector streams are
    preserved. The resulting triggers therefore provide an empirical
    estimate of the background distribution.
    
\begin{figure}
        \centering
        \includegraphics[width=\linewidth]{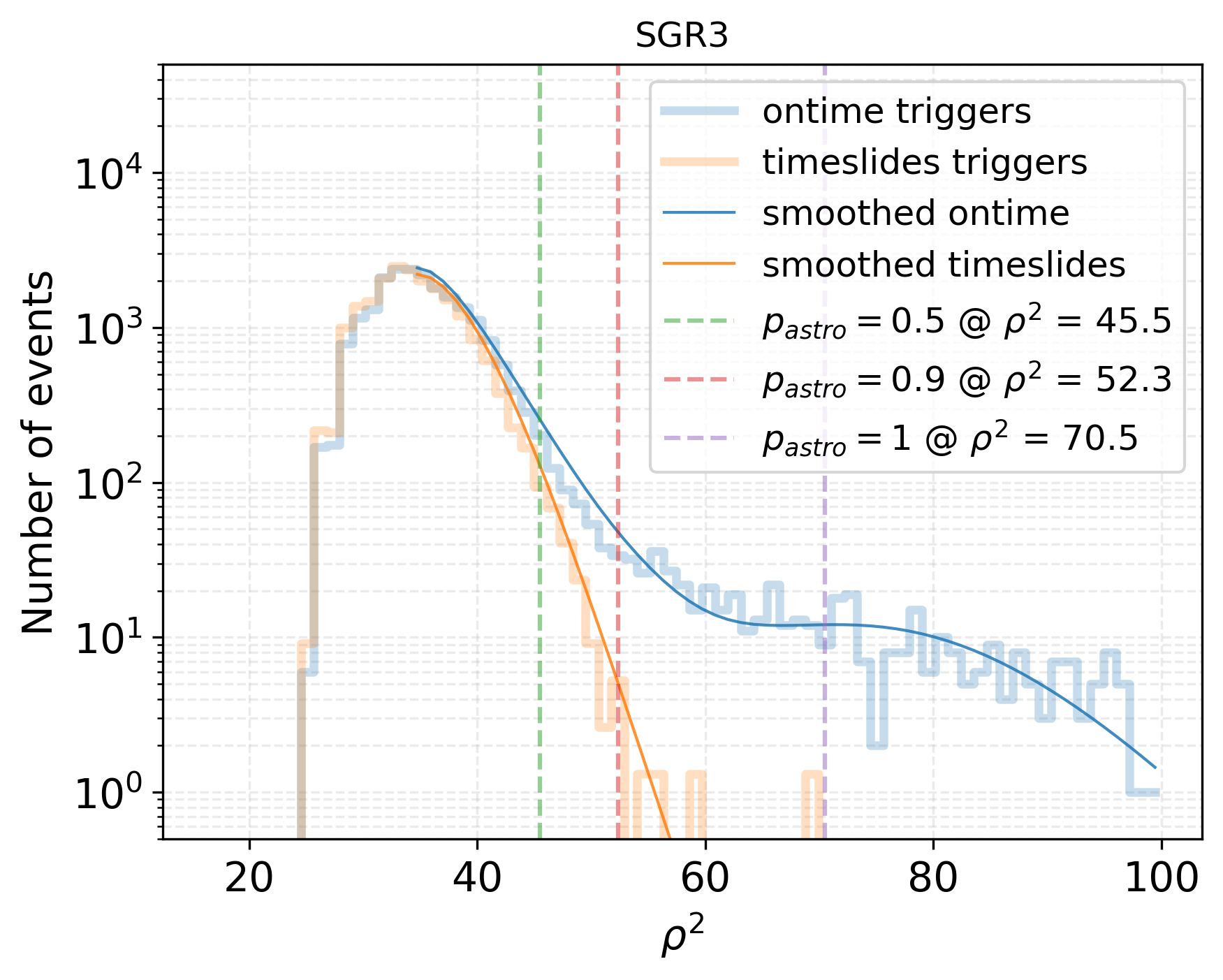}
        \caption{
        Distribution of the calibrated detection statistic for triggers
        in the \texttt{SGR3} class. The on-time distribution is compared
        with the empirical background distribution obtained from
        time-slid GBM data. Their relative rates are used to determine the
        class-dependent astrophysical probability $p_{\rm astro}$.
        }
        \label{fig:sgr_pastros}
    \end{figure}

    For each source class, we compare the on-time and time-slid trigger
    rate distributions and define
    
    \begin{equation}
    p_{\rm astro}(\rho^2)
    =
    \frac{
    p(\rho^2|\mathcal{H}_1) p(\mathcal{H}_1)
    }{
    p(\rho^2|\mathcal{H}_0) p(\mathcal{H}_0)
    +
    p(\rho^2|\mathcal{H}_1) p(\mathcal{H}_1)
    },
    \end{equation}
    where $\rho$ is the calibrated detection SNR, and $ p(\mathcal{H}_i)$ the prior of hypothesis $i$.
 The signal contribution
    is estimated from the excess of the on-time distribution above the
    time-slid background distribution. Thus, $p_{\rm astro}$ provides an
    empirical, class-dependent estimate of the probability that a trigger
    is associated with a real transient. Figure~\ref{fig:sgr_pastros}
    shows the on-time and time-slid distributions used to determine
    $p_{\rm astro}$ for the \texttt{SGR3} class.
    
    \subsection{Swift/BAT follow-up}
    
    Finally, each GBM trigger was followed up with a targeted search of
    the continuous Swift/BAT rate data. The search uses a boxcar temporal
    template matched to the GBM burst duration and evaluates the statistic
    
    \begin{equation}
    \label{eq:BAT_statistic}
    \mathcal{S}_{B}(t_0)
    =
    \sum_{t=t_0-\Delta/2}^{t_0+\Delta/2}
    \frac{
    (d_t-\hat{b}_t)T_t
    }{
    \sigma_{t_0}\sqrt{\Delta}
    },
    \end{equation}
    where, $d_t$ is the observed counts rate, $T_t$ is the boxcar template, $\Delta$ is the burst duration, $\hat{b}_t$ is the locally
    estimated BAT background, and $\sigma_{t_0}$ characterizes its
    fluctuations. As for the GBM search, the statistic is corrected for
    local background drift. The same analysis is then repeated at
    time-shifted locations to construct an empirical null distribution.
    Comparison of the on-time and shifted distributions yields
    $p_{\rm bat}$, the probability that the BAT rate excess associated
    with a GBM trigger represents a real coincident signal rather than a
    statistical fluctuation.

\section{Data Selection} \label{sec:data_selection}
    The detection and parameter-estimation pipeline described above provides, for each GBM trigger, posterior estimates of its sky position and spectral parameters, a refined peak-flux duration, a phenomenological source classification, a class-dependent probability of astrophysical origin
    $p_{\rm astro}$, and an independent Swift/BAT association probability
    $p_{\rm bat}$.
    
    For the analysis presented here, we extend the sample beyond the
    specific SGR classes defined in our previous work
    \citep{perera_expanding_2026}. Those classes were constructed to
    maximize the separation between different transient populations and to
    enable a robust, class-dependent assignment of the astrophysical
    probability $p_{\rm astro}$, rather than to provide a complete
    selection of all possible magnetar bursts. In the present analysis,
    each trigger retains the $p_{\rm astro}$ value derived from its
    original class-dependent calibration in
    \citet{perera_expanding_2026}, even though the broader selection used
    here spans multiple classes.
    
    Here, we instead select all triggers in our catalog that satisfy
    $E_{\rm peak}<60~{\rm keV}$, have a peak-flux duration shorter than
    $1~{\rm s}$, are not classified as solar flares or terrestrial
    gamma-ray flashes, and have $p_{\rm astro}\geq0.5$.
    These cuts define a broader magnetar-like sample while retaining an
    interpretable probabilistic threshold and removing the transient
    populations most likely to contaminate the analysis. The resulting sample consists of 3675 triggers.

\section{Associating GBM Bursts with Astrophysical Sources}\label{sec:association_with_sources_methods}

\subsection{Motivation}
    As discussed in the Introduction, the localization precision of Fermi/GBM is generally insufficient to associate an individual weak burst uniquely with an astrophysical source. Precise localizations from another instrument can resolve this ambiguity, but such measurements are available for only a fraction of GBM events.
    
\begin{figure*}
        \centering
        \includegraphics[width=1.0\linewidth]{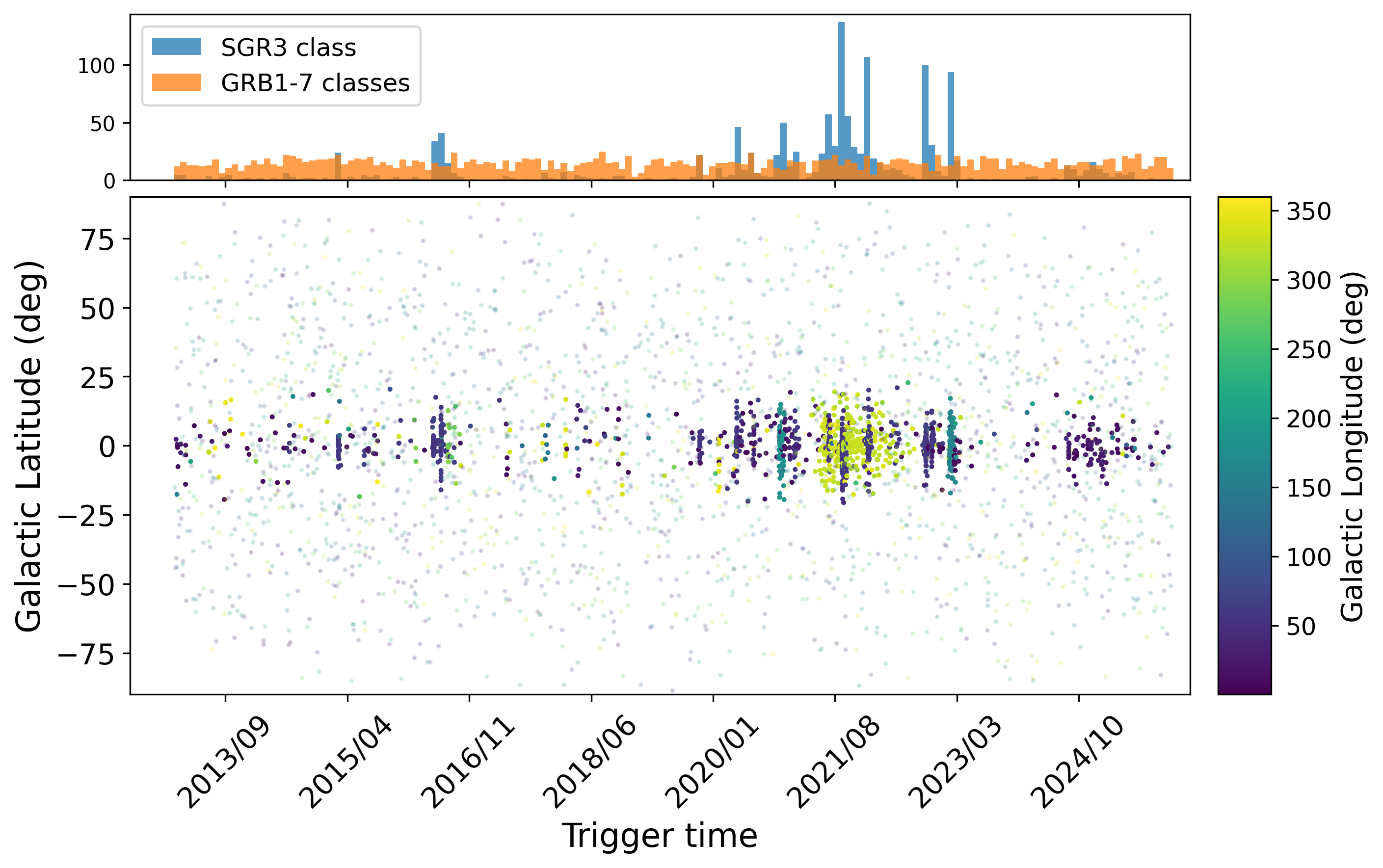}
        \caption{\textbf{Temporal distribution of magnetar-like and GRB-like triggers,
                reproduced from \citet{perera_expanding_2026}}. Magnetar-like events
                exhibit pronounced temporal clustering and burst-storm activity,
                whereas GRB-like triggers are distributed much more uniformly in
                time. This clustering behavior motivates the source-association
                method developed in this work.}
        \label{fig: sgr bursts from prev paper}
    \end{figure*}

    Magnetars, however, may provide an additional handle. Their bursts are
    often emitted during episodes of enhanced activity, producing temporal
    clusters or ``burst storms'' \citep{gogus_temporal_2001,cheng_statistical_2020,kaspi_magnetars_2017}, although isolated bursts also occur. This behavior is illustrated in
    Fig.~\ref{fig: sgr bursts from prev paper}, reproduced from
    \citet{perera_expanding_2026}, where magnetar-like triggers are seen to
    cluster strongly in time, in contrast to the more uniformly distributed
    GRB population.
    
    We exploit this temporal clustering to extend precise source associations to otherwise poorly localized GBM bursts. If a statistically significant cluster contains an event that can be confidently localized and associated with an astrophysical source, that event can serve as an anchor for the remaining bursts in the cluster. The other events can then be associated with the same source provided that their timing and GBM localization information are consistent with membership in the cluster.

    We obtain these source anchors through a targeted \textit{Swift}/BAT imaging search at the times and sky locations of GBM triggers. Because the search is conditioned on the GBM trigger time and localization, it can recover BAT counterparts that would not necessarily be identified in an untargeted image search. The resulting BAT localizations are then incorporated into the association framework to identify the astrophysical source responsible for the clustered bursts.
    
    
    The following subsections describe the BAT imaging search, the clustering analysis, and the criteria used to associate GBM bursts with known and candidate magnetar sources.

\subsection{Swift BAT Localization of GBM Triggers}
    The Swift Burst Alert Telescope (BAT) is a coded-aperture instrument
    designed to detect and localize gamma-ray transients in the
    15-350~keV energy range \citep{barthelmy_burst_2005}. Its field of
    view depends on the coding fraction and reaches $1.94~{\rm sr}$ at
    10\% partial coding \citep{krimm_swiftbat_2013}. For sources within
    the coded field of view, a sky image can be reconstructed by
    deconvolving the detector-plane count distribution with the coded-mask
    pattern. For sufficiently well-coded and significant events, BAT can
    localize sources to an accuracy of approximately $4$ arcmin.
    
    
    BAT imaging products are available in two forms relevant to this
    analysis. When BAT produces an onboard trigger, event data containing
    the arrival time, energy, and detector position of individual photons
    are typically telemetered around the trigger time. These events can be
    accumulated over a user-defined interval to construct a detector-plane
    image and subsequently a sky image. At other times, event data may not
    be available, but the BAT archive can contain scaled maps, which are
    detector-plane count histograms accumulated over longer predefined
    intervals. A transient occurring within the coded field of view may
    therefore leave a detectable coded-mask signature in a scaled map even
    when event-level data are unavailable.
    
    In a blind BAT image search, a relatively high detection threshold is
    required because a large number of independent sky positions are
    tested. Consequently, moderate-significance image peaks that would not
    be compelling in an all-sky or full-image search can become
    statistically meaningful when both the burst time and an approximate
    sky position are supplied independently by GBM. Our search therefore
    uses the GBM trigger information to reduce the effective trials factor
    and probe BAT image significances below the threshold normally adopted
    for blind source detection. The statistical significance of these
    targeted associations is quantified in Section~\ref{sec: statistical associations}.
    
    We apply the imaging search to every GBM trigger (the full trigger list obtained in \cite{perera_expanding_2026}) with
    $p_{\rm bat}\geq0.5$. When a scaled map covers the
    trigger time, we reconstruct a sky image from the detector-plane map
    using \texttt{batfftimage} and search for sources using
    \texttt{batcelldetect}; in this case the effective integration time is
    set by the accumulation interval of the scaled map.

    When BAT event data are available, we additionally construct a
    detector-plane image using \texttt{batbinevt} in the 15-350~keV band,
    apply the detector quality mask with \texttt{batdetmask}, and form a
    sky image using the same imaging and source-detection procedure. The
    events are accumulated over a time interval of duration
    $T_{\rm img}$ centered on the GBM trigger time. We set
    $T_{\rm img}$ equal to the refined peak-flux duration obtained from the
    GBM parameter-estimation stage, thereby approximately matching the BAT
    integration interval to the duration over which the transient signal
    is strongest. Both the event-data and scaled-map searches are
    performed whenever the corresponding products are available.

\subsection{Statistical Association Framework} \label{sec: statistical associations}

    After performing the BAT image search around each GBM trigger, we retain all sources with image SNR $\rho>4$ and compute their angular separation $\Delta$ from the maximum-likelihood GBM localization. This produces a two-dimensional distribution of candidate detections, $\{\rho_i,\Delta_i\}$, where $\rho_i$ is the image SNR of candidate $i$ and $\Delta_i$ is its angular offset from the corresponding GBM localization.
    
    The distribution of candidates found in images centered on GBM triggers is denoted by $p(\rho,\Delta|\mathcal{H}_1)$, where $\mathcal{H}_1$ represents the hypothesis that the sample contains both true associations and accidental coincidences. To estimate the contribution from chance coincidences, we repeat the analysis on randomly time-shifted images, obtaining the background distribution $p(\rho,\Delta|\mathcal{H}_0)$. The background distribution is obtained by repeating the exact same BAT imaging search on real event data at times randomly offset from each GBM trigger (using a different Swift observation), then reprojecting every detection detector-plane coordinates through the spacecraft attitude at the original trigger time so that the source is placed in the burst-time BAT field of view.
    
    The probability that an individual candidate is a genuine astrophysical association is then (for a full derivation, see \cite{perera_multi-mission_2026})
    \begin{equation} \label{eq: psource}
        p_{\rm src} \equiv p(S|\rho,\Delta)
        =
        \frac{n(\rho,\Delta|\mathcal{H}_1)-\alpha n(\rho,\Delta|\mathcal{H}_0)}
        {n(\rho,\Delta|\mathcal{H}_1)},
    \end{equation}
    where $n(\rho,\Delta|\mathcal{H})$ denotes the measured number density of candidates in the corresponding sample, and $\alpha$ accounts for the difference in the total number of observations and event rates between the signal and background samples. $n$ is estimated by binning the raw 2D points, smoothing with a Gaussian kernel, and then evaluating on a bivariate cubic spline. Probabilities are set to zero when $n(\rho,\Delta|\mathcal{H}_1)<\alpha n(\rho,\Delta|\mathcal{H}_0)$. 
    
    A related quantity is the ratio of ontime to background density
    \begin{equation}
        \mathcal{R} = \frac{1}{1-p(S|\rho,\Delta)}
    \end{equation}
    can be derived from Eq.~\ref{eq: psource}, which provides an alternative description that is useful for interpreting the resulting probabilities in terms of relative occurrence rates.

\begin{figure}
        \centering
        \includegraphics[width=1.0\linewidth]{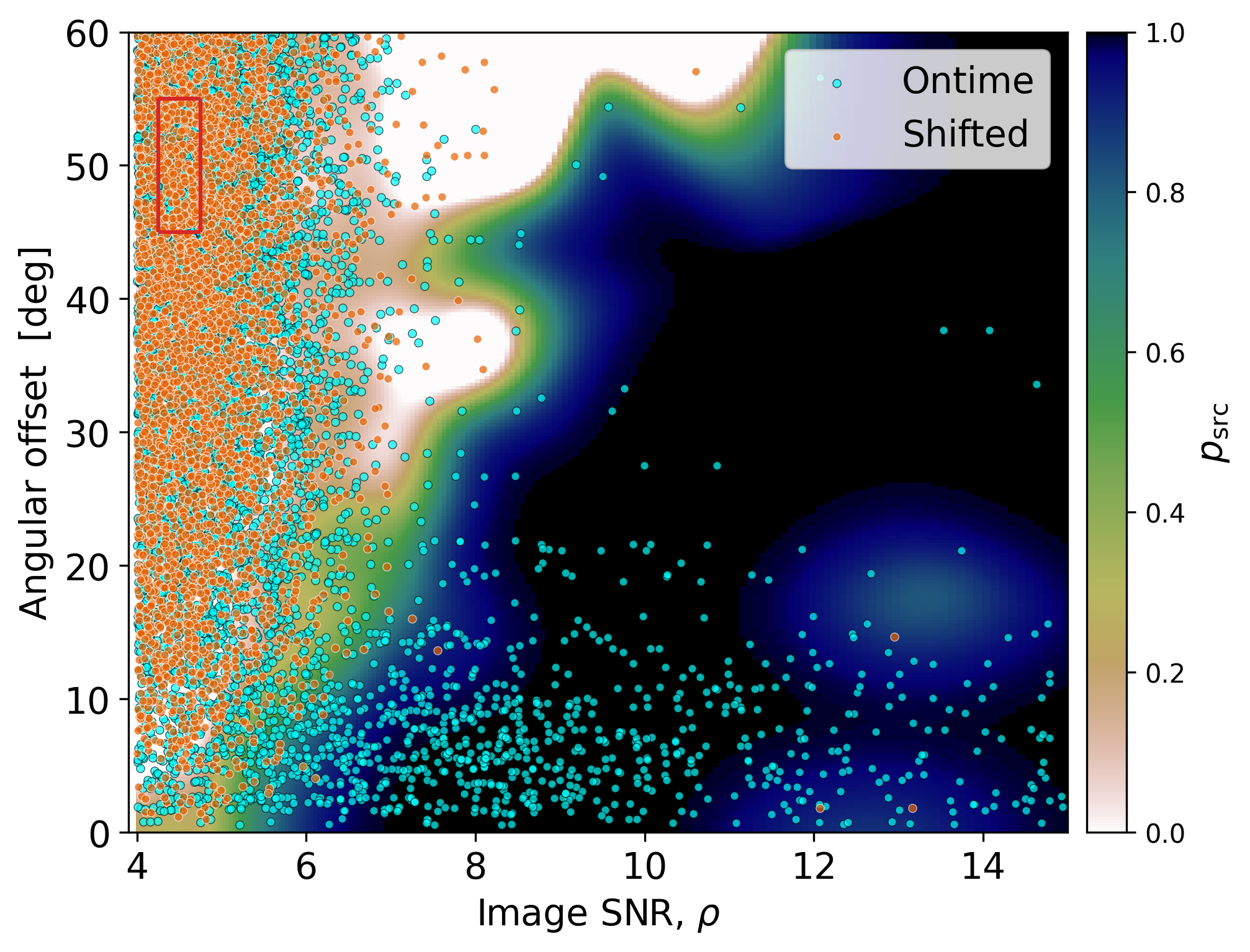}
        \caption{\textbf{Source probability distribution.} The estimated source probability, $p_{\rm src}$, as a function of SNR and angular separation, with the individual on-time and shifted candidates overlaid. A clear overdensity of on-time candidates is visible at small angular separations and is reflected in the increased source probability in this region. Because the on-time and shifted samples contain different numbers of observations, their number densities cannot be compared directly. We therefore introduce the normalization factor $\alpha$ in Eq.~(\ref{eq: psource}), estimated from the background-dominated region highlighted in red. This gives $\alpha=1.74$. Consequently, a region may contain more on-time than shifted candidates while still having a low source probability once the relative normalization of the two samples is taken into account.
        }
        \label{fig: psrc}
    \end{figure}

    We estimate $\alpha$ from the relative number of events in a background-dominated region, chosen here as $\mathrm{SNR}\in[4.25,4.75]$ and angular separation $\Delta\theta\in[45^\circ,55^\circ]$. Fig.~\ref{fig: psrc} shows the resulting two-dimensional distribution of $p_{\rm src}$, overlaid with the individual data points used in its estimation. For clarity, the region beyond $\Delta=60^\circ$ and $\rho=15$ is not shown, as candidates in this regime can be interpreted without requiring the probabilistic classification described here.    
    
    Since multiple candidates may be detected in a single BAT image, while at most one can be the counterpart to the GBM trigger, a second inference step is required to determine which candidate is the most probable association. Let $A_i$ denote the hypothesis that candidate $i$ is the true counterpart. Conditioning on the presence of a true counterpart in the image, taking a uniform prior on $\rho,\Delta$, and noting that $A_i$ does not depend on $\Omega$, Bayes' theorem gives
    \begin{equation} \label{eq: within image prob}
        p_{\rm assoc} \equiv p(A_i|\{\rho,\Delta,\Omega\})
        =
        \frac{\pi(\Omega_i)\,p_{\rm src}^{\,i}}
        {\sum_j \pi(\Omega_j)\,p_{\rm src}^{\,j}},
    \end{equation}
    where $p_{\rm src}^{\,i}$ is the source probability of candidate $i$, given by Eq. (\ref{eq: psource}), and $\pi(\Omega_i)$, with $\Omega=(\alpha,\delta)$, is a sky prior. The sky prior incorporates knowledge of persistent or repeating sources. This prior is estimated from the observed rates of repeating sources in the high-significance, small-offset region of the parameter space. Specifically, we use $\mathrm{SNR}\in[10,50]$ and angular separation $\Delta\theta\in[1^\circ,10^\circ]$.
    The sky prior modifies only the relative association probabilities between candidates within the same image and does not alter the astrophysical probability assigned to an individual candidate.
    
    The sky-position prior, $\pi(\Omega)$, is estimated self-consistently from the data using the high-source-probability population to determine both the fraction of events associated with known sources and their relative occurrence rates. For a position associated with a known source, the prior is assigned as the probability mass of the BAT point-spread function (PSF) region surrounding that source, weighted by the relative occurrence rate $w(\Omega)$ of that source. We conservatively adopt a PSF radius of $6'$, such that
    $ p = \Omega_{\rm PSF} / 4\pi $
    where $\Omega_{\rm PSF}$ is the solid angle subtended by a circular region of radius $6'$. Positions not associated with a known source are assigned a uniform prior over the remaining sky, denoted by $q$. The resulting mixture prior, expressed as a probability per PSF-sized sky cell, is
    \begin{equation}
    \pi(\Omega)
    =
    f_{\rm source} w(\Omega) p
    +
    (1-f_{\rm source}) q
    \end{equation}
    where $f_{\rm source}$ is the fraction of events associated with known sources and $w(\Omega)$ is the relative occurrence rate of the source associated with sky position $\Omega$.
    
 The fraction of events associated with known sources is estimated as
    \begin{equation}
    \hat{f}_{\rm source}
    =
    \frac{N_{\rm known}}{N},
    \end{equation}
    where $N_{\rm known}$ is the number of events associated with known sources and $N$ is the total number of events used to estimate the prior. For a sky position within the PSF of a known source, its weight is estimated as
    \begin{equation}
    \hat{w}(\Omega)
    =
    \frac{N_{\rm src}(\Omega)}{N_{\rm known}},
    \end{equation}
    where $N_{\rm src}(\Omega)$ is the number of bursts associated with the source at that position.
    
    For each image, we identify the most likely counterpart as the source with the largest association probability, $p_{\rm assoc}$, and use the corresponding $p_{\rm src}$ in the subsequent analysis presented in this paper. In addition, we restrict the sample to events with angular offsets smaller than $30^\circ$ and image exposures less than $10$~s, irrespective of their $p_{\rm src}$. This selection provides additional robustness against anomalous events and unmodeled systematics, which can lead to unreliable probability estimates in sparsely populated regions of the parameter space. We therefore impose these cuts to improve the reliability of the final sample.

    Overall, we identify 543 triggers with $p_{\rm src}>0.5$, of which 456 have $p_{\rm src}>0.9$. The majority of localized events are X-ray binaries that exhibit bursts longer than those of Magnetars. Constraining burst durations to be shorter than one second results in 112 events, which are more likely to be Magnetar bursts or GRBs. The relatively similar number triggers for these two probabilities can be explained by the fact that BAT typically saves event data or scaled maps in response to an external trigger, introducing a selection effect that favors observations containing a localizable astrophysical source. Consequently, once a BAT image is available, the probability of identifying a credible source counterpart is relatively high.

    Given the small BAT localization uncertainty, typically $\sim4'$, we associate a localized source with a magnetar if its position lies within $4'$ of a magnetar listed in the McGill Magnetar Catalog \citep{olausen_mcgill_2014}\footnote{\url{http://www.physics.mcgill.ca/~pulsar/magnetar/main.html}}. We additionally include two magnetars identified by \textit{Swift} after the most recent update of the catalog \citep{palmer_swift_2021,page_swift_2020,gogus_discovery_2020,younes_nicer_2020}, as well as the radio pulsar PSR J1119, which has exhibited magnetar-like bursts \citep{blumer_psr_2017}. The resulting list of localized magnetar sources is presented in Table~\ref{tab: localized magnetars}. Throughout the remainder of this work, we refer collectively to the sources in the McGill Magnetar Catalog and these additional sources as "catalog" or "known" magnetars.

    \begin{deluxetable}{lcc}
    \tablecaption{Number of unique GBM triggers associated with known magnetars based on the localization method discussed in Sec.~\ref{sec: statistical associations}.
    \label{tab: localized magnetars}}
    \tablehead{
      \colhead{Magnetar} & \multicolumn{2}{c}{$N$} \\
      \colhead{} & \colhead{$p_{\rm src} > 0.5$} & \colhead{$p_{\rm src} > 0.9$}
    }
    \startdata
      SGR 1935$+$2154 & 39 & 33 \\
      Swift J1555.2$-$5402 & 9 & 6 \\
      1E 1841$-$045 & 6 & 4 \\
      SGR 1806$-$20 & 4 & 4 \\
      PSR J1119$-$6127 & 3 & 2 \\
      4U 0142$+$61 & 2 & 1 \\
      CXOU J164710.2$-$455216 & 2 & 1 \\
      SGR 1830$-$0645 & 2 & 2 \\
      1E 1547.0$-$5408 & 1 & 1 \\
      Swift J1818.0$-$1607 & 1 & 0 \\
      \hline
      Total & 69 & 54 \\
    \enddata
    \tablecomments{Counts are unique triggers, taking the localization with the highest $p_{\rm assoc}$ per trigger. Only triggers with angular offset $<30^\circ$ and BAT image exposure $<10$\,s are included.}
    \end{deluxetable}


\subsection{Clustering Analysis}\label{sec: clustering}
    Our goal is to identify groups of bursts that are clustered in time and are consistent with originating from a common sky position. The data constitute an inhomogeneous mixture of source populations with potentially different temporal behaviors: some sources emit in clustered episodes, others produce sporadic bursts, and some exhibit both. Moreover, the characteristic durations and recurrence timescales of these emission episodes may differ substantially and may overlap in time (see Fig.~\ref{fig: sgr bursts from prev paper}). The data also contain background events that need not be associated with any astrophysical source (we include triggers with $0.5<p_{\rm astro}<1$). We therefore require a clustering method that can identify clusters without specifying their number or shape a priori, while allowing individual events to remain unassigned as noise.
    
    For this purpose, we use the Density-Based Spatial Clustering of Applications with Noise algorithm (DBSCAN; \citealt{ester_density-based_1996}). DBSCAN is defined by two parameters: a neighborhood scale, $\epsilon$, and a minimum number of points, $N_{\min}$, required to define a locally dense region. A point is classified as a core point if its $\epsilon$-neighborhood contains at least $N_{\min}$ points. Points that do not satisfy this criterion but lie within the $\epsilon$-neighborhood of a core point are classified as border points. Clusters are constructed from connected sets of core points together with their associated border points, while points that are not density-reachable from any core point are classified as noise.

    In our case, the metric used for the neighborhood scale should incorporate both temporal and spatial information. A direct combination of these quantities requires defining characteristic scales for each, for example, $d^2 = (\Delta t/\tau)^2 + (\Delta\theta/\sigma_{\theta})^2$, where $\tau$ and $\sigma_{\theta}$ characterize the relevant temporal and angular scales respectively. However, the bursts in our sample span a wide range of recurrence timescales, while the significance of a given angular separation depends on the distribution of source positions and localization uncertainties. A single choice of temporal and angular scales is therefore not well suited to the full data set.

    To address this, we apply DBSCAN hierarchically in two stages. In the first stage, the clustering is performed solely in time, using the temporal separation between bursts as the distance measure. This step identifies coarse temporal clusters based on temporal proximity alone and therefore allows a single cluster to contain bursts originating from multiple sources. In the second stage, DBSCAN is applied independently to the events within each temporal cluster, where instead of a distance metric, we use a probability distribution derived from the data (discussed below) as the measure. It incorporates both temporal separation and consistency in sky position. Thus, the first stage identifies candidate emission episodes, while the second separates activity within those episodes into subclusters that are consistent with originating from the same source.

    The probability distribution used is the probability that two triggers are associated with the same source. Specifically, we define $d=1-p(T|\Delta t,\Delta\theta)$, where $p(T|\Delta t,\Delta\theta)$ is the estimated probability that triggers with time separation $\Delta t$ and angular offset $\Delta\theta$ originate from the same source. This quantity is constructed analogously to $p_{\rm src}$ introduced in Sec.~\ref{sec: statistical associations}, Eq.~\ref{eq: psource}, and in our work \citep{perera_multi-mission_2026}. We estimate it as
    \begin{equation} \label{eq: pmetric}
        p(T|\Delta t, \Delta\theta) = \frac{n(\Delta t, \Delta\theta | \mathcal{H}_1) - n(\Delta t, \Delta\theta | \mathcal{H}_0)}{n(\Delta t, \Delta\theta | \mathcal{H}_1)},
    \end{equation}
    where $n(\Delta t,\Delta\theta|\mathcal{H}_1)$ denotes the foreground distribution of observed trigger pairs and $n(\Delta t,\Delta\theta|\mathcal{H}_0)$ denotes the corresponding background distribution expected for unrelated triggers. The foreground distribution is constructed by computing the temporal and angular separations between all trigger pairs with $\Delta t\leq3$ days. The background distribution is obtained by shuffling the sky positions of different triggers while preserving their $\Delta t$, thereby removing real spatial associations while retaining the temporal properties of the data. Because the foreground and background samples contain the same number of pairs, their relative normalization is unity, and hence the normalization factor introduced in Eq.~\ref{eq: psource} is $\alpha=1$.

\begin{figure}
        \centering
        \includegraphics[width=1.0\linewidth]{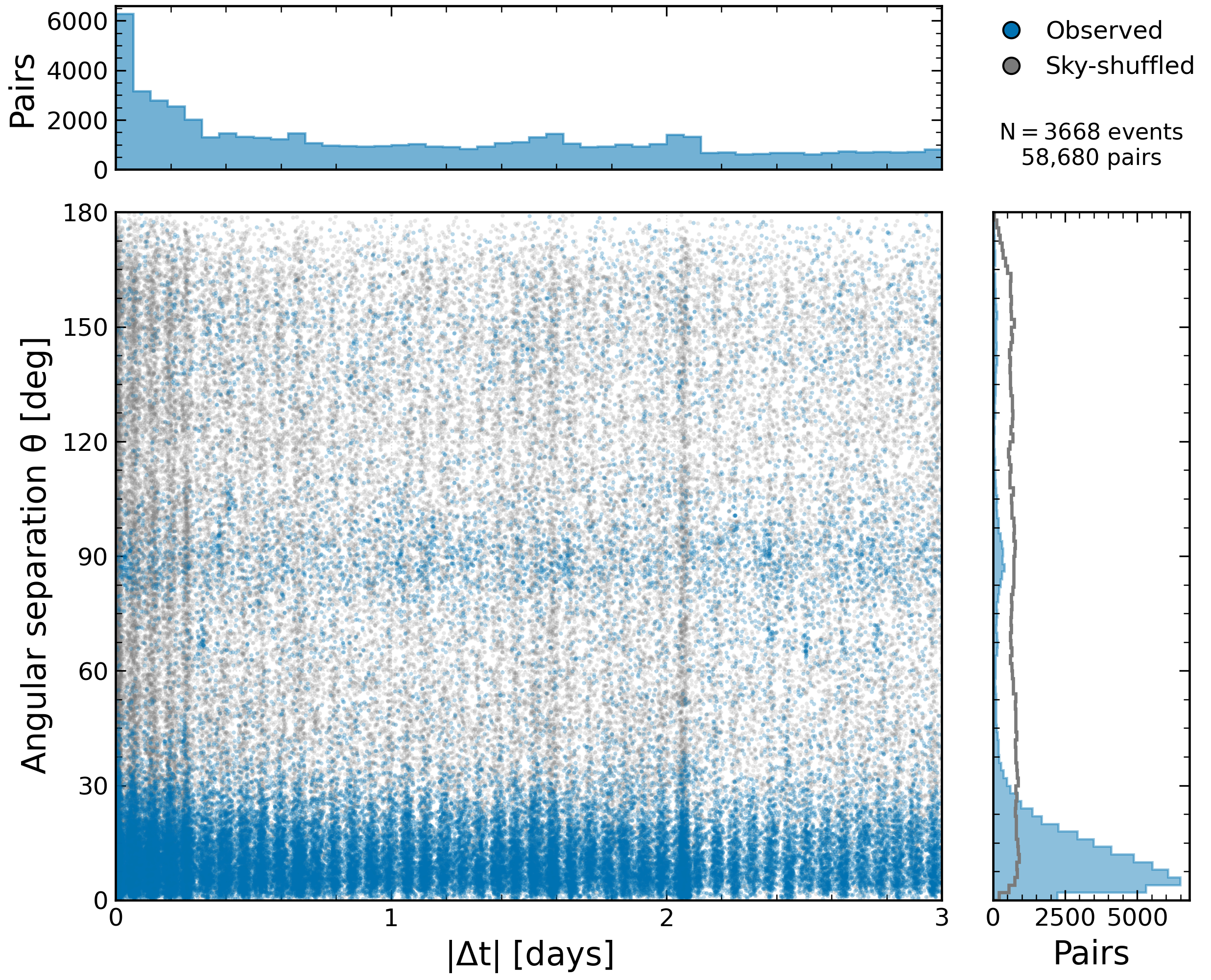}
        \includegraphics[width=1.0\linewidth]{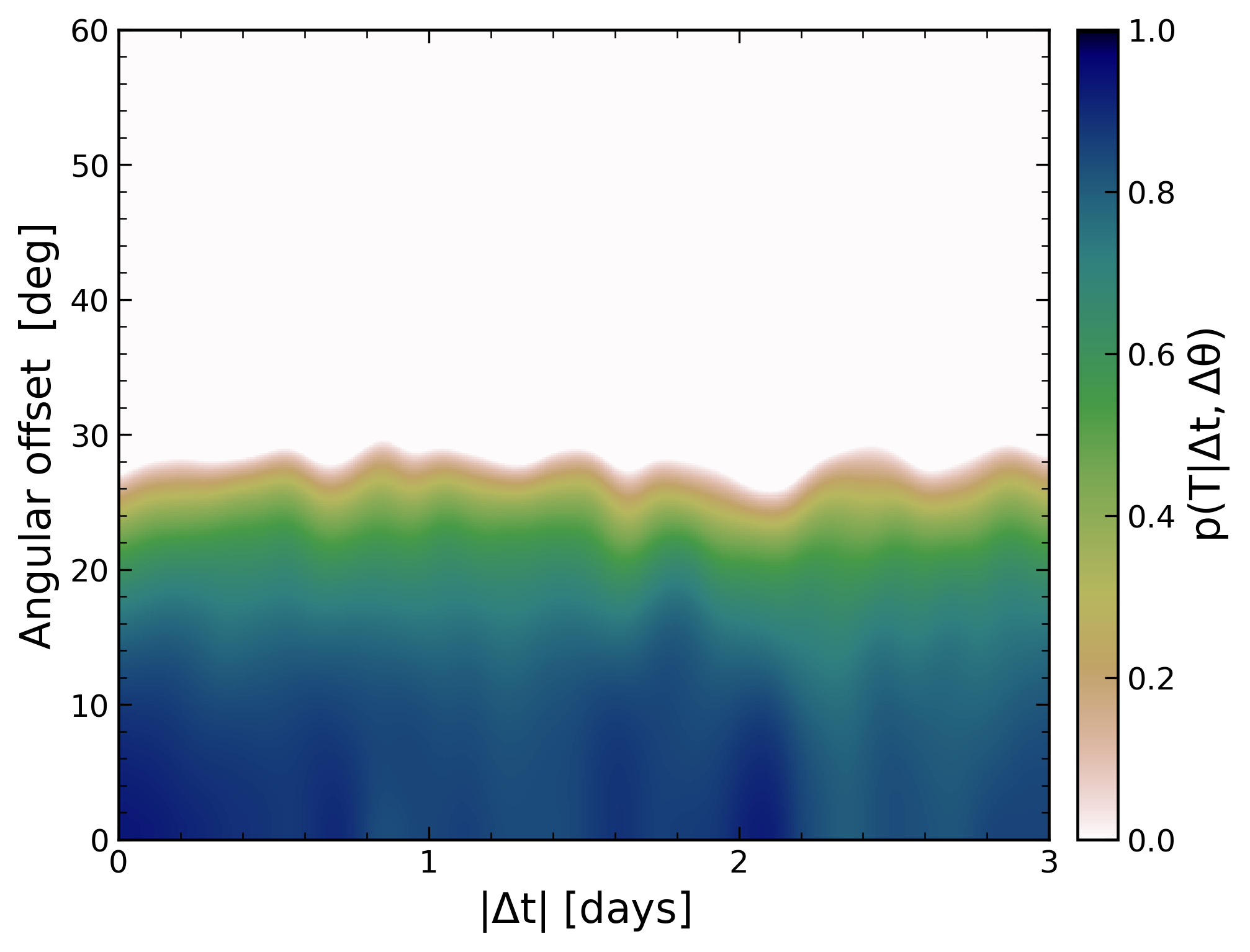}
        \caption{\textbf{Construction of the clustering association probability.} \textit{Top:} Temporal separation, $\Delta t$, versus angular separation, $\Delta\theta$, for all trigger pairs with $\Delta t\leq3$ days, together with their marginal distributions. Blue points and histograms show the observed pairs (foreground), while gray points and histograms show pairs obtained by shuffling the trigger sky positions (background). The excess near $\Delta\theta\sim90^{\circ}$ arises from two active sources that produced numerous bursts over overlapping time intervals; this feature is further investigated in Appendix \ref{appendix: clustering supporting figures}. At angular separations $\Delta\theta\gtrsim30^{\circ}$, the background becomes comparable to or exceeds the foreground, driving the inferred association probability toward zero. \textit{Bottom:} Association probability, $p(T | \Delta t,\Delta\theta)$ (Eq.~\ref{eq: pmetric}), obtained from the foreground and background distributions in the top panel after smoothing and interpolation (see Appendix \ref{appendix: clustering supporting figures}). The association probability depends strongly on angular separation but only weakly on temporal separation over the 3-day maximal separation considered here. This probability is used to define the neighborhood criterion in the second stage of the clustering analysis.}
        \label{fig: metric}
    \end{figure}

    We restrict the construction of $p(T|\Delta t, \Delta\theta)$ to pairs with $\Delta t\leq3$ days. This limit substantially reduces the computational cost of constructing the pair distributions, while extending the calculation to longer temporal separations adds little information. The foreground and background distributions in the $(\Delta t,\Delta\theta)$ plane are shown in the upper panel of Fig.~\ref{fig: metric}, and the resulting association probability is shown in the lower panel. A prominent feature near $\Delta\theta\sim90^\circ$ is produced by two active sources producing a large number of bursts during the same time span (specifically, these are SGR 1935+2154 and Swift J1555.2-5402). This feature disappears when the analysis is restricted to data obtained before 2020 (see Appendix \ref{appendix: clustering supporting figures}). Because the inferred association probability becomes negligible for $\Delta\theta>60^\circ$, we restrict the construction of the second-stage dissimilarity to angular separations below $60^\circ$. 

    The association probability $p(T|\Delta t,\Delta\theta)$ shows relatively little dependence on temporal separation over the 3-day interval considered here. In contrast, it depends strongly on angular separation, with a high association probability for pairs separated by less than $\sim15^{\circ}$. This behavior is similar to that found for the associations with BAT discussed in Sec.~\ref{sec: statistical associations} and reflects the angular uncertainties of our GBM localizations.

    When applying the two-stage DBSCAN procedure described above, we use $N_{\min}=3$ in both stages. This choice represents the smallest cluster size that provides some robustness against chance pairwise associations; that is, we require at least three mutually connected events rather than allowing pairs alone to define a cluster. In the first stage, we set the temporal neighborhood scale to $\epsilon_t=1$ day. This choice is motivated by the characteristic timescales of magnetar bursting activity and is sufficiently broad to group bursts belonging to the same episode of enhanced activity, while limiting the probability of grouping three or more unrelated events within a single temporal cluster.
    
\begin{figure}
        \centering
        \includegraphics[width=1.0\linewidth]{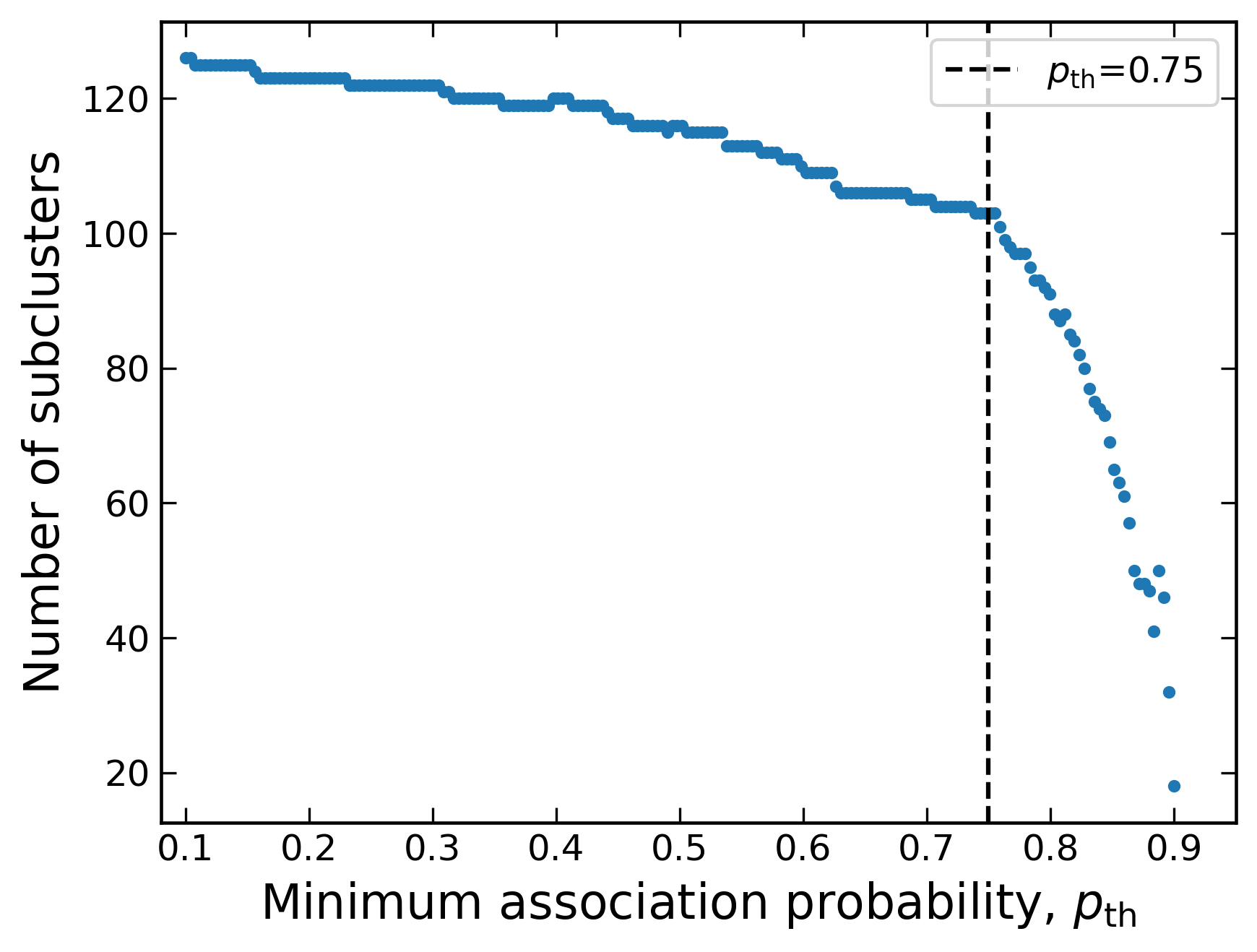}
        \caption{\textbf{Number of subclusters selection}. The figure shows the number of subclusters identified by the second-stage DBSCAN analysis as a function of the minimum association probability, $p_{\rm th}$, required for two events to be considered neighbors. The number of recovered subclusters varies only gradually at lower thresholds but decreases rapidly for $p_{\rm th}\gtrsim0.75$ as the association criterion becomes increasingly restrictive. The vertical dashed line marks the adopted value, $p_{\rm th}=0.75$, chosen near the onset of this decline.}
        \label{fig: nclusts}
    \end{figure}

    In the second stage, the DBSCAN neighborhood criterion is defined in terms of the association probability introduced above. We consider two events to be neighbors when $p(T|\Delta t,\Delta\theta)\geq0.75$, again adopting $N_{\min}=3$. To select this threshold, we examine the number of identified subclusters as a function of the minimum association probability (Fig.~\ref{fig: nclusts}). The number of subclusters remains relatively stable over a range of thresholds before declining sharply as the criterion becomes more restrictive. We therefore adopt $p(T|\Delta t,\Delta\theta)=0.75$, near the onset of this decline, as a compromise between retaining a large number of candidate subclusters and requiring a high probability of association among neighboring events. 

\begin{figure*}
        \centering
        \includegraphics[width=1.0\linewidth]{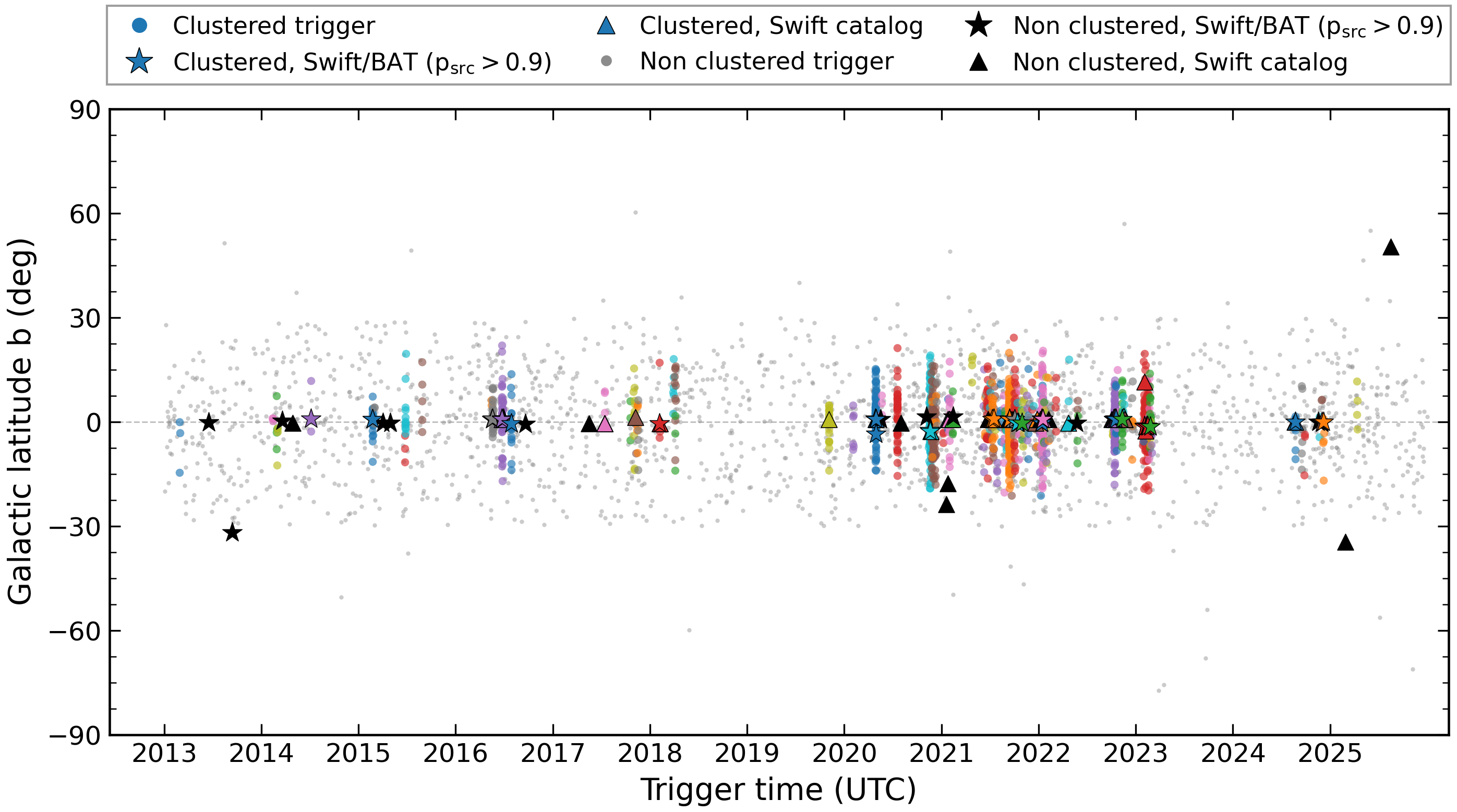}
        \caption{\textbf{Results of the clustering and source-association procedure.} Triggers passing the selection criteria described in Section~\ref{sec:data_selection} are shown as a function of time and sky position after application of the two-stage clustering procedure. Colors distinguish different subclusters. Precise localizations of individual triggers are overlaid when available: stars indicate localizations obtained from BAT associations and triangles indicate localizations taken from the GBM catalog when a Swift/BAT or XRT association is available. Black markers denote precisely localized triggers that are not assigned to any subcluster. Gray points show the remaining unassociated triggers, which may include both GRBs and SGR bursts that could not be clustered.} 
    \label{fig: SGR clusters}
    \end{figure*}

    The resulting clusters are illustrated in Fig.~\ref{fig: SGR clusters}, with individual source associations distinguished by color. Precise localizations of individual triggers are overlaid, with marker shapes indicating whether the localization was obtained using BAT associations or taken from an external catalog. Some clusters are difficult to distinguish visually because of the long time span covered by the data, particularly between 2020 and mid-2022, when several sources were simultaneously active. The remaining gray points represent triggers that are not assigned to any identified source cluster and may include both GRBs and unassociated SGR bursts. Precise localizations for these unclustered events are also indicated when available.

    \subsection{Associating bursts to clusters}\label{sec:assoc_bursts_to_clusters}
        The subclusters identified in the second stage of our clustering procedure are constructed to contain bursts consistent with originating from a common source. We therefore use independent precise localizations of individual subcluster members, when available, to associate the entire subcluster with a known source. These localizations include the Swift/BAT localizations obtained from scaled maps and event-data localizations using the statistical framework described in Sec.~\ref{sec: statistical associations}. We additionally use events that appear in the GBM Trigger Catalog\footnote{\url{https://heasarc.gsfc.nasa.gov/w3browse/fermi/fermigtrig.html}} and have an associated Swift/BAT or Swift/XRT localization. If a member of a subcluster can be confidently associated with a known source through one of these localizations, we assign the remaining members of that subcluster to the same source. Precisely localized triggers that are not assigned to any subcluster are retained in the catalog as individual events.
        
        Following this initial assignment, we manually inspect all subclusters to verify the plausibility of the source associations and to identify unassociated or adjacent subclusters that may originate from the same source. As part of this inspection, we combine the localization posteriors of the events within each subcluster by multiplying the individual posteriors and extracting the maximum-likelihood position of the resulting combined posterior. This procedure can substantially reduce the localization region, particularly for subclusters containing many events.
 
\begin{figure*}
            \centering
            \includegraphics[scale=0.65]{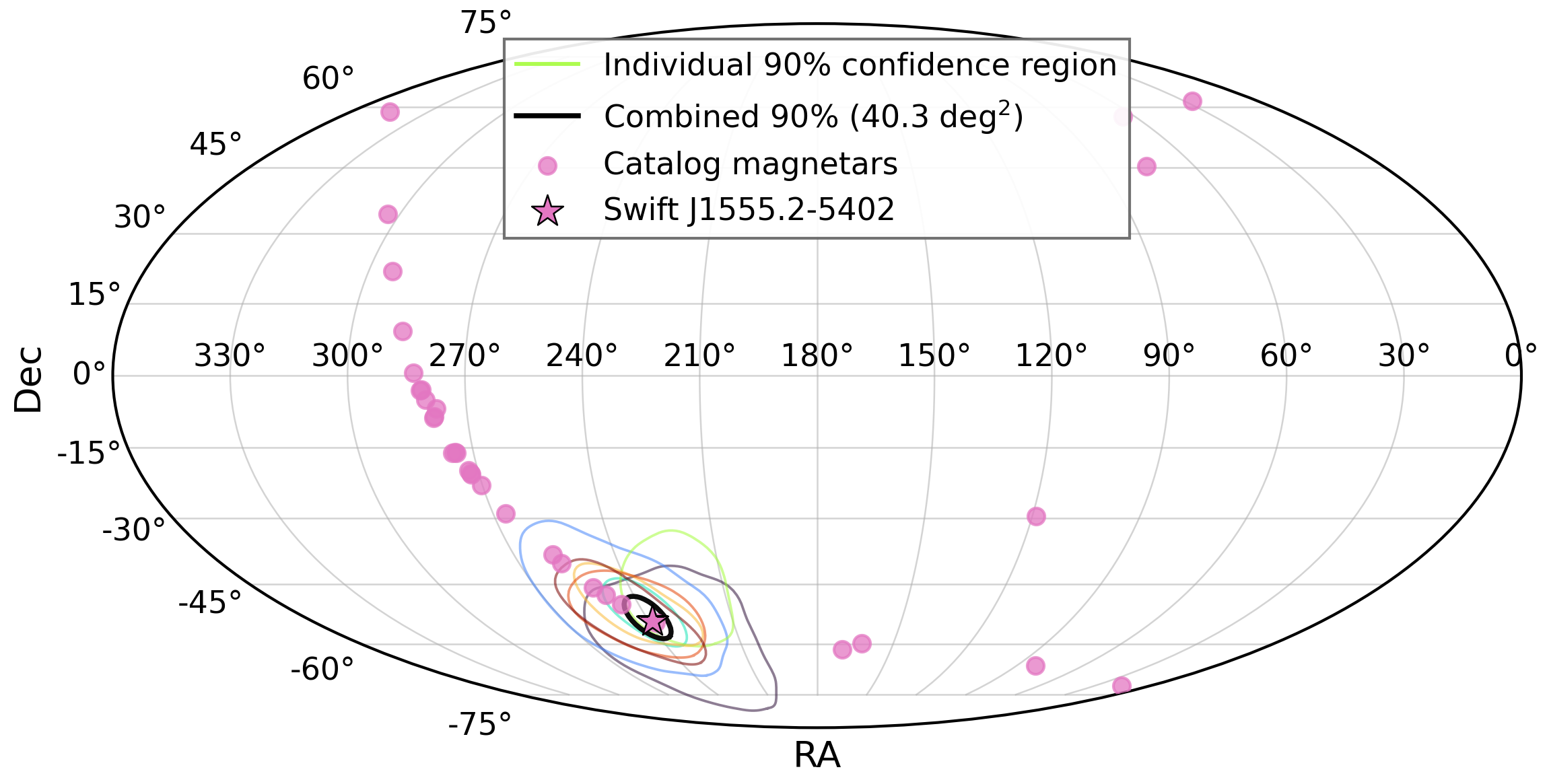}
            \caption{\textbf{Example of a combined localization}. The individual localization posteriors of the bursts in subcluster 57 are multiplied to obtain a combined posterior for the common source position. The contours show the $90\%$ credible regions of the individual bursts and the resulting combined posterior $90\%$ credible region, illustrating the reduction in the localization region obtained by combining multiple events.}
            \label{fig: join contours}
        \end{figure*}

        
        We also use the combined localization information to identify adjacent subclusters that are spatially consistent and likely represent separate episodes of activity from the same source. Such fragmentation can occur when an active source ceases bursting and subsequently resumes after a quiescent interval longer than the temporal scale used in the first clustering stage. The two episodes may then be assigned to separate temporal clusters despite being consistent with the same sky position. When neighboring subclusters have mutually consistent localizations, their combined posterior further supports a common sky position, and their temporal behavior is compatible with distinct episodes of activity from the same source, we merge them into a single source association. A complete list of all subclusters, together with their properties, combined best-estimated positions, and final source associations following this manual inspection, is provided in Appendix~\ref{appendix: all subclusters}, Table~\ref{tab:subcluster_nearest_magnetars}. Associations established or modified through the manual inspection described above are indicated in italics in the table. Fig.~\ref{fig: join contours} illustrates an example of this procedure, showing how combining the individual localization posteriors reduces the $90\%$ credible region and provides a more tightly constrained estimate of the common source position.
        
        A notable example is the extended bursting episode of Swift J1555.2$-$5402, which was divided into multiple subclusters by the temporal clustering procedure. Although most of these subclusters were separated by only a few days, some gaps were as long as approximately 20 days. One subcluster contained a large number of bursts and yielded a well-constrained combined localization consistent with Swift J1555.2$-$5402 (e.g. subcluster 52). We observe that temporally neighboring subclusters (47-52 in this example) are consistent with the same source and therefore interpret them as separate episodes within a single extended period of activity from Swift J1555.2-5402 and associate them with the same source.
        
        
        Conversely, we reject an association when the localization information is inconsistent with a common origin. This occurred for subcluster 99, for which the remaining events were spatially inconsistent with a precisely localized event initially assigned to the same subcluster. We therefore removed the association of the other events and retained only the precisely localized event as an individual source association.
        
        During the manual inspection, we also identified a small number of subclusters associated with bursting X-ray binaries rather than magnetars. Both low and high-mass X-ray binary systems can produce transient emission on timescales of several seconds, and sufficiently bright peaks within these events can trigger our search pipeline. We identify these cases during the manual inspection and remove them from the magnetar burst sample. These non-magnetar subclusters are also indicated in Table \ref{tab:subcluster_nearest_magnetars}.

        After these assignments, mergers, and removals, we identify 1170 bursts that can be associated with 11 different sources. An additional 121 bursts belong to subclusters but cannot be confidently associated with a specific source, and a further 429 bursts pass our selection cuts (Sec.~\ref{sec:data_selection}), have $p_{\rm astro}>0.9$, and are classified as \texttt{SGR3}; these are therefore interpreted as magnetar bursts. This gives a total of 1720 identified magnetar bursts.
        
 The number of bursts associated with each source, together with the corresponding subtotals, is summarized in Table \ref{tab: magnetar_burst_counts}. The bursts that were successfully associated with astrophysical sources and are listed in this table, therefore, constitute our burst catalog.

        \begin{deluxetable}{lccc}
        \tablecaption{The number of bursts found for each known magnetar} \label{tab: magnetar_burst_counts} \tablehead{
            \colhead{Magnetar} & \colhead{R.A.} & \colhead{Decl.} & \colhead{Bursts} \\
            \colhead{} & \colhead{(J2000)} & \colhead{(J2000)} & \colhead{}
            }
            \startdata
            CXOU J010043.1-721134 & 01:00:43 & $-$72:11:34 & 0 \\
            4U 0142+61 & 01:46:22 & 61:45:03 & 12 \\
            SGR 0418+5729 & 04:18:34 & 57:32:23 & 0 \\
            SGR 0501+4516 & 05:01:07 & 45:16:34 & 0 \\
            SGR 0526-66 & 05:26:01 & $-$66:04:36 & 0 \\
            SGR 0755-2933 & 07:55:42 & $-$29:33:49 & 0 \\
            1E 1048.1-5937 & 10:50:07 & $-$59:53:21 & 0 \\
            PSR J1119-6127 & 11:19:14 & $-$61:27:50 & 17 \\
            1E 1547.0-5408 & 15:50:54 & $-$54:18:24 & 1 \\
            Swift J1555.2-5402 & 15:55:09 & $-$54:03:08 & 213 \\
            PSR J1622-4950 & 16:22:45 & $-$49:50:53 & 0 \\
            SGR 1627-41 & 16:35:52 & $-$47:35:23 & 0 \\
            CXOU J164710.2-455216 & 16:47:10 & $-$45:52:17 & 8 \\
            1RXS J170849.0-400910 & 17:08:47 & $-$40:08:52 & 0 \\
            CXOU J171405.7-381031 & 17:14:06 & $-$38:10:31 & 0 \\
            SGR J1745-2900 & 17:45:40 & $-$29:00:30 & 0 \\
            SGR 1801-23 & 18:01:00 & $-$22:56:48 & 0 \\
            SGR 1808-20 & 18:08:11 & $-$20:38:49 & 0 \\
            SGR 1806-20 & 18:08:39 & $-$20:24:40 & 5 \\
            XTE J1810-197 & 18:09:51 & $-$19:43:52 & 0 \\
            Swift J1818.0-1607 & 18:18:04 & $-$16:07:32 & 1 \\
            AX J1818.8-1559 & 18:18:51 & $-$15:59:23 & 0 \\
            Swift J1822.3-1606 & 18:22:18 & $-$16:04:27 & 0 \\
            SGR 1830-0645 & 18:30:42 & $-$06:45:16 & 2 \\
            SGR 1833-0832 & 18:33:44 & $-$08:31:08 & 0 \\
            Swift J1834.9-0846 & 18:34:52 & $-$08:45:56 & 0 \\
            1E 1841-045 & 18:41:19 & $-$04:56:11 & 59 \\
            AX J1845.0-0258 & 18:44:55 & $-$02:56:53 & 0 \\
            PSR J1846-0258 & 18:46:25 & $-$02:58:30 & 1 \\
            3XMM J185246.6+003317 & 18:52:47 & 00:33:18 & 0 \\
            SGR 1900+14 & 19:07:14 & 09:19:20 & 0 \\
            SGR 1935+2154 & 19:34:56 & 21:53:48 & 851 \\
            SGR 2013+34 & 20:13:57 & 34:19:48 & 0 \\
            1E 2259+586 & 23:01:08 & 58:52:44 & 0 \\
            \tableline
            Associated & \nodata & \nodata & 1170 \\
            \tableline
            Unassociated & \nodata & \nodata & 121 \\
            \enddata
            \tablecomments{The Associated row sums only named magnetar associations. Unassociated denotes clustered bursts without a clear magnetar identification.}
            \end{deluxetable}

\section{Identification of Previously Unknown Sources}\label{sec:identifying_unknown_sources}

    \subsection{New Candidates in the BAT image search}

        The association framework described in Section~\ref{sec: statistical associations}, which quantifies the probability that a GBM trigger and a source detected in a BAT image originate from the same astrophysical event, provides a means of localizing faint transients and searching for previously unknown sources. In particular, an associated BAT excess whose localization is inconsistent with any known source may indicate a new transient source.

        Table~\ref{tab:new_magnetar_candidates} shows bursts of candidate new magnetars along with other bursts of similar odds ratios. Four of the events are associated with known magnetars based on membership in temporally dense subclusters that contain more localized bursts. There is also one trigger that occurs in a dense burst storm that localizes to a position inconsistent with the subcluster's associated magnetar, implying false association. This self-consistent validation suggests that our probabilities provide only a lower limit on the actual associations, and hence should be regarded as conservative.

        Finally, three triggers are not associated with any identified subcluster but remain viable candidates for new magnetar sources. Their occurrence ratios show excesses of 43\%, 25\%, and 25\%, respectively.

        
        This interpretation is further supported by the short integration times selected for the scaled maps in two of the three events. In both cases, GBM and BAT independently identified a transient on comparably short timescales of only a few milliseconds. The temporal agreement between the two instruments strengthens the case that the associated excesses arise from the same short-duration astrophysical event rather than from unrelated background fluctuations.

\begin{figure*}
            \centering
            \includegraphics[width=0.49\linewidth]{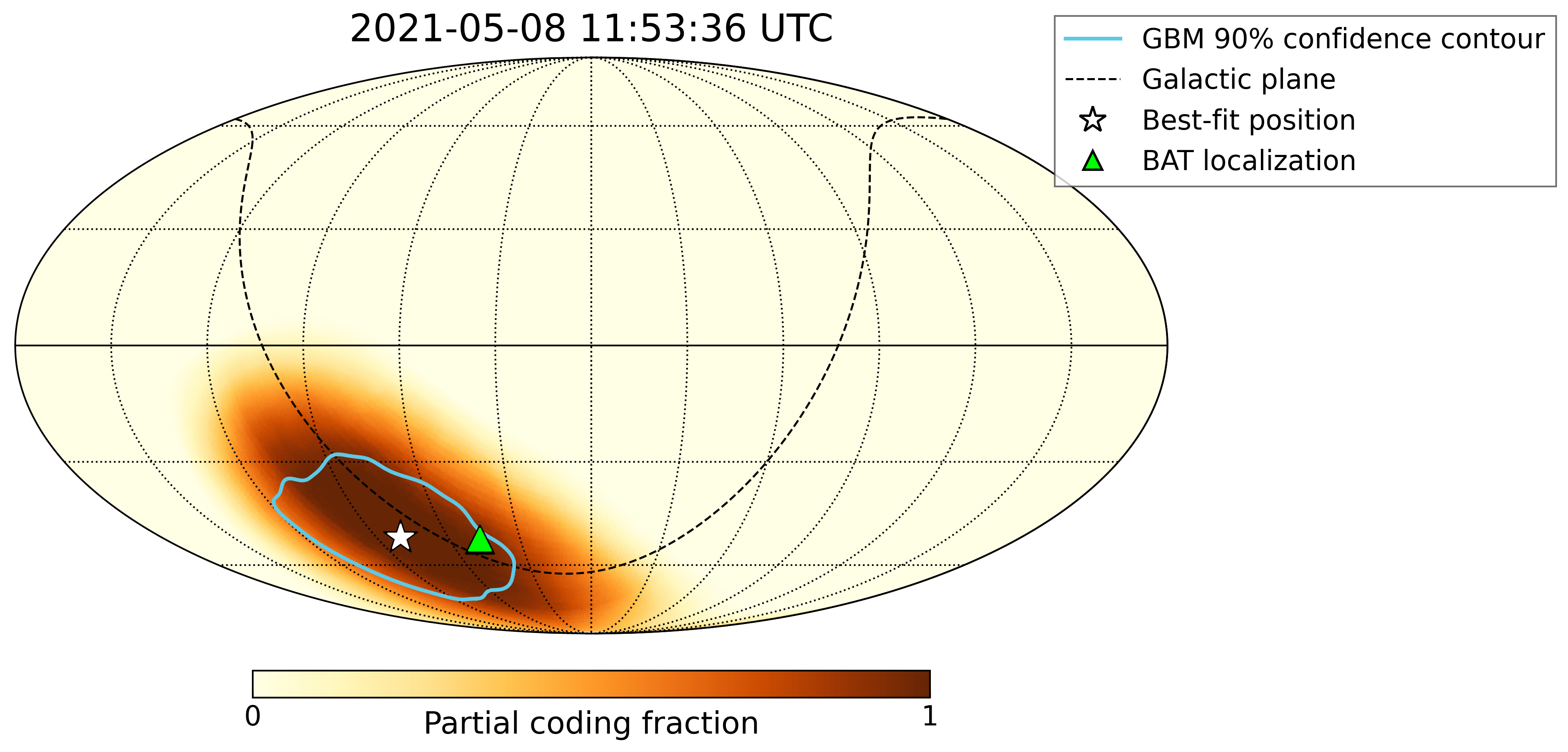}
            \includegraphics[width=0.49\linewidth]{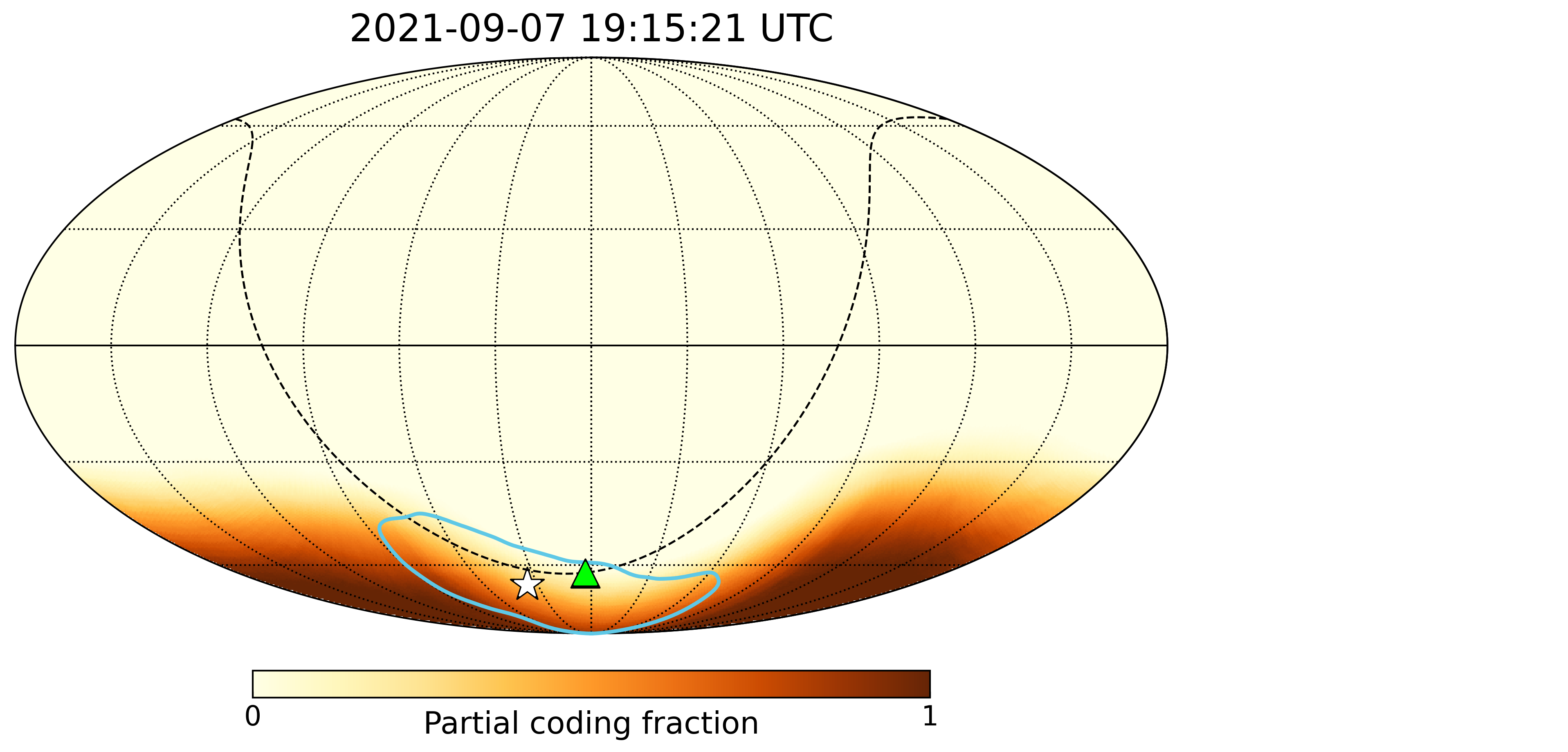}
            \includegraphics[width=0.49\linewidth]{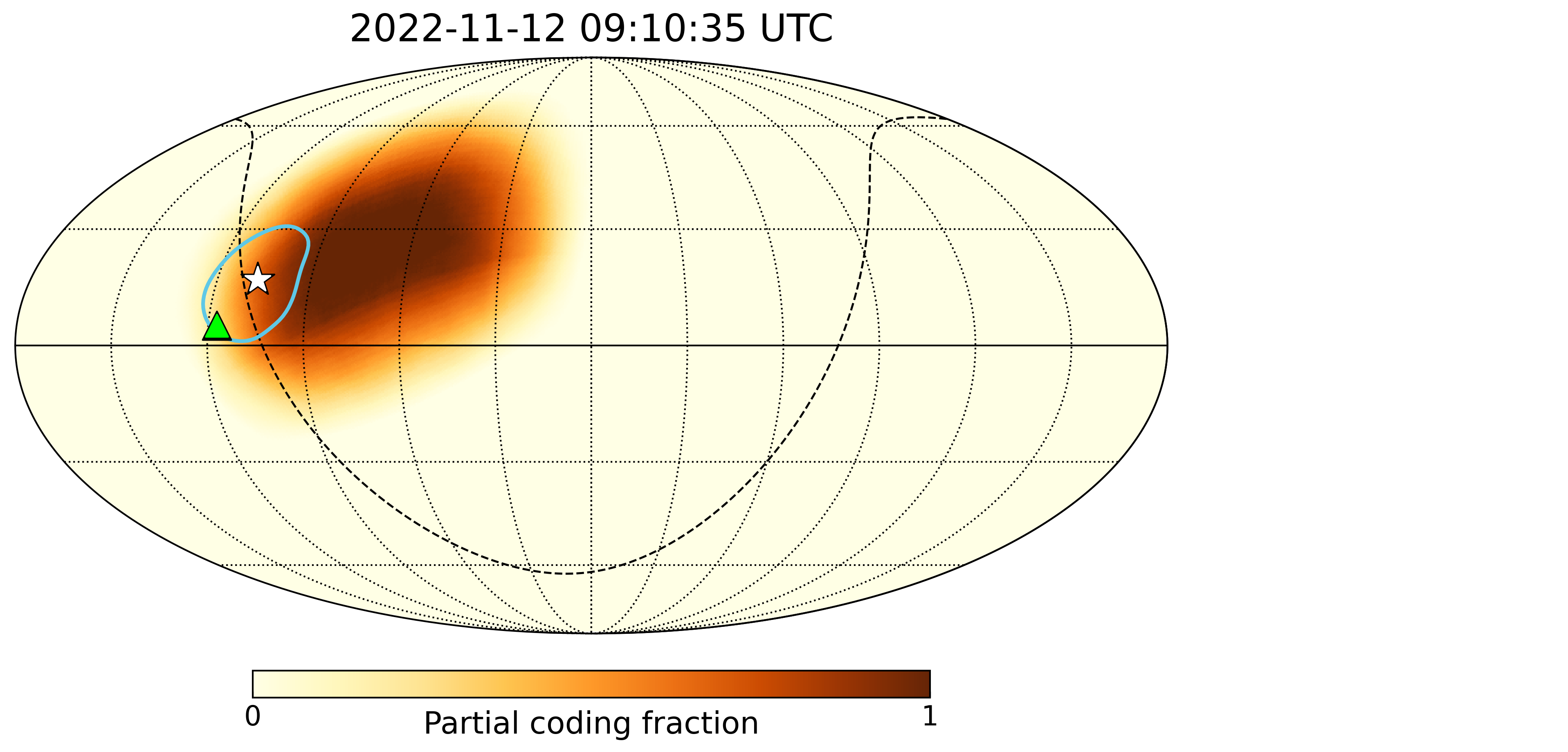}
            \caption{
            \textbf{BAT image localizations of three unassociated SGR-like candidates.}
            The panels show the events on 2021 May 8, 2021 September 7, and 2022 November 12.
            The background shading indicates the BAT partial-coding fraction at the time of each trigger,
            with darker regions corresponding to greater coded exposure.
            Cyan contours show the GBM 90\% posterior credible regions,
            white stars mark the GBM best-fit positions,
            and green triangles indicate the BAT image localizations.
            The dashed black curve denotes the Galactic plane.
            In each case the BAT localization lies within, or is consistent with,
            the broader GBM 90\% region and is not coincident with a known magnetar.
            }
            \label{fig:image_associations}
        \end{figure*}

        Figure~\ref{fig:image_associations} shows the BAT localizations for the three candidates together with the BAT partial coding fraction and the corresponding GBM 90\% confidence regions. In each case, the BAT localization lies within, or is consistent with, the broader GBM localization region.

\begin{deluxetable*}{llcccccccccccccc}
\tabletypesize{\scriptsize}
\tablewidth{\textwidth}
\setlength{\tabcolsep}{2.5pt}
\tablecaption{Image-based new magnetar candidates and bursts with similar odds ratios.\label{tab:new_magnetar_candidates}}
\tablehead{
\colhead{Trigger Time} & \colhead{Magnetar} & \colhead{Subcluster} & \colhead{$p_{\rm src}$} & \colhead{$\mathcal{R}$} & \colhead{$\theta$} & \colhead{$\rho_{\rm img}$} & \colhead{$\rho_{\rm GBM}$} & \colhead{$t_{\rm exp}$} & \colhead{$\Delta t$} & \colhead{$E_{\rm peak}$} & \colhead{BAT R.A.} & \colhead{BAT Decl.} & \colhead{BAT $b$} & \colhead{$f_{\rm code}$} & \colhead{$r_{\rm err}$} \\
\colhead{(UTC)} & \colhead{} & \colhead{} & \colhead{} & \colhead{} & \colhead{(deg)} & \colhead{} & \colhead{} & \colhead{(s)} & \colhead{(s)} & \colhead{(keV)} & \colhead{(deg)} & \colhead{(deg)} & \colhead{(deg)} & \colhead{} & \colhead{(deg)}
}
\startdata
2016-07-28 01:30:02.476 & PSR J1119 & 13 & 0.473 & 1.90 & 10.64 & 5.79 & 7.82 & 0.016 & 0.016 & 24.3 & 169.72 & -61.51 & -0.60 & \nodata & 0.033 \\
\textit{2021-09-12 05:37:34.099} & \textit{\nodata} & \textit{53} & \textit{0.399} & \textit{1.66} & \textit{20.27} & \textit{6.29} & \textit{7.44} & \textit{0.256} & \textit{0.133} & \textit{30.7} & \textit{285.57} & \textit{3.99} & \textit{-0.58} & \textit{0.999} & \textit{0.031} \\
2021-07-04 09:45:01.882 & Swift J1555.2 & \nodata & 0.394 & 1.65 & 22.14 & 6.37 & 10.77 & 0.032 & 0.022 & 54.8 & 238.78 & -54.06 & -0.33 & 0.448 & 0.030 \\
2022-01-14 08:16:01.345 & SGR 1935 & 70 & 0.358 & 1.56 & 8.59 & 5.28 & 11.41 & 0.032 & 0.054 & 27.6 & 293.67 & 21.86 & 0.85 & 0.305 & 0.037 \\
\textbf{2022-11-12 09:10:35.772} & \textbf{\nodata} & \textbf{\nodata} & \textbf{0.302} & \textbf{1.43} & \textbf{16.24} & \textbf{6.04} & \textbf{9.77} & \textbf{1.000} & \textbf{0.022} & \textbf{28.8} & \textbf{297.35} & \textbf{5.01} & \textbf{-10.48} & \textbf{0.274} & \textbf{0.032} \\
\textbf{2021-09-07 19:15:21.078} & \textbf{\nodata} & \textbf{\nodata} & \textbf{0.202} & \textbf{1.25} & \textbf{24.70} & \textbf{6.04} & \textbf{9.04} & \textbf{0.008} & \textbf{0.005} & \textbf{23.2} & \textbf{183.13} & \textbf{-62.67} & \textbf{-0.13} & \textbf{0.053} & \textbf{0.032} \\
\textbf{2021-05-08 11:53:36.859} & \textbf{\nodata} & \textbf{\nodata} & \textbf{0.201} & \textbf{1.25} & \textbf{17.13} & \textbf{5.65} & \textbf{6.88} & \textbf{0.016} & \textbf{0.007} & \textbf{20.1} & \textbf{227.22} & \textbf{-51.94} & \textbf{5.39} & \textbf{0.961} & \textbf{0.034} \\
2014-07-05 09:41:06.149 & SGR 1935 & 4 & 0.145 & 1.17 & 20.16 & 5.49 & 8.24 & 0.040 & 0.040 & 37.6 & 293.76 & 21.97 & 0.83 & 1.0 & 0.035 \\
\enddata
\tablecomments{Rows are ordered by decreasing $p_{\rm src}$. Bold rows have no catalogued magnetar near the BAT image position. The italicized row marks a clear false association. Magnetar is the nearest catalogued magnetar to the BAT image position, truncated before the coordinate sign (empty if none). Subcluster is the subcluster number (see Table~\ref{tab:subcluster_nearest_magnetars}). $p_{\rm src}$ is the probability that the BAT source is associated with the GBM trigger. $\mathcal{R}$ is the corresponding occurrence ratio. $\theta$ is the angular offset between the GBM localization and the BAT image position. $\rho_{\rm img}$ is the BAT image detection SNR. $\rho_{\rm GBM}$ is the GBM search SNR. $t_{\rm exp}$ is the BAT image exposure. $\Delta t$ is the pipeline search timescale. $E_{\rm peak}$ is the Band-function peak energy at the maximum-likelihood sample. BAT R.A., BAT Decl., and BAT $b$ are the BAT image right ascension, declination, and Galactic latitude. $f_{\rm code}$ is the BAT partial-coding fraction at the image position. $r_{\rm err}$ is the BAT image error radius.}
\end{deluxetable*}

    \subsection{New Candidates in the unassociated Subclusters}
        Bursts that form clusters but cannot be associated with a known source can arise for several reasons. First, the cluster may not contain a burst with a precise enough localization to establish an association with a known source. Second, the combined localization of the cluster may be relatively small, but fall in a region of the Galaxy where several known sources are close enough that multiple associations are possible. Conversely, clusters containing only a small number of bursts may have broad combined localization regions, making a significant association difficult. Finally, a cluster may be localized to a region where no known magnetar is present. Such a cluster may therefore represent a previously unidentified magnetar source.

\begin{figure*}
            \centering
            \includegraphics[width=0.49\linewidth]{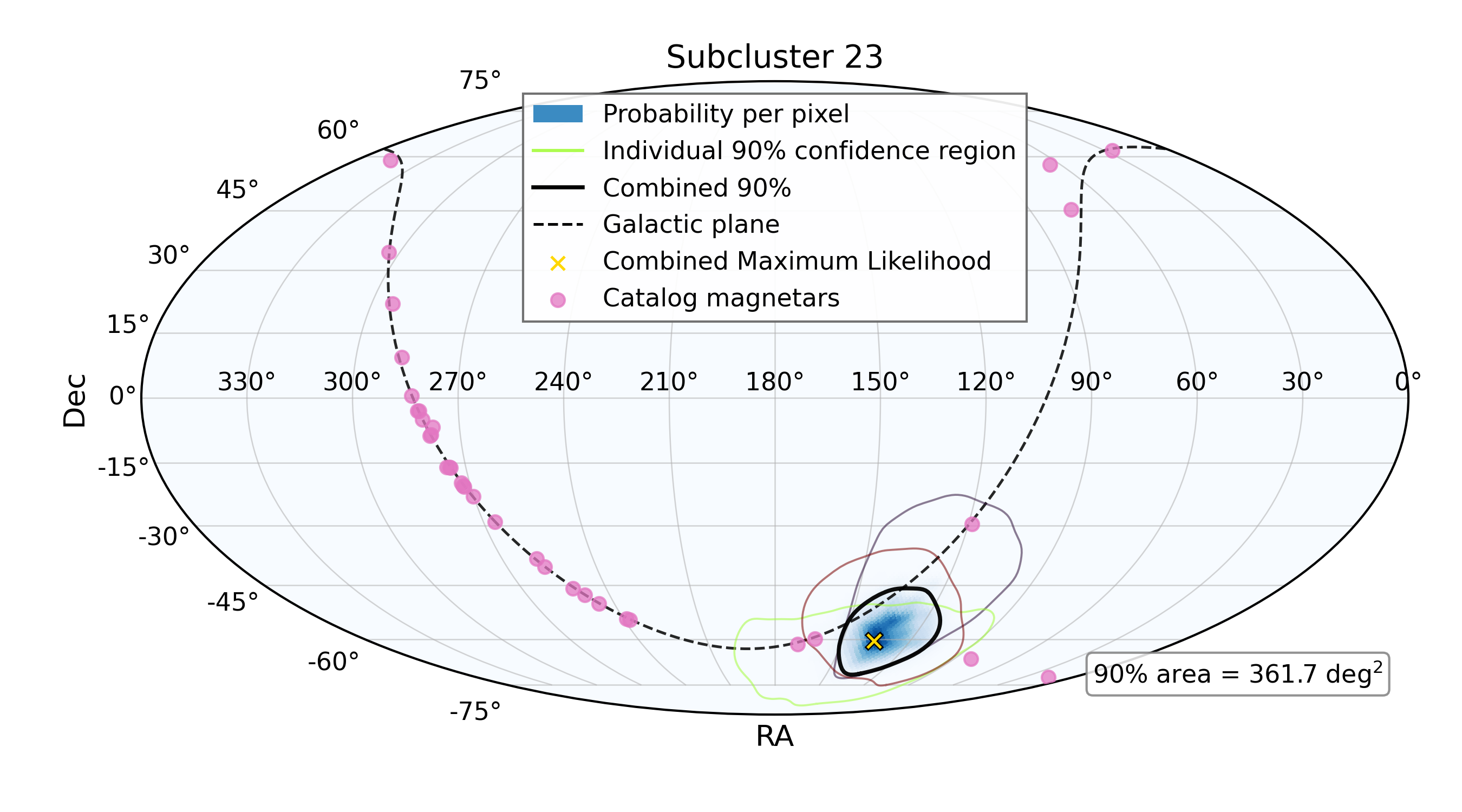}
            \includegraphics[width=0.49\linewidth]{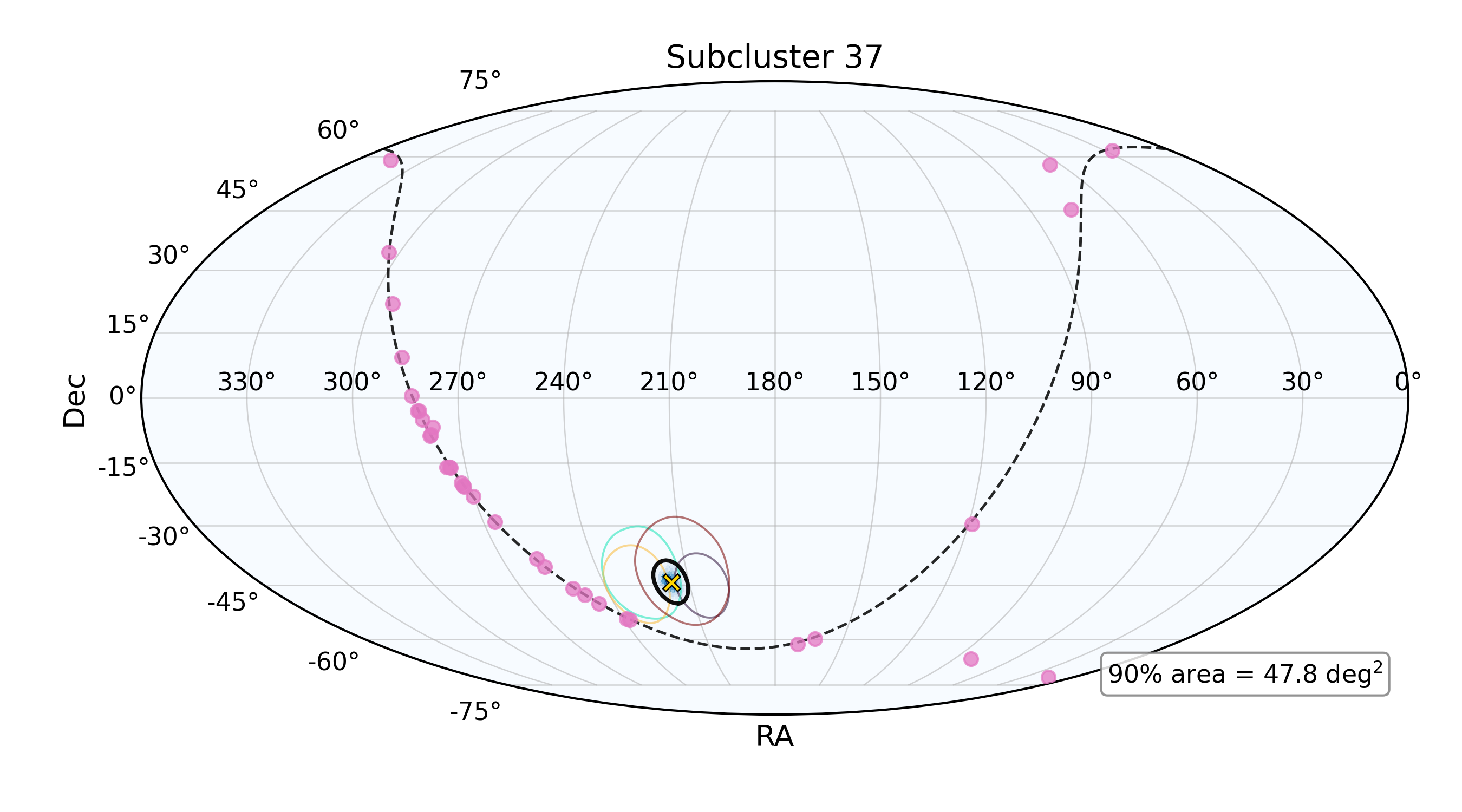}
            \includegraphics[width=0.49\linewidth]{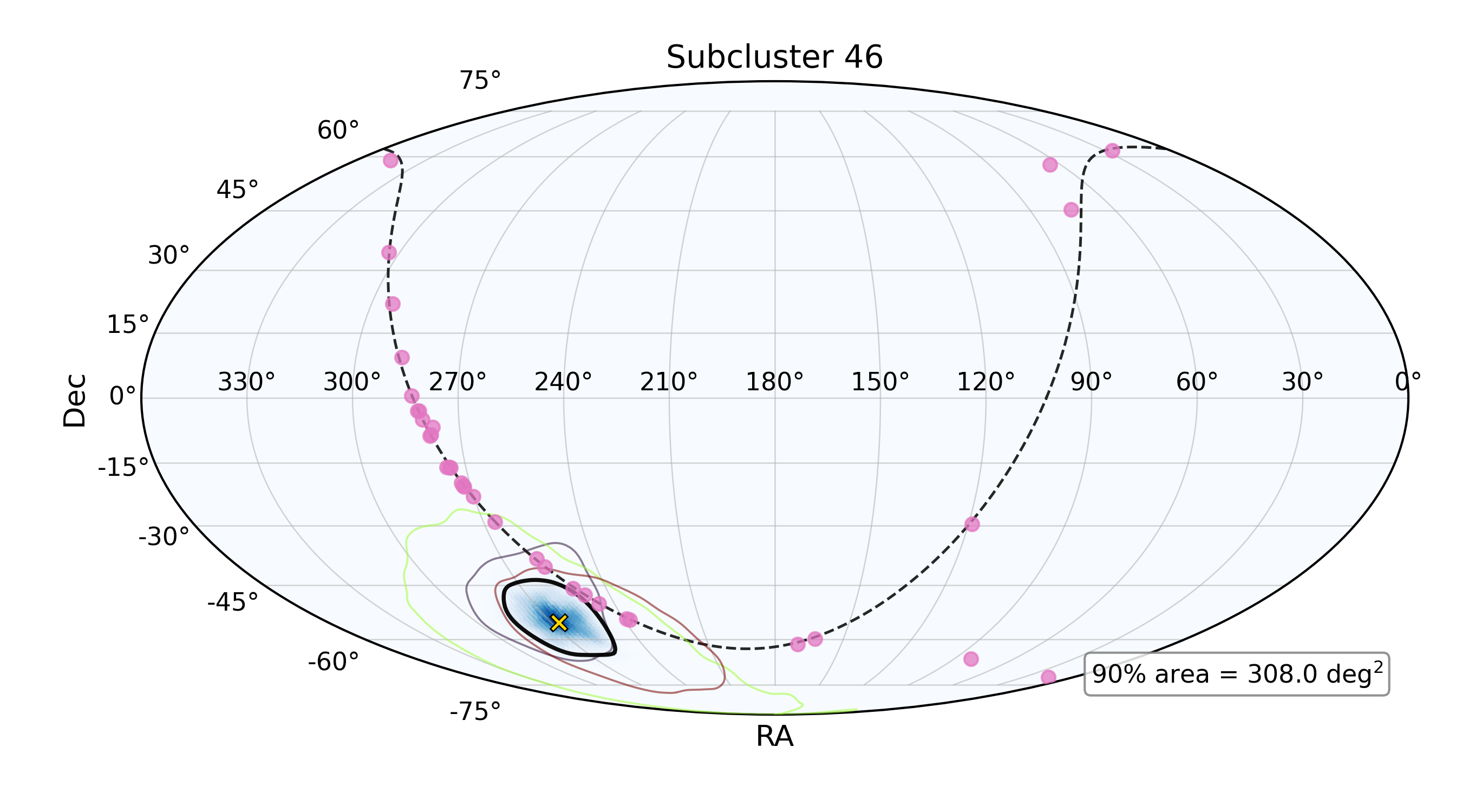}
            \includegraphics[width=0.49\linewidth]{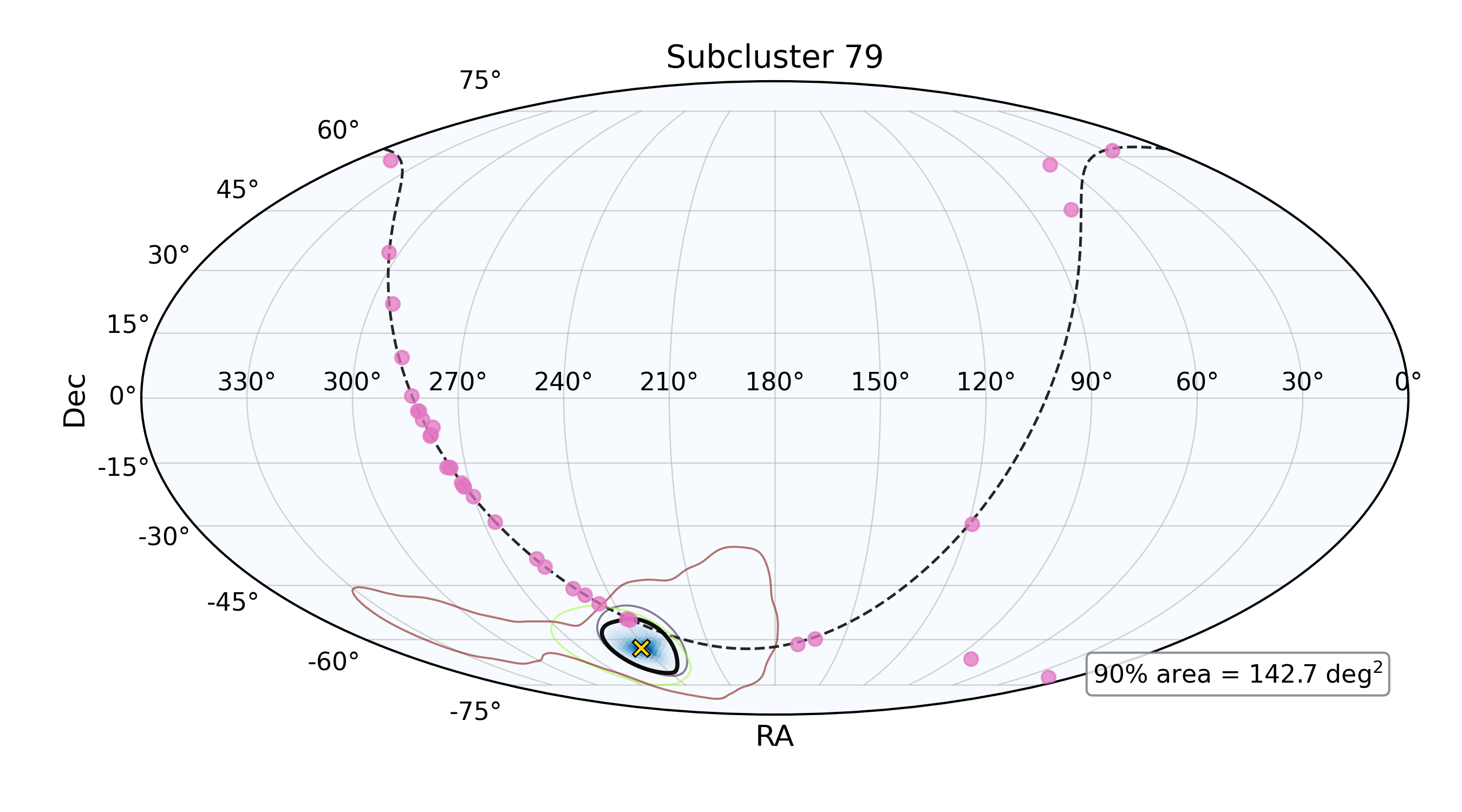}
            \caption{Same as Fig.~\ref{fig: join contours}, but for four unassociated subclusters with no catalog magnetar association (23, 37, 46, and 79). Colored curves show member 90\% GBM contours; the blue map and black contour are the product posterior and its 90\% region; the gold cross is the combined maximum-likelihood position. Magenta points mark known magnetars. Combined 90\% areas are annotated in each panel.}
            \label{fig:new_candidates_subclusters}
        \end{figure*}

        We examined the combined posteriors of all unassociated subclusters to search for previously unidentified magnetar sources. We identify four candidate new magnetars in subclusters 23, 37, 46, and 79. Figure \ref{fig:new_candidates_subclusters} shows the individual 90\% confidence regions of the bursts in each subcluster, together with the combined posterior confidence region. All the subcluster properties are summarized in Appendix~\ref{appendix: all subclusters}, Table~\ref{tab:subcluster_nearest_magnetars}. We note, however, that we do not have a false alarm rate on the occurrence of clusters and hence although we suggest these four candidates, they should be followed up and examined carefully before claiming new detections.
        
        Of the four candidates, one has a significant fraction of its combined posterior within the Galactic plane and in a region without any known cataloged magnetar, making it our most promising candidate for a new magnetar source using this discovery channel. Unfortunately, the relatively large localization region makes confirmation and follow-up challenging. The other three candidates are offset from the Galactic plane. Although Galactic magnetars outside the plane are not unprecedented, their locations make a magnetar interpretation less compelling. These candidates also suffer from the same limitation of broad localization regions, making confirmation and follow-up difficult.

        Nevertheless, the evidence from their clustered GBM bursts makes them interesting candidates that deserve further analysis and follow-up.

    \subsection{Unclustered Bursts}
        The unclustered bursts, shown as the background events in
        Fig.~\ref{fig: SGR clusters}, may also contain evidence for previously
        unidentified sources if multiple otherwise unrelated bursts
        localize to the same region of the sky. To investigate this possibility, we
        selected unclustered triggers whose classification contained \texttt{SGR} and
        examined their sky localizations for several thresholds in
        $p_{\mathrm{astro}}$. The constituent classes and their corresponding SNR thresholds are listed in Table~\ref{tab:pastro_snr}.

        \begin{deluxetable}{lcccccc}
        \tablecaption{SNR thresholds for the $p_{\rm astro}$ cuts used in the unclustered SGR skymaps\label{tab:pastro_snr}}
        \tablehead{
          \colhead{} & \multicolumn{2}{c}{$p_{\rm astro} \geq 0.5$} & \multicolumn{2}{c}{$p_{\rm astro} \geq 0.9$} & \multicolumn{2}{c}{$p_{\rm astro} = 1$} \\
          \colhead{Class} & \colhead{$\rho^2$} & \colhead{$N$} & \colhead{$\rho^2$} & \colhead{$N$} & \colhead{$\rho^2$} & \colhead{$N$}
        }
        \startdata
          SGR1 & 55.4 & 5 & 73.4 & 1 & 91.3 & 1 \\
          SGR2 & 43.1 & 965 & 62.0 & 34 & 137.8 & 11 \\
          SGR3 & 45.5 & 924 & 52.3 & 449 & 70.5 & 255 \\
          SGR4 & 51.3 & 26 & 53.9 & 21 & 57.8 & 18 \\
          (GRB/SGR)2 & 57.0 & 1 & 100.8 & 1 & 128.1 & 1 \\
        \enddata
        \tablecomments{$N$ is the number of unclustered SGR triggers in each class passing the corresponding $p_{\rm astro}$ cut. $\rho^2$ is the class-dependent SNR$^2$ threshold on that $p_{\rm astro}$ curve.}
        \end{deluxetable}
        
\begin{figure*}
    \centering
    \includegraphics[width=0.49\linewidth]{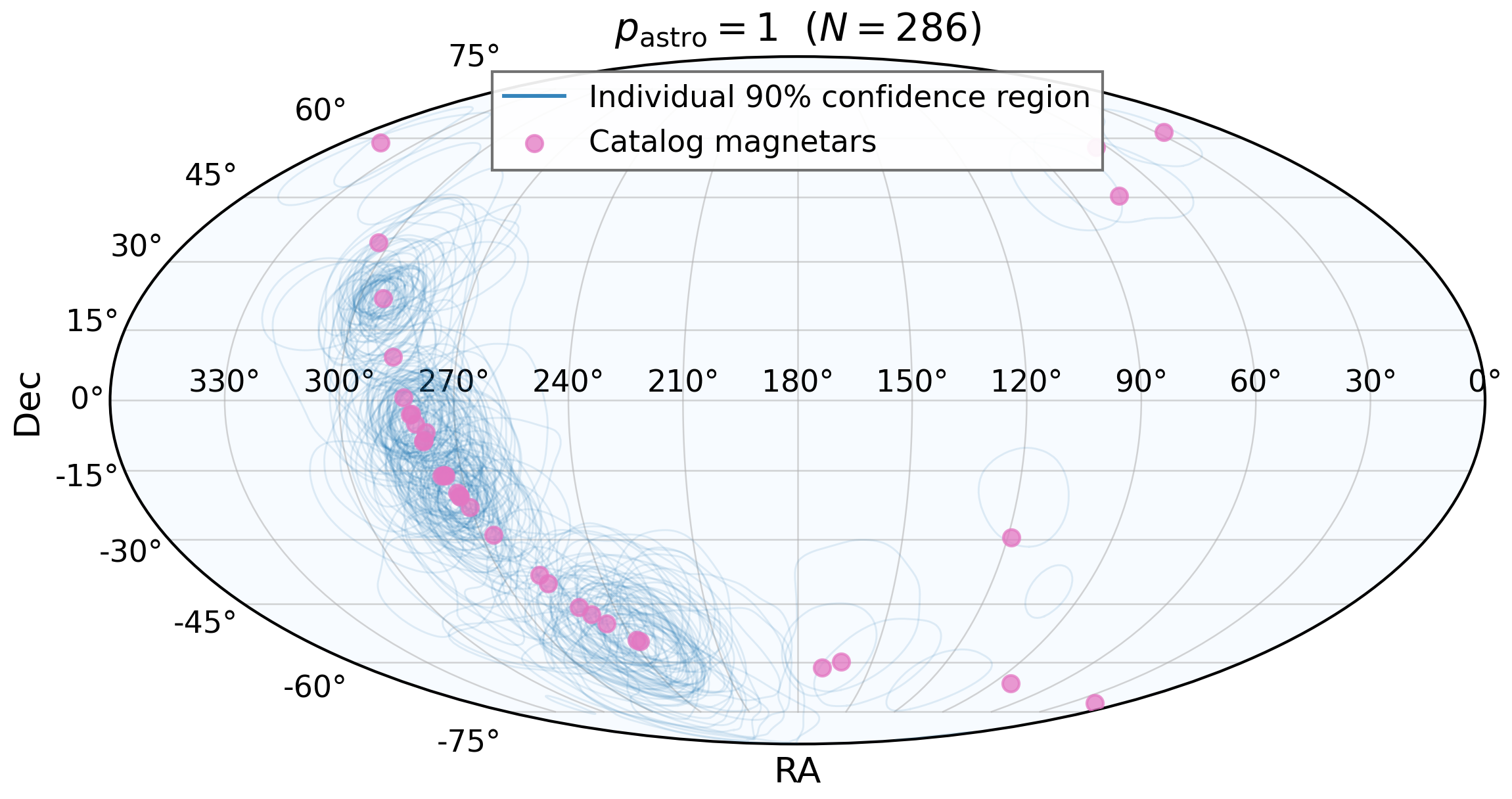}
    \includegraphics[width=0.49\linewidth]{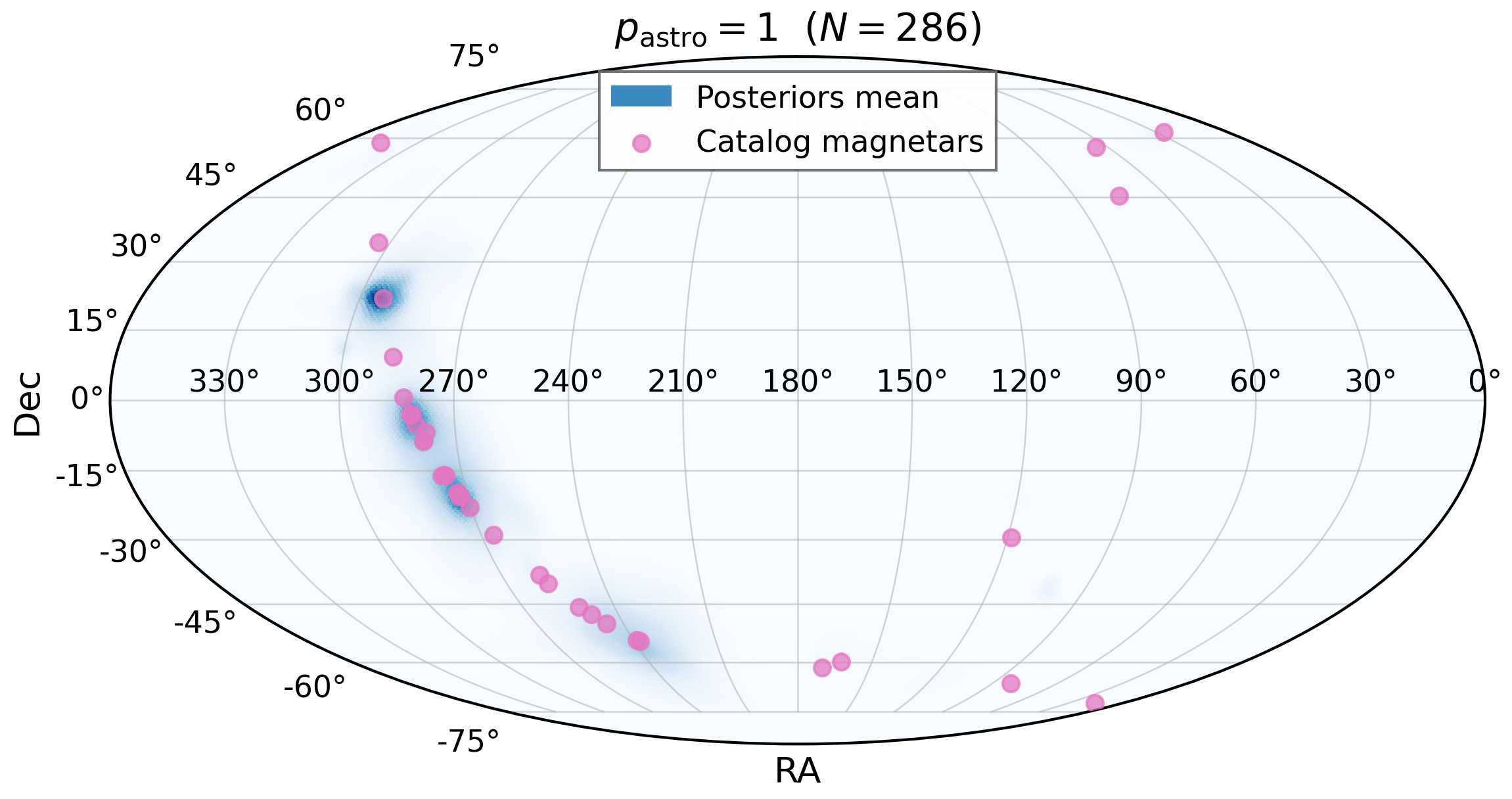}
    \includegraphics[width=0.49\linewidth]{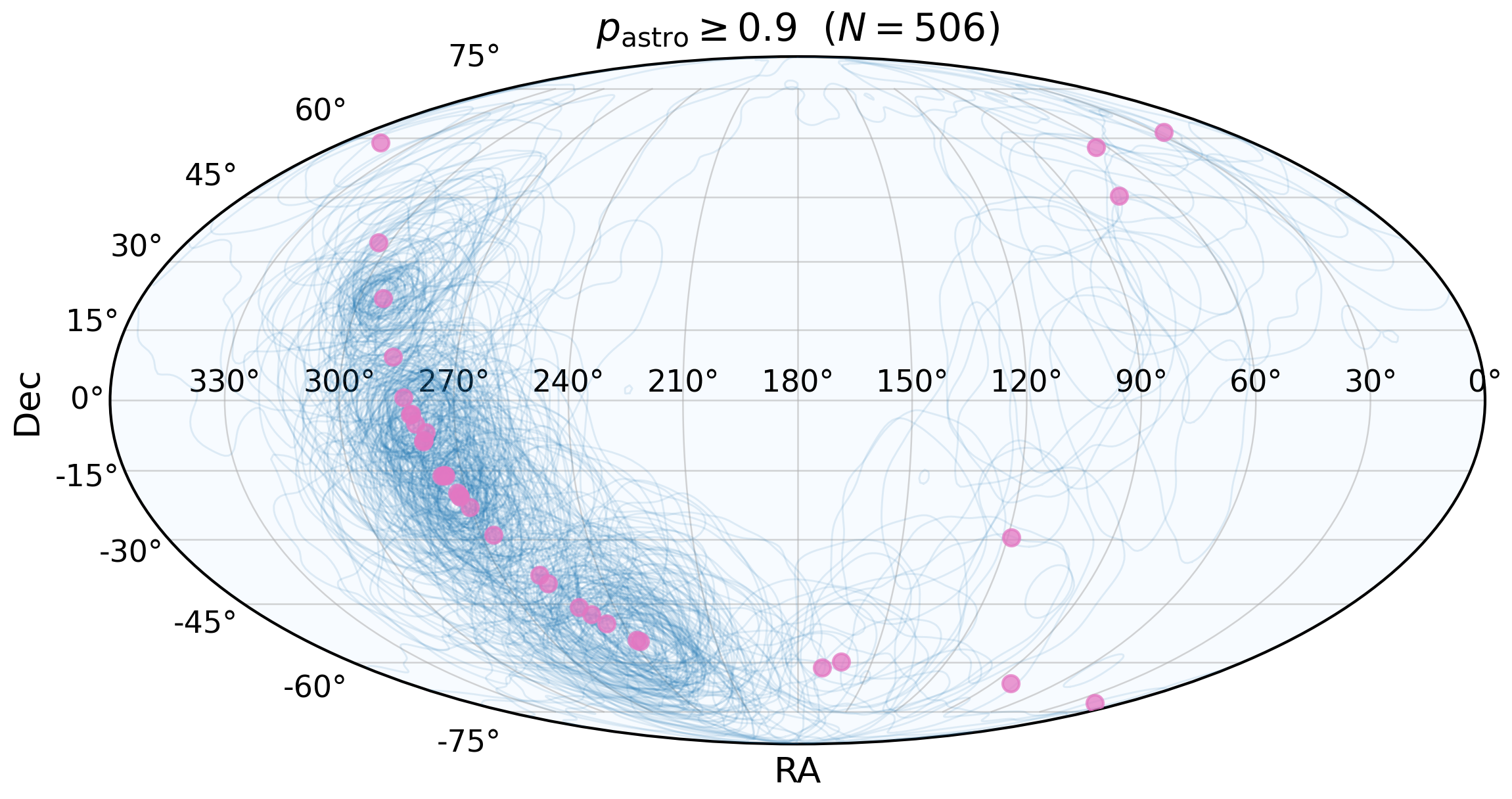}
    \includegraphics[width=0.49\linewidth]{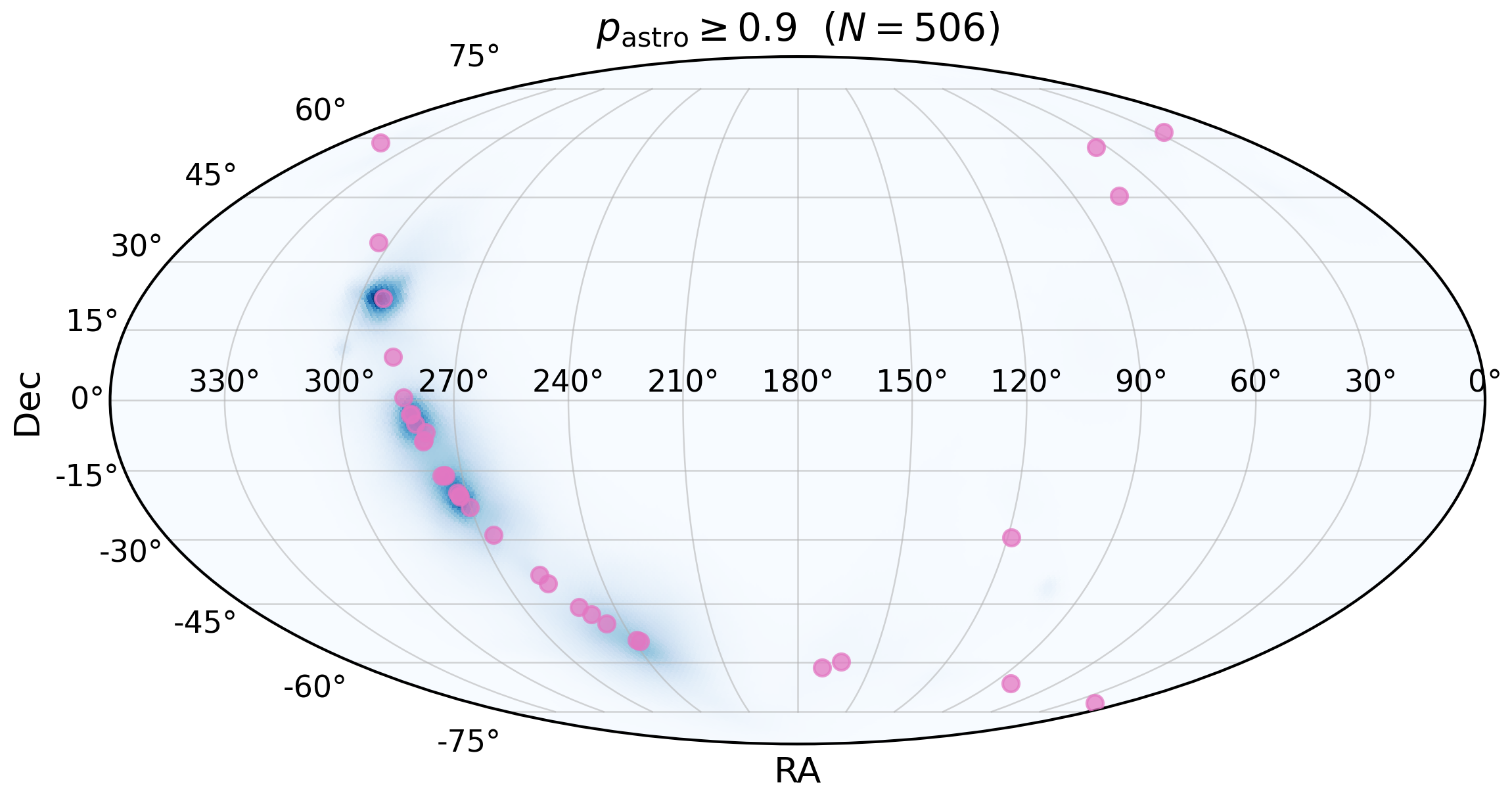}
    \includegraphics[width=0.49\linewidth]{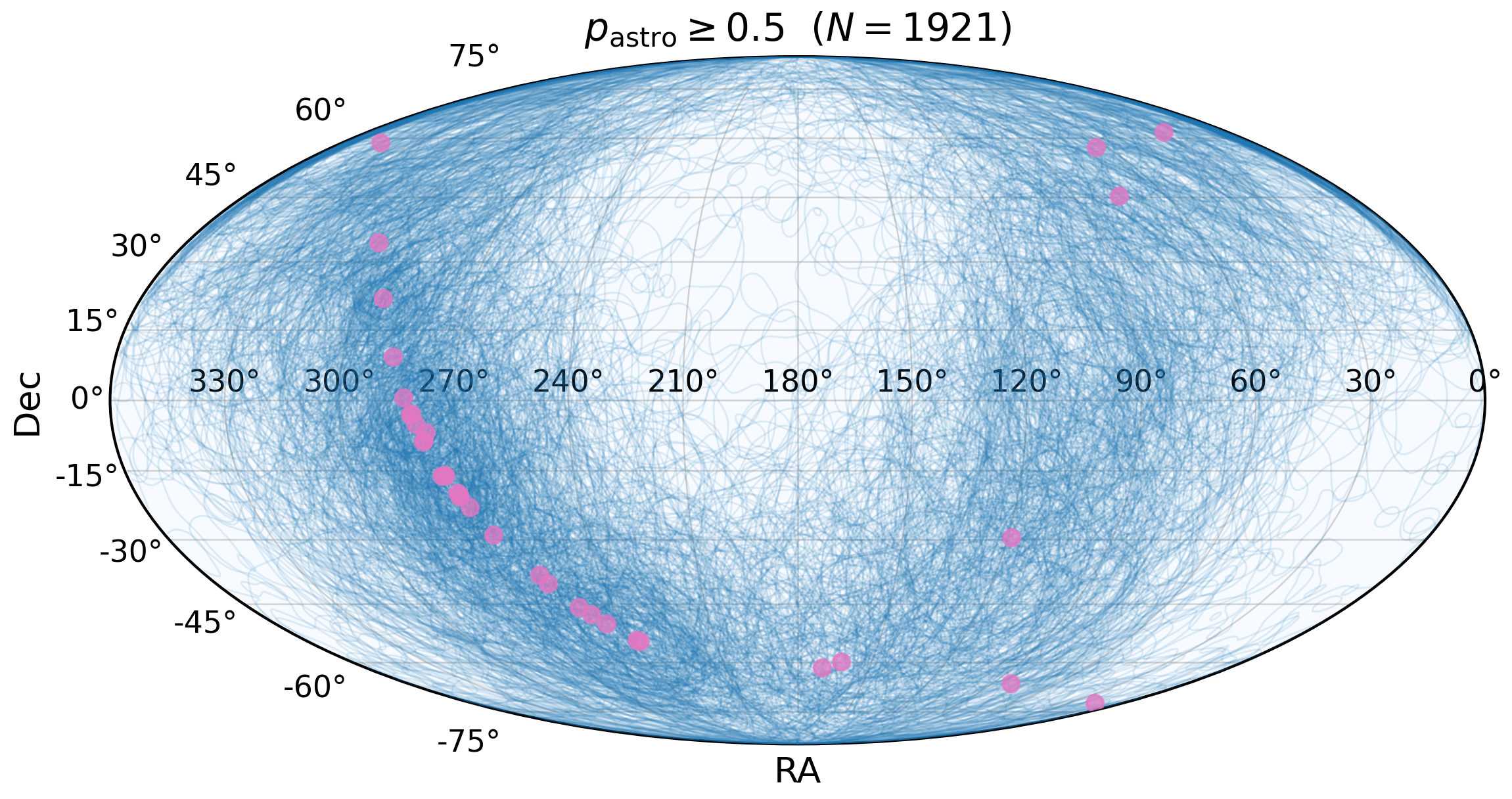}
    \includegraphics[width=0.49\linewidth]{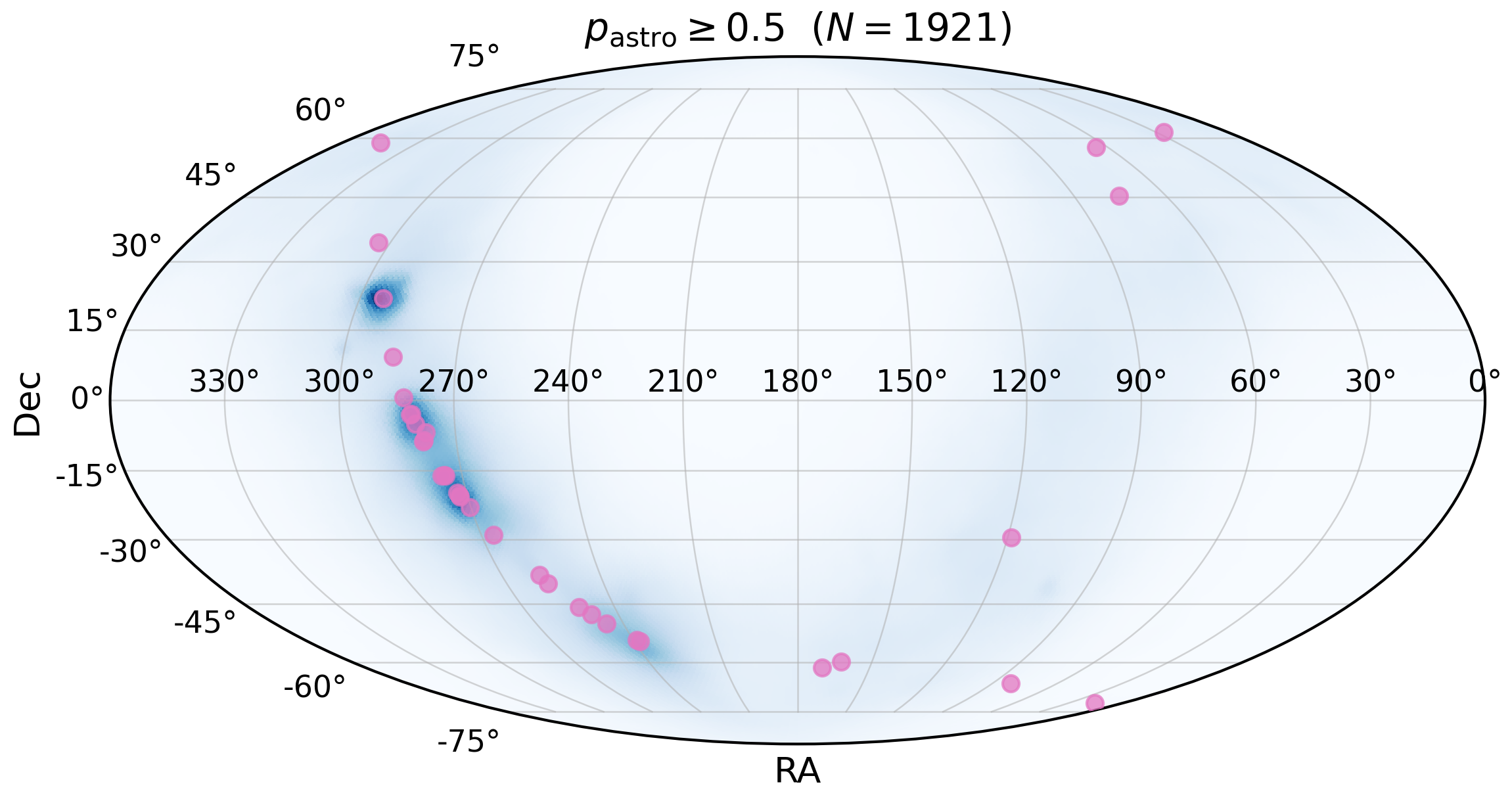}
    \caption{
    \textbf{Sky distribution of unclustered triggers classified as containing an
    \texttt{SGR} component for different $p_{\mathrm{astro}}$ thresholds.}
    From top to bottom, the panels correspond to
    $p_{\mathrm{astro}}=1$, $p_{\mathrm{astro}}\geq0.9$, and
    $p_{\mathrm{astro}}\geq0.5$, containing 286, 506, and 1921 events,
    respectively. Left: individual 90\% confidence regions of the burst
    localizations. Right: mean sky posterior obtained by averaging the
    localization posteriors of all selected events. Magenta circles indicate the
    positions of cataloged magnetars. The stacked posterior shows enhanced
    probability along portions of the Galactic plane, but the broad individual
    localizations prevent the identification of additional resolved sources.
    }
    \label{fig:unclustered_sgr}
\end{figure*}

        The left panels of Fig.~\ref{fig:unclustered_sgr} show the individual 90\% localization regions of the selected bursts. Since no association between individual bursts is assumed, we also combined their sky posteriors by averaging the posterior probability density at each sky position. The resulting maps, shown in the right panels, provide an incoherent stacking of the localization information and highlight regions that are repeatedly favored by different events.
                
        The stacked maps show enhanced probability along portions of the Galactic plane, partly reflecting the selection cuts applied to the \texttt{SGR} classes, including regions that are not centered on cataloged magnetars. However, the individual burst localizations are generally too broad for these enhancements to be resolved into distinct point sources. We therefore do not identify any additional magnetar candidates from the unclustered population using this method. Nevertheless, the concentration of posterior probability along the Galactic plane suggests that some of these events may originate from unresolved Galactic sources.
        
        A substantial fraction of the unclustered bursts are also consistent with the positions of known magnetars, suggesting our trigger list contains more bursts than what was analyzed in our catalog (see Table \ref{tab: localized magnetars} and Sec.~\ref{sec:burst_catalog}).

\section{Burst Catalog}\label{sec:burst_catalog}

    \subsection{Analysis Methods}
    
        All bursts assigned to a particular source in the previous section were subjected to a temporal and spectral analysis. For the temporal analysis, in addition to the search timescale $\Delta$ (the duration of the boxcar template that yielded the highest signal-to-noise ratio; see Sec.~\ref{sec: detection}), we also measured $T_{90}$ so that our durations can be compared with previous magnetar catalogs. $T_{90}$ is the interval over which $5\%$-$95\%$ of the burst fluence accumulate \citep{kouveliotou_identification_1993}. These durations are defined on the deconvolved energy fluence. The fluence is calculated over the $8-200$ keV range.
        
        To isolate each burst we extracted a window extending $\max(20\Delta t,\,2\,\mathrm{s})$ before and $\max(30\Delta t,\,3\,\mathrm{s})$ after the trigger time and applied the Bayesian Blocks algorithm \citep{scargle_studies_1998,scargle_studies_2013} to the count light curve to identify segments of locally constant rate. For each block, we computed $\mathrm{SNR}=(N_{\rm obs}-N_{\rm bkg})/\sqrt{N_{\rm bkg}}$ and grouped consecutive blocks exceeding $3\sigma$. The contiguous group that contains the trigger or the nearest group, provided it lies within $\max(2\Delta t,\,0.1\,\mathrm{s})$ (so that
        neighboring events during SGR burst storms do not contaminate the fluence) is selected.
        This interval is then padded by $25\%$ of its length (or five time bins, whichever is larger) on each side to form the accumulation window.
        
        Within that window, we built a deconvolved light curve by folding the
        best-fit spectral model through the detector response (of the sky-position to the known source, generated for each burst at the time of the event) and taking the per-bin matched-filter amplitude (the maximum-likelihood number of source photons of that spectrum, defined in \citealt{perera_new_2025}). The same accumulation yields the $8$--$200\,\mathrm{keV}$ energy fluence. Uncertainties are estimated with a bootstrap that resamples the matched-filter amplitude noise.
    
        Because the inferred deconvolved fluence depends on the assumed photon spectrum, each burst is fitted with six spectral models commonly used to describe magnetar bursts \citep{collazzi_five_2015,kaspi_magnetars_2017}: a power law (PL), a Comptonized or cutoff power-law model (COMPT), a blackbody (BB), the sum of two blackbodies (BB+BB), optically thin thermal bremsstrahlung (OTTB), and the Band function \citep{band_batse_1993}. The Band function is included primarily for consistency with the coherent search and localization framework, where it is used as the spectral template, rather than because it is generally preferred as a magnetar-burst continuum model. The functional forms of all six models are listed in Table~\ref{tab:spec_models}.
            
        All spectral fits are time-integrated over the accumulation window and
        use the parameter-estimation framework described in
        Section~\ref{sec: pe} and is publicly available online in the \href{https://github.com/PeAriel/grpype}{GitHub repository} \texttt{grpype} \citep{ariel_perera_pearielgrpype_2026}. For each model, the photon spectrum is folded
        through the detector responses of all GBM detectors and energy
        channels. The overall model amplitude is maximized and is
        therefore not sampled in the MCMC. The remaining spectral parameters
        are sampled numerically.
        
        For each burst, we calculate $T_{90}$ and the $8$--$200\,\mathrm{keV}$
        energy fluence separately for every spectral model. The catalog values
        are taken from the preferred model, defined as the model with the
        lowest Bayesian information criterion
        \citep[BIC;][]{schwarz_estimating_1978},
         \begin{equation}
            \mathrm{BIC}
            =
            k\ln n
            -
            2\ln\hat{\mathcal{L}},
        \end{equation}
        where $\hat{\mathcal{L}}$ is the maximum likelihood, $n$ is the number
        of detector-energy bins entering the likelihood, and $k$ is the total
        number of free model parameters. Since the overall amplitude is maximized rather than sampled, we take
        $k=n_{\mathrm{spec}}+1$, where $n_{\mathrm{spec}}$ is the number of
        sampled spectral parameters listed in Table~\ref{tab:spec_models}. We
        also report $\Delta\mathrm{BIC}$ for the alternative spectral models
        fitted to each burst, with higher values representing a less likely model. It should be noted that $\Delta\mathrm{BIC}$ is not itself a log-likelihood ratio, but rather a penalized difference in the maximized log likelihoods. Moreover, it does not generally follow a $\chi^2$ distribution, since the models being compared are non-nested and non-linear. Rather, $\Delta\mathrm{BIC}$ is commonly interpreted as an approximation to Bayesian model comparison \citep{kass_bayes_1995}.

        Posterior samples for all 6 models for every burst in our catalog are available in the Zenodo repository: \href{https://doi.org/10.5281/zenodo.22884088} {10.5281/zenodo.22884088} \citep{perera_catalog_2026}

        \begin{deluxetable}{lccc}
        \tablecaption{Spectral models used in the catalog fits.
        $N(E)$ is the photon spectrum (photons\,cm$^{-2}$\,s$^{-1}$\,keV$^{-1}$);
        $E_{\mathrm{piv}}=100\,\mathrm{keV}$. The amplitude $A$ is maximized. $k$ is the number of free parameters entering the BIC
        ($n_{\mathrm{spec}}+1$).
        \label{tab:spec_models}}
        \tablehead{
        \colhead{Model} &
        \colhead{Photon spectrum $N(E)$} &
        \colhead{Sampled parameters} &
        \colhead{$k$}
        }
        \startdata
        PL &
        $A\,(E/E_{\mathrm{piv}})^{\alpha}$ &
        $\alpha$ &
        2 \\
        COMPT &
        $A\,(E/E_{\mathrm{piv}})^{\alpha}\exp\!\left[-(\alpha+2)E/E_{\mathrm{peak}}\right]$ &
        $\alpha$, $E_{\mathrm{peak}}$ &
        3 \\
        BB &
        $A\,E^{2}/(e^{E/kT}-1)$ &
        $kT$ &
        2 \\
        BB+BB\tablenotemark{a} &
        $A\left[f\,\tilde{B}(E;kT_{\mathrm{c}})+(1-f)\,\tilde{B}(E;kT_{\mathrm{h}})\right]$ &
        $kT_{\mathrm{c}}$, $kT_{\mathrm{h}}$, $f$ &
        4 \\
        OTTB &
        $A\,(E/E_{\mathrm{piv}})^{-1}\exp(-E/kT)$ &
        $kT$ &
        2 \\
        Band &
        $A\,(E/E_{\mathrm{piv}})^{\alpha}\exp\!\left[-(\alpha+2)E/E_{\mathrm{peak}}\right]$,
        $E<E_{\mathrm{c}}$ &
        $\alpha$, $\beta$, $E_{\mathrm{peak}}$ &
        4 \\
         &
        $A\,(E/E_{\mathrm{piv}})^{\beta}e^{\beta-\alpha}
        \left[({\alpha-\beta})E_{\mathrm{peak}}/(E_{\mathrm{piv}}(\alpha+2))\right]^{\alpha-\beta}$,
        $E\ge E_{\mathrm{c}}$ &
         &
         \\
        \enddata
        \tablenotetext{a}{$\tilde{B}(E;kT)$ is a blackbody normalized to unit
        bolometric energy flux; $f$ is the cool component's fraction of the total
        energy flux. The prior enforces $kT_{\mathrm{c}}<kT_{\mathrm{h}}$.}
        \end{deluxetable}

    \subsection{Results and Discussion}
        In the following sections, we present the catalog results for SGR~1935+2154, Swift~J1555.2$-$5402, 1E~1841$-$045,
        PSR~J1119$-$6127, and 4U~0142+61, for which we have a sufficiently large number of bursts to construct meaningful per-source distributions. The full set of burst parameters and derived results for all events in our catalog is publicly available in the Zenodo repository \href{https://doi.org/10.5281/zenodo.22884088} {10.5281/zenodo.22884088} \citep{perera_catalog_2026} and available in a machine-readable form in the online version of this paper. In Appendix \ref{sec:appendix_catalog} we provide a full description of the catalog entries and include an example in Table \ref{tab:catalog_stub}.
    
        \subsubsection{Model Selection} \label{sec:model_selection}

\begin{figure}
                \centering
                \includegraphics[width=1.0\linewidth]{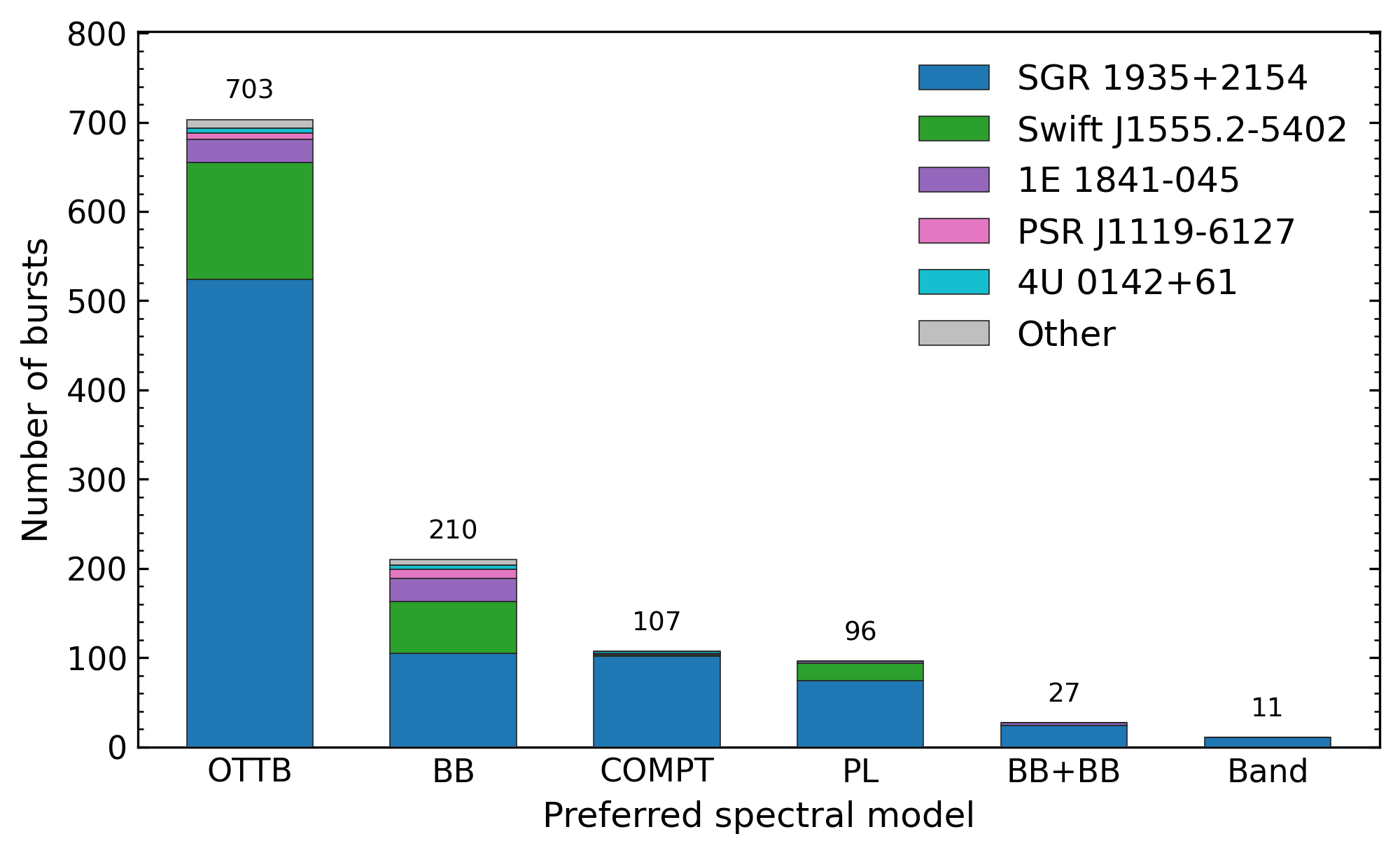}
                \caption{\textbf{Preferred spectral model for the magnetar burst sample.} Each bar shows the number of bursts for which the corresponding model has the lowest BIC, with the bars subdivided by the associated magnetar source.}
                \label{fig:model_preference}
            \end{figure}

        The choice of spectral model affects the burst search sensitivity as well as the inferred fluence and $T_{90}$, since both quantities are calculated from the deconvolved photon spectrum. The distribution of the preferred spectral models is shown in Fig.~\ref{fig:model_preference}. The OTTB model is preferred for 703 of the 1154 bursts in our sample ($\sim61\%$), followed by BB, COMPT, and PL. However, the preferred model does not necessarily imply that it is uniquely favored by the data. The similarity between the several spectral models when convolved through the detector response, and especially when the SNR and photon counts are low, can provide similarly good descriptions of the observed counts. In addition, the Band function converges to COMPT when the high energy spectral index is decreased, which itself is identical to OTTB when the spectral index $\alpha$ equals $-1$.
        
\begin{figure}
                \centering
                \includegraphics[width=1.0\linewidth]{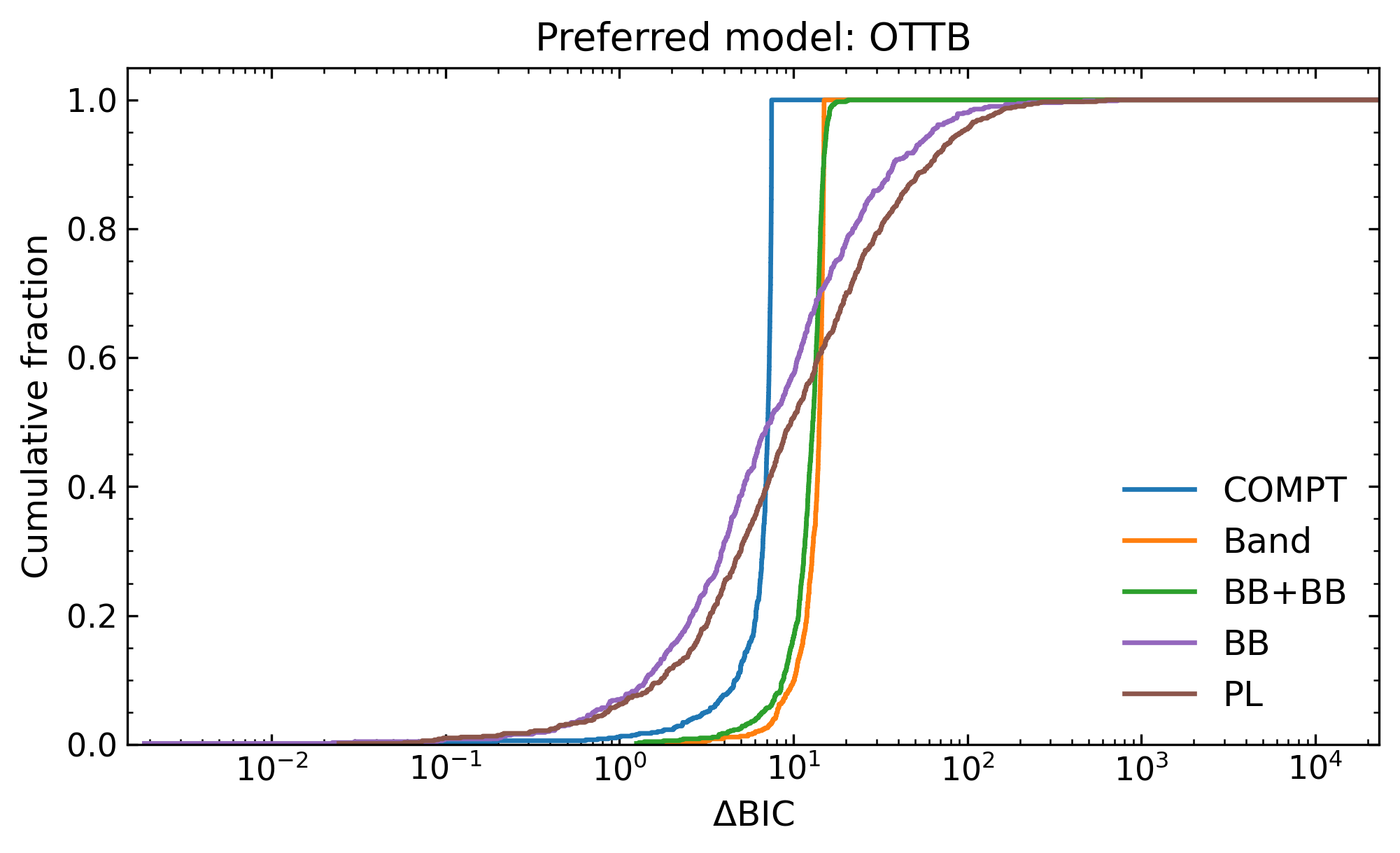}
                \caption{\textbf{Cumulative distribution of $\Delta\mathrm{BIC}$ for the alternative spectral models relative to OTTB, for bursts whose preferred model is OTTB.} For each burst, $\Delta\mathrm{BIC}$ is defined relative to the OTTB BIC, with higher values representing a less likely model. The curves show that many alternative models have small $\Delta\mathrm{BIC}$ and therefore provide fits of comparable quality, illustrating the spectral-model degeneracy of the burst sample.
}
                \label{fig:dbic_ottb}
            \end{figure}

        This degeneracy is illustrated in Fig.~\ref{fig:dbic_ottb}, which shows the cumulative distribution of $\Delta\mathrm{BIC}$ relative to the OTTB model for bursts in which OTTB is the preferred model. A large fraction of the alternative models have small $\Delta\mathrm{BIC}$, indicating that they provide fits of comparable quality to the preferred OTTB model. Similar behavior is found when the other spectral models are preferred; the corresponding distributions are shown in Appendix~\ref{appendix: bics}, Fig.~\ref{fig:rest_of_bics}. Thus, although a single model is selected for reporting the catalog parameters, the data often do not strongly distinguish between several spectral descriptions.
        
        This degeneracy is also relevant to the interpretation of the search template bank. The coherent search uses a broad range of Band-function spectral parameters spanning a large region of spectral parameter space. The fact that the detected bursts can subsequently be described by several different spectral models, whose functional forms can be recovered as limiting cases of the Band function, highlights the sensitivity achieved by our search. In all of the $\Delta\mathrm{BIC}$ distributions presented above, the Band function rises rapidly and saturates at $\Delta\mathrm{BIC}\sim10$. This indicates that, even when Band is not the preferred model, it generally provides a good description of the observed spectra. Since the Band function is used to construct the search template bank, this agreement suggests that our template bank is well-suited for detecting magnetar bursts.

        \subsubsection{Burst Durations}
\begin{figure}
                \centering
                \includegraphics[width=1.0\linewidth]{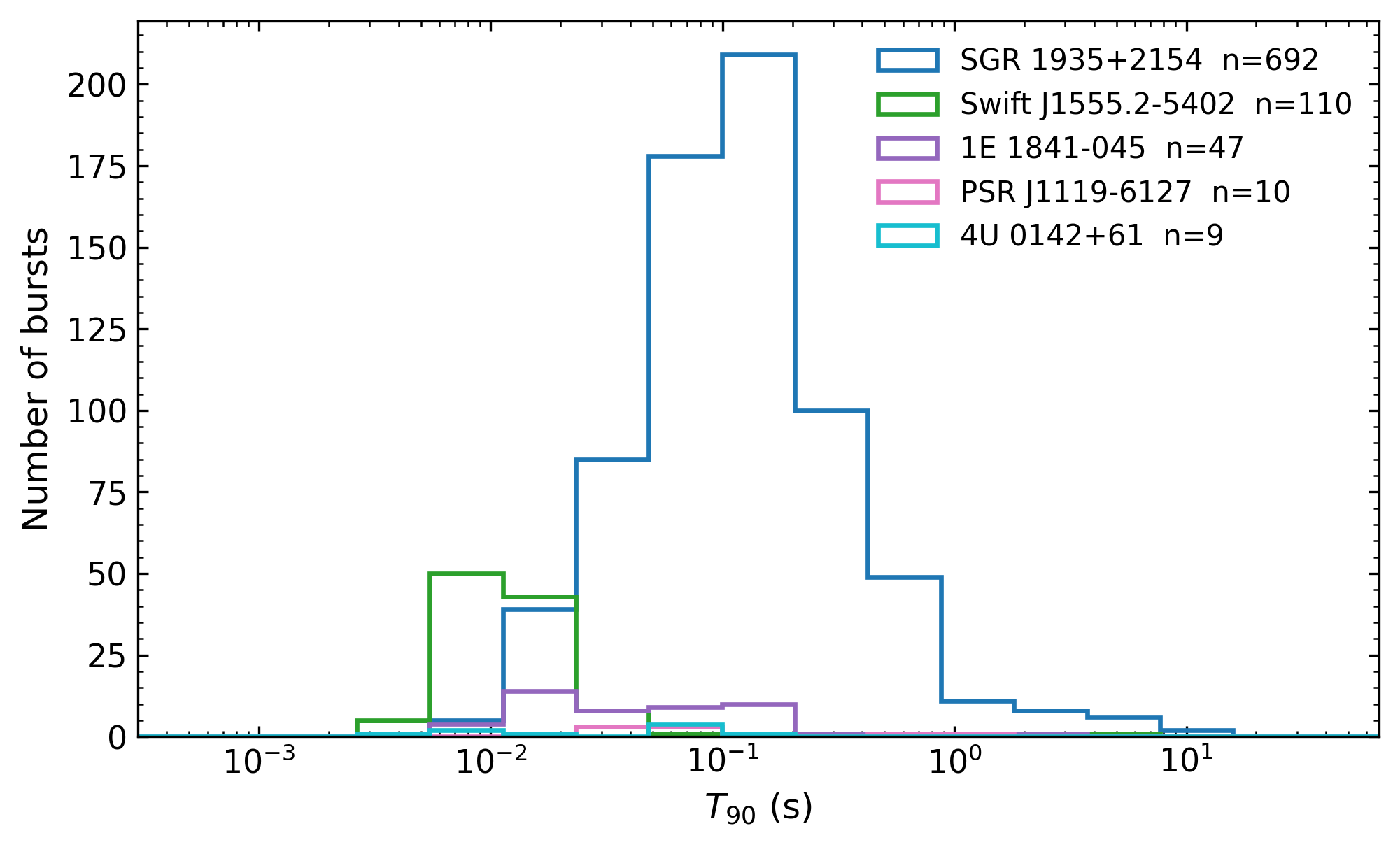}
                \includegraphics[width=1.0\linewidth]{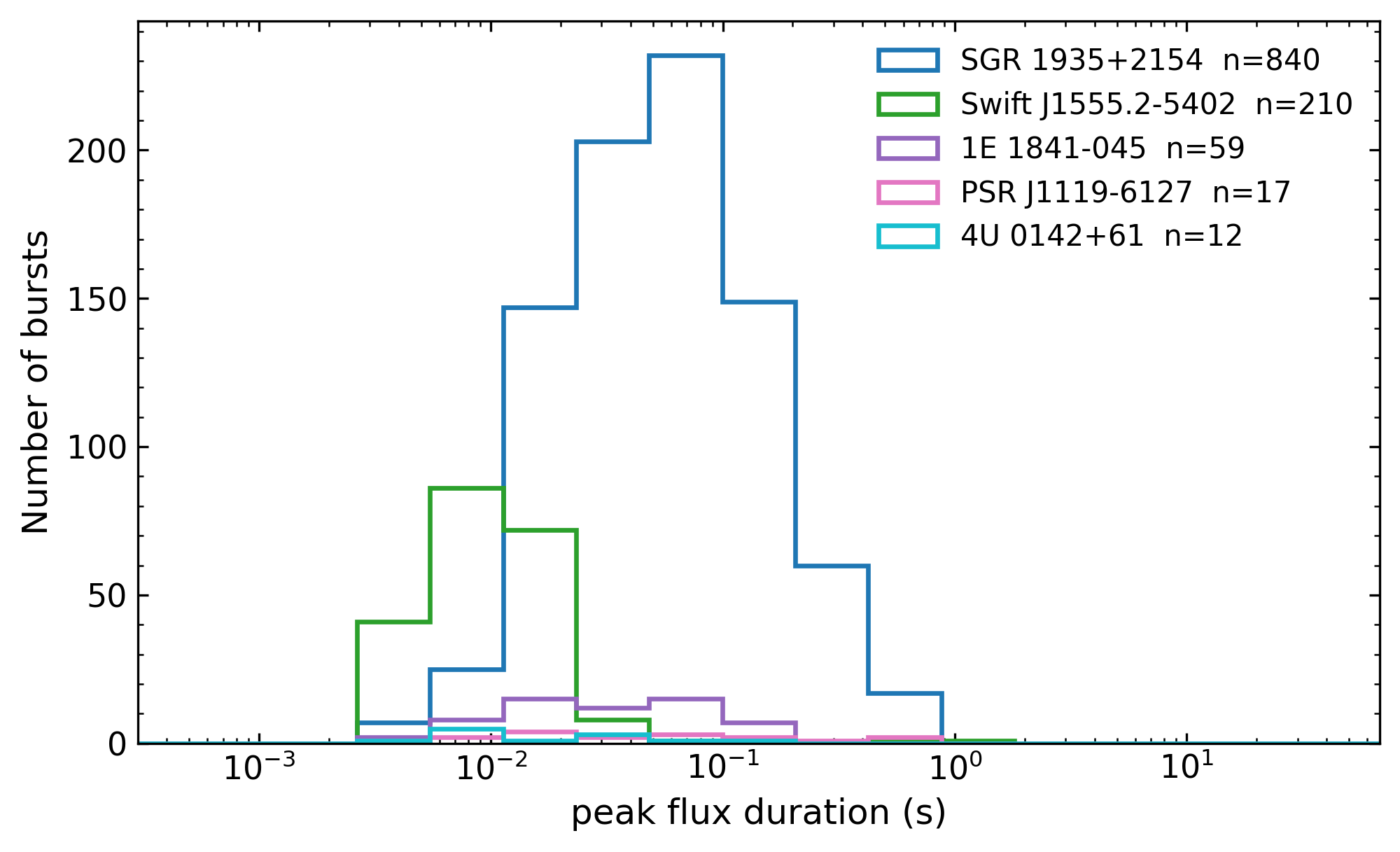}
                \caption{\textbf{Distributions of burst duration for the magnetar sample.} The top panel shows the $T_{90}$ duration, measured from the cumulative deconvolved energy fluence, while the bottom panel shows the peak-flux duration selected by the GBM detection pipeline. The legend indicates the number of bursts in each distribution. The lower cutoff near $2$~ms reflects the minimum duration searched by the detection pipeline.}
                \label{fig:duration_dist}
            \end{figure}

            The distributions of the $T_{90}$ durations and the peak-flux durations measured by the detection pipeline are shown in Fig.~\ref{fig:duration_dist}. The number of bursts in the $T_{90}$ distribution is smaller than that in the peak-flux duration distribution because the $T_{90}$ calculation fails for a fraction of the bursts, particularly at low SNR and on the shortest burst durations.
            
            Several features are apparent in the duration distribution. First, there is a small systematic offset between the two duration measures, with the peak-flux duration distribution shifted toward shorter durations. A similar difference between peak-flux timescales and $T_{90}$ has been observed in previous studies of GRBs \citep{burgess_bayesian_2019,poolakkil_fermi-gbm_2021,perera_expanding_2026}. The distributions also show a sharp lower cutoff at approximately $2$~ms, which is a consequence of the minimum duration searched by our detection pipeline. This demonstrates that the present sample does not constrain the population of bursts below this timescale. Searches optimized for shorter durations could therefore reveal an additional population of very short magnetar bursts. This possibility is particularly interesting in light of the recent NICER magnetar burst catalog, which reports bursts on microsecond timescales \citep{chu_nicer_2026}, although NICER operates at lower energies than GBM.

            The duration distributions also show differences between individual sources. SGR 1935+2154 exhibits systematically longer bursts than Swift J1555.2$-$5402. For the remaining sources, the smaller number of bursts makes a detailed comparison difficult, although their duration distributions appear broadly consistent with one another and generally fall between those of SGR 1935+2154 and Swift J1555.2$-$5402. Overall, the distributions seems to be consistent with those reported by the \textit{INTEGRAL}/IBIS magnetar burst catalog \citep{pacholski_integral_2025}, which operates over a similar energy range, as well as with the duration distribution measured by NICER \citep{chu_nicer_2026} which operates in a different energy range.

        \subsubsection{Spectral Parameters}
\begin{figure*}
                \centering
                \includegraphics[width=0.45\linewidth]{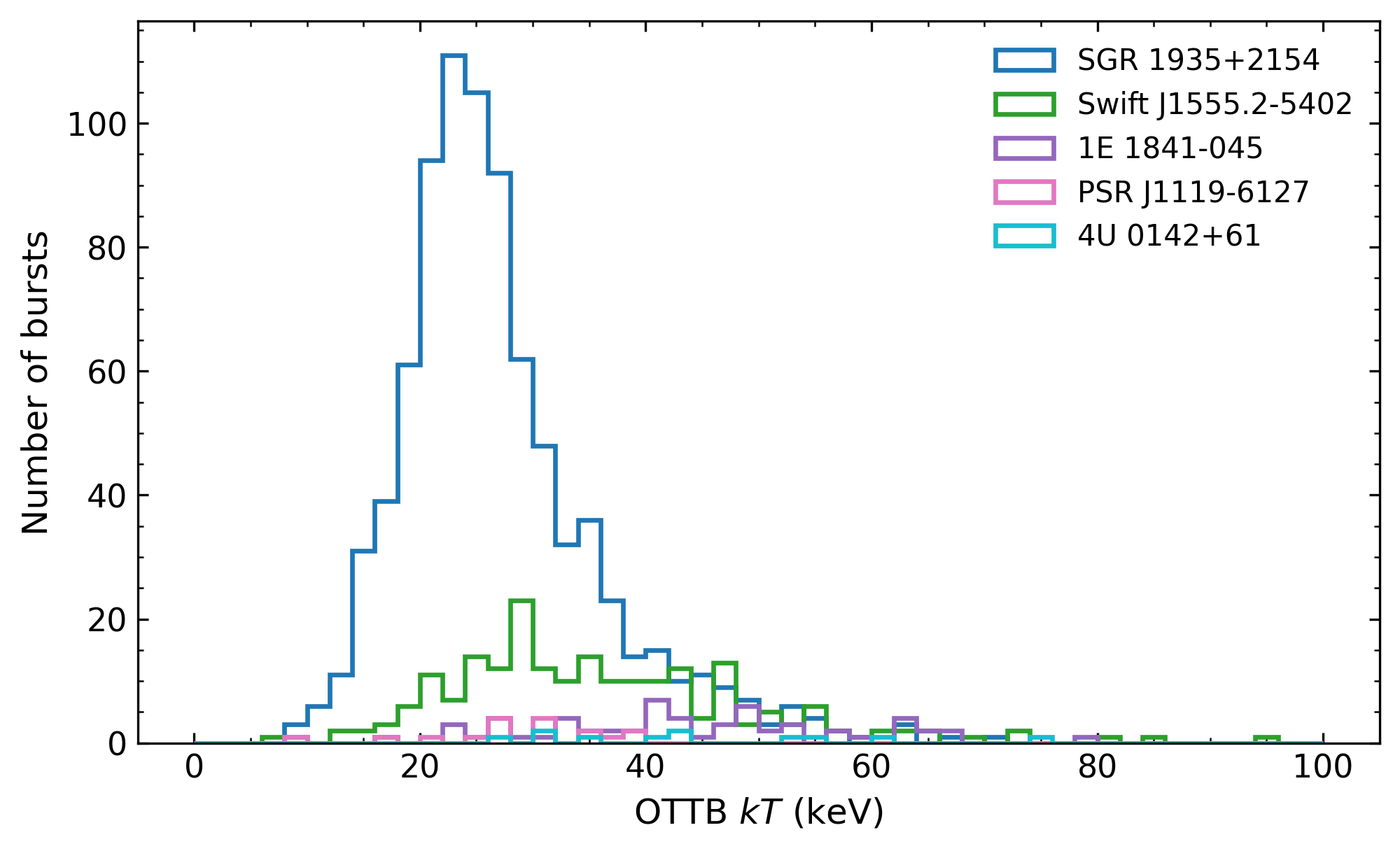}
                \includegraphics[width=0.45\linewidth]{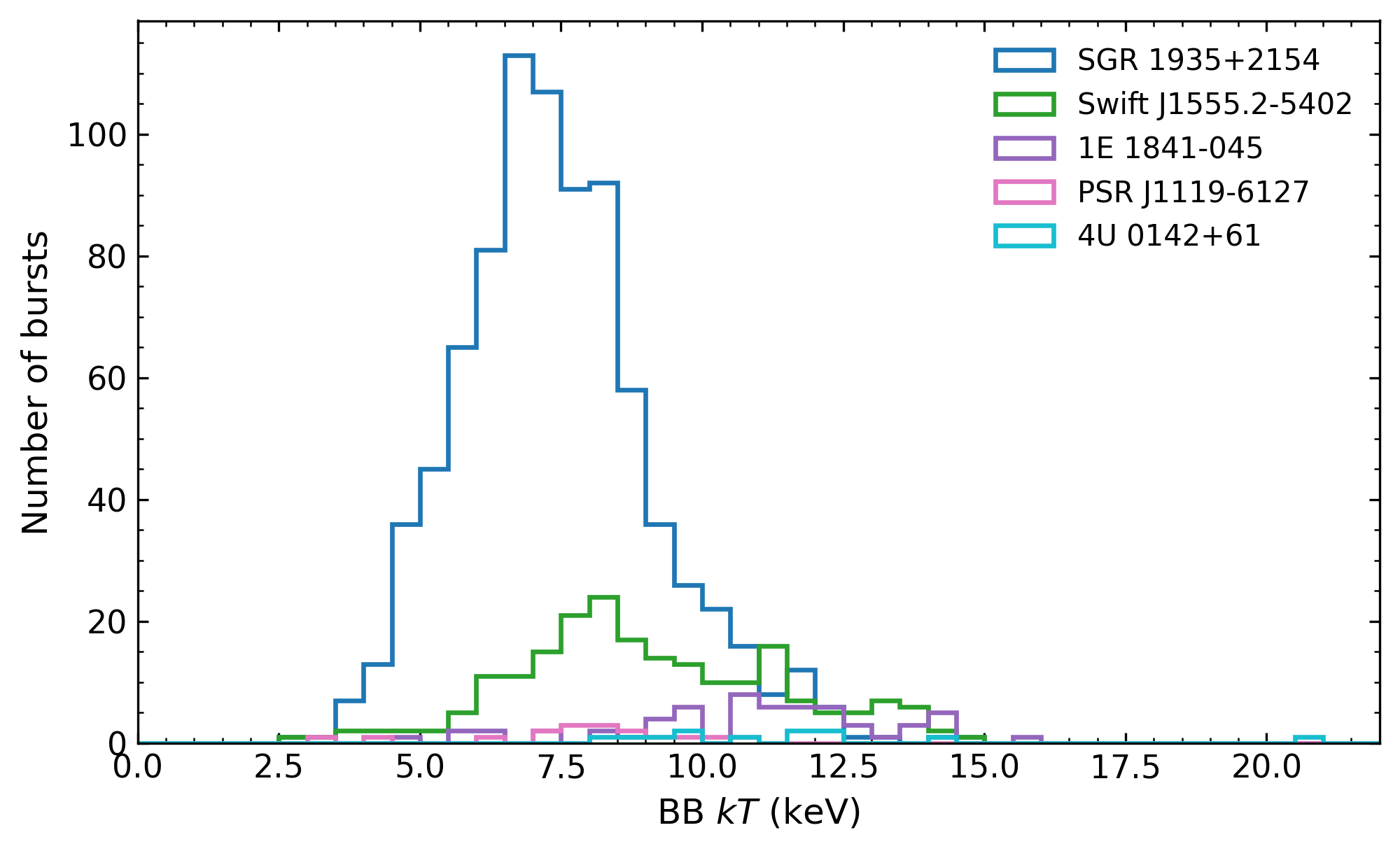}
                \includegraphics[width=0.45\linewidth]{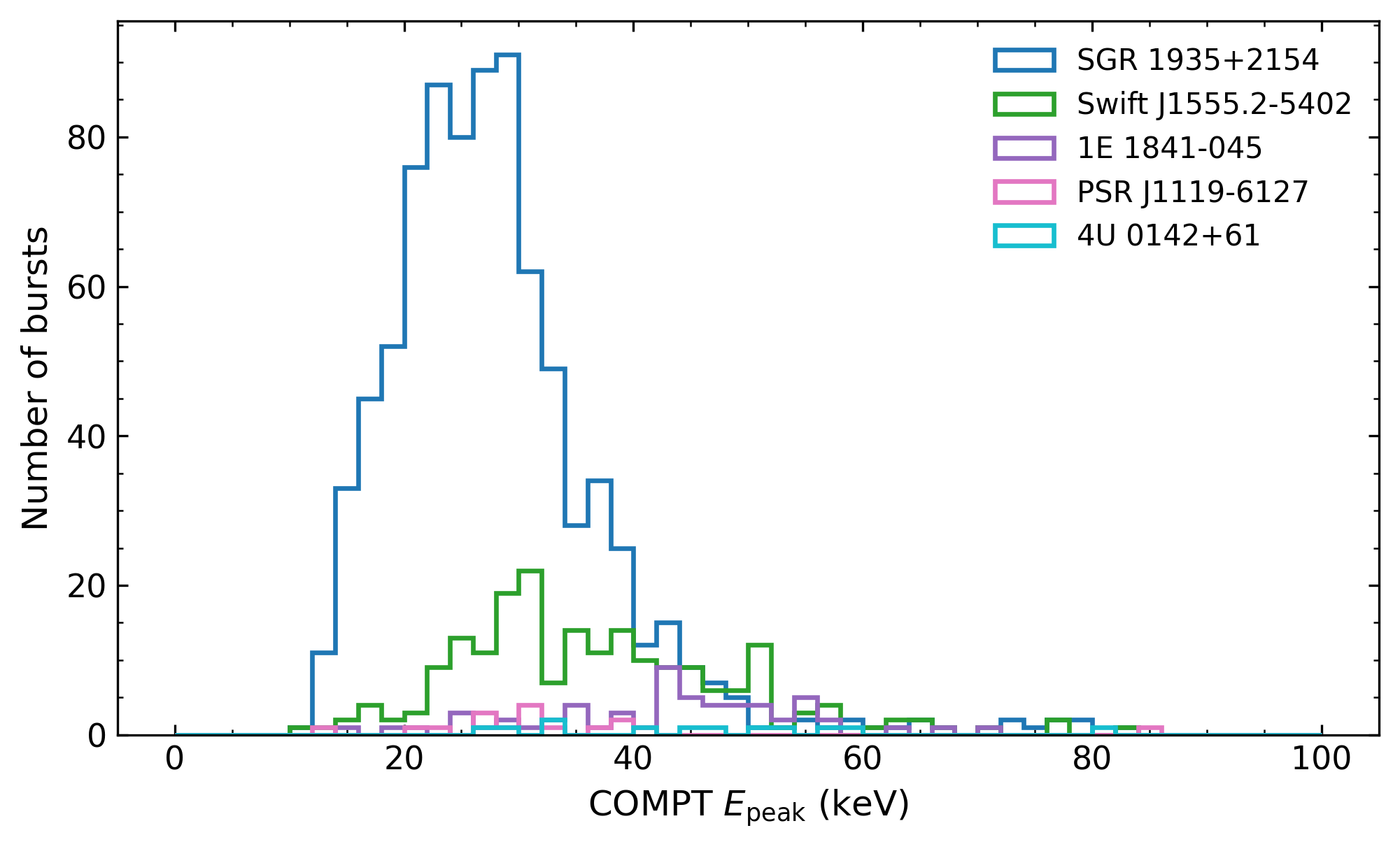}
                \includegraphics[width=0.45\linewidth]{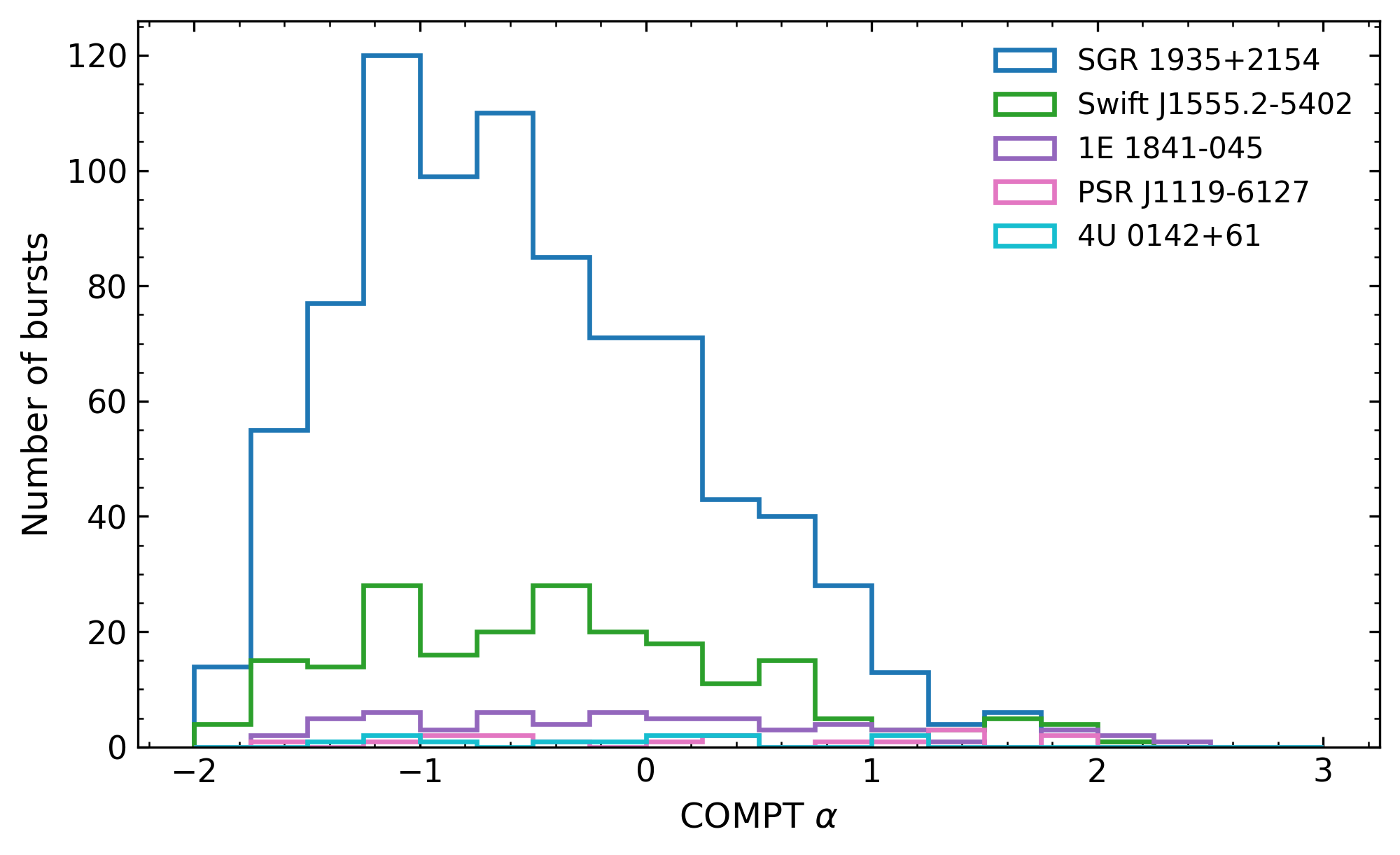}
                \caption{\textbf{Spectral parameter distributions for the several spectral models.} The panels show the fitted OTTB temperature $kT$, BB temperature $kT$, and COMPT parameters $E_{\rm peak}$ and $\alpha$ for the magnetar burst sample. Different colors denote the individual magnetar sources. The similarity between the OTTB $kT$ and COMPT $E_{\rm peak}$ distributions, together with the concentration of COMPT $\alpha$ near $-1$, reflects the limiting relationship between the two models.}
                \label{fig:spectral_parameters}
            \end{figure*}

            We next compare the spectral parameter distributions for the three most frequently preferred models: OTTB, BB, and COMPT. Figure \ref{fig:spectral_parameters} shows the distributions of their fitted parameters. The OTTB temperature and COMPT $E_{\rm peak}$ distributions span very similar ranges, consistent with the relationship between the two models discussed in Section~\ref{sec:model_selection}. In particular, COMPT becomes identical to OTTB for $\alpha=-1$ with $E_{\rm peak}=kT$.
             
            This relationship is also reflected in the $\Delta\mathrm{BIC}$ distributions. For bursts for which OTTB is the preferred model, the COMPT $\Delta\mathrm{BIC}$ distribution approaches unity at $\Delta\mathrm{BIC}\sim8$, indicating that COMPT generally provides a comparably good fit. Since COMPT has one additional free parameter, OTTB is nevertheless preferred by the BIC in these cases. The converse is not generally true. For bursts for which COMPT is preferred, the OTTB $\Delta\mathrm{BIC}$ distribution is substantially broader, with approximately $70\%$ of these bursts having $\Delta\mathrm{BIC}_{\rm OTTB}>10$. Thus, while OTTB is often sufficient to describe bursts that can also be fitted by COMPT, the additional freedom in $\alpha$ is important for a substantial fraction of bursts for which COMPT is preferred. 
            
            The COMPT $\alpha$ distribution provides a further indication of this relationship, with a substantial fraction of bursts having $\alpha\approx-1$, the value for which COMPT reduces to the OTTB functional form.

            The parameter distributions are broadly similar across the different sources. SGR 1935+2154 appears to have somewhat lower OTTB temperatures, COMPT $E_{\rm peak}$ values, and BB temperatures than the other sources, while 1E~1841$-$045 shows some indication of slightly higher values. Given the smaller burst samples of 1E~1841$-$045, however, these differences should be regarded as suggestive.

        \subsubsection{Fluence}

\begin{figure}
                \centering
                \includegraphics[width=1.0\linewidth]{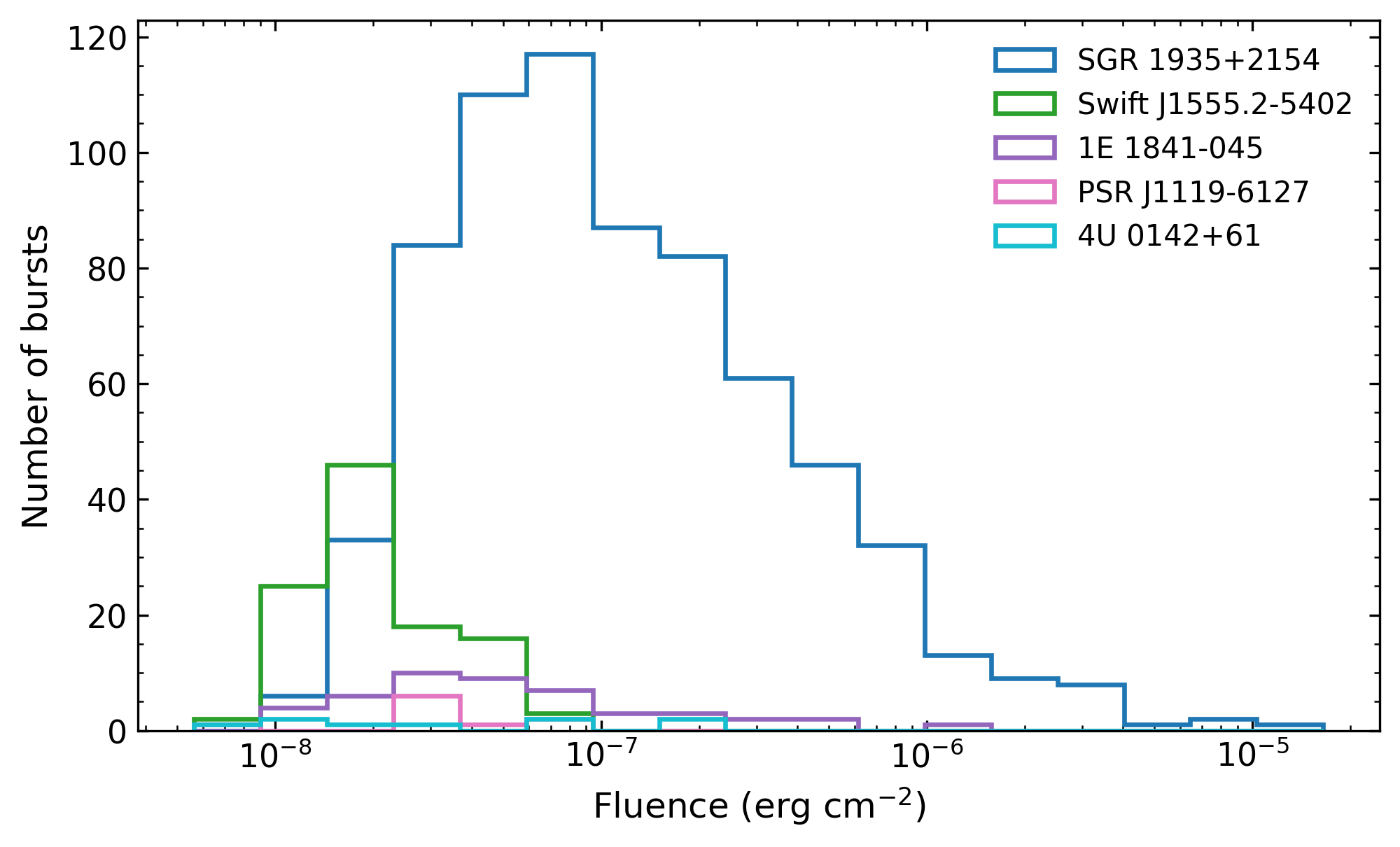}
                \caption{\textbf{Fluence distributions of the magnetar burst sample.} The distributions are shown separately for each associated magnetar source. The sample extends to fluences below $10^{-8}\,\mathrm{erg\,cm^{-2}}$, illustrating the sensitivity of the coherent GBM search to very faint bursts. The fluence is calculated over the $8-200$ keV range.}
                \label{fig:fluence_dist}
            \end{figure}

            The fluence distribution provides information on the range of bursts contained in our catalog. As shown in Fig.~\ref{fig:fluence_dist}, the catalog contains bursts with fluences below $10^{-8},\mathrm{erg,cm^{-2}}$, extending to lower fluences than those reported in other GBM-based magnetar burst catalogs \citep{godwin_17_2026}. We do not characterize the completeness of the search as a function of fluence, however, and the turnover toward lower fluences should therefore not be interpreted as a measure of the search sensitivity.
            
            The fluence distributions also differ between sources. SGR~1935+2154 shows the broadest distribution and contains the largest number of high-fluence bursts. These differences may partly reflect the differing burst-duration distributions, since longer bursts can accumulate greater total fluence. A similar trend may contribute to the somewhat higher fluences observed for 1E~1841$-$045 compared with Swift~J1555.2$-$5402, as the former also shows a larger fraction of longer bursts (Fig.~\ref{fig:duration_dist}). These comparisons should be interpreted cautiously, however, because the fluence distributions are also affected by the source-dependent selection function and, for comparisons between sources, by their differing distances. The smaller samples available for the other sources make more detailed comparisons difficult.

\section{Summary and Outlook}\label{sec:summary}
    In this paper, we applied our coherent detection and parameter-estimation pipeline to 13 years of continuous Fermi/GBM TTE data (2013--2025) to search for magnetar bursts. This produced a broad sample of short, soft, magnetar-like triggers, each with a posterior localization, spectral parameters, and an astrophysical probability $p_{\rm astro}$.

    Building on this sample, we developed a targeted Swift/BAT imaging search around our GBM triggers, capable of localizing bursts to a precision of ${\sim}4'$. The framework assigns each candidate counterpart an association probability, $p_{\rm src}$, that quantifies the probability that a source detected in a BAT image is the counterpart of the GBM trigger, based on its image SNR and angular offset from the GBM localization. Because the search is conditioned on the GBM trigger time and position, it recovers BAT counterparts that would otherwise be below the detection threshold required for blind image searches. This method allows precise localizations not only of magnetar bursts but also of other transients, such as GRBs, in our full trigger list.
    
    We further developed a two-stage clustering scheme that identifies magnetar bursts occurring in temporal and spatial proximity, isolating episodes of source activity that are subsequently associated with specific sources using the precise localizations obtained from BAT and from external catalogs. Combined, these methods yielded 1720 magnetar bursts, out of which 1170 are confidently associated with 11 different sources. An additional 121 clustered bursts could not be assigned to a specific source, either for lack of a precisely localized anchor or because their combined localization regions remained too broad. Non-clustered magnetar-like bursts constitute another 429 bursts.

    Based on these methods, we defined and applied three channels for identifying new magnetars, revealing 7 candidate sources. The first channel searches for BAT image counterparts of GBM triggers that are localized to ${\sim}4'$ but are inconsistent with the position of any known bursting source; it revealed 3 new candidates. The second channel identifies burst clusters whose combined localization is consistent with a common sky position at which no known magnetar is present; it revealed 4 candidates, although their localization regions remain broad even after combining the localization information of all cluster members. The third channel incoherently stacks the localization posteriors of all unclustered magnetar-like triggers in an attempt to identify a dominant sporadically emitting source; while the stacked maps show enhanced probability along the Galactic plane, this channel did not yield resolved candidates. Nevertheless, continued GBM observations may resolve individual sources as more trigger data is collected and analyzed.

    The 1170 associated bursts constitute our magnetar burst catalog, the largest over this observation span (2013-2025; compared to 735 bursts from \cite{godwin_17_2026}), extending to the fluences smaller than $10^{-8}~\mathrm{erg~cm^{-2}}$.
    
    In addition to the associated bursts, our search revealed a substantial population of unclustered magnetar-like triggers, suggesting that the true number of magnetar bursts in the sample is larger still. These faint events broaden the observed burst distributions and provide a valuable sample for population studies of magnetar burst energetics.

    Our spectral analysis shows that most bursts are adequately described by several spectral models, with optically thin thermal bremsstrahlung preferred most often. This degeneracy should be taken into account when fitting and interpreting magnetar burst spectra. These findings are also relevant for burst detection. Our search templates were based on the Band function, and the fact that the detected bursts are well described by models that are limiting cases of it shows that the template bank was flexible enough to capture them. At the same time, a template bank built directly from the observed magnetar spectral distributions could capture even fainter events in future coherent searches similar to ours.

    The new candidates motivate follow-up observations. The three BAT-localized candidates can be followed up with sensitive focusing X-ray telescopes such as Chandra and XMM-Newton, which could confirm the sources through detection of persistent emission or pulsations. The localization regions of the other four candidates are too broad for such pointed observations, but the Wide-field X-ray Telescope on the Einstein Probe \citep{yuan_science_2025} can assist in the efforts of finding these sources. Finally, the coded-mask camera ECLAIRs on SVOM \citep{godet_eclairs_2026} operates over an energy range similar to that of BAT. As a result, ECLAIRs could provide a complementary and potentially future alternative for applying the methodology developed here to localize and identify magnetar bursts.
    
    We emphasize that the methodology developed here of combining the all-sky coverage and continuous exposure of one instrument with the precise localization capability of another, through statistical association, is general and transferable to combinations of high-energy transient missions. Applying it across current and future missions may reveal additional rare sources, sharpen our understanding of known ones, and provide the most complete picture of the high-energy transient sky achievable with the available instrumentation.

\section*{Data Availability}
    The full burst catalog, together with the posterior samples for all localization and spectral fits, is publicly available in the Zenodo repository \href{https://doi.org/10.5281/zenodo.22884088} {10.5281/zenodo.22884088} \citep{perera_catalog_2026}, and the catalog is also available in a machine-readable format in the online version of this paper. The detection and parameter estimation pipeline we developed is publicly available at the \href{https://github.com/PeAriel/grpype}{GitHub repository} \texttt{grpype} \citep{ariel_perera_pearielgrpype_2026}: https://github.com/PeAriel/grpype. Any other data generated by this study is available upon reasonable request.

\section*{Acknowledgments}
    We thank Jimmy Delaunay for valuable information and insightful discussions about the BAT instrument and data products.
    This research was supported by grant no 2022136 from the United States - Israel Binational Science Foundation
    (BSF), Jerusalem, Israel.
    BZ is supported by a research grant from the Willner Family Leadership Institute for the Weizmann Institute of Science. 
    TV additionally acknowledges support from NSF grants 2012086 and 2309360, the Alfred P. Sloan Foundation through grant number FG-2023-20470, and the Hellman Family Faculty Fellowship during the time this work was done.

\appendix
    \section{Clustering Analysis Supporting Figures}\label{appendix: clustering supporting figures}
        Here we present additional figures supporting the clustering analysis described in Sec.~\ref{sec: clustering}, with particular emphasis on the construction of the association metric shown in Fig.~\ref{fig: metric}.

\begin{figure}
            \centering
            \includegraphics[width=0.5\linewidth]{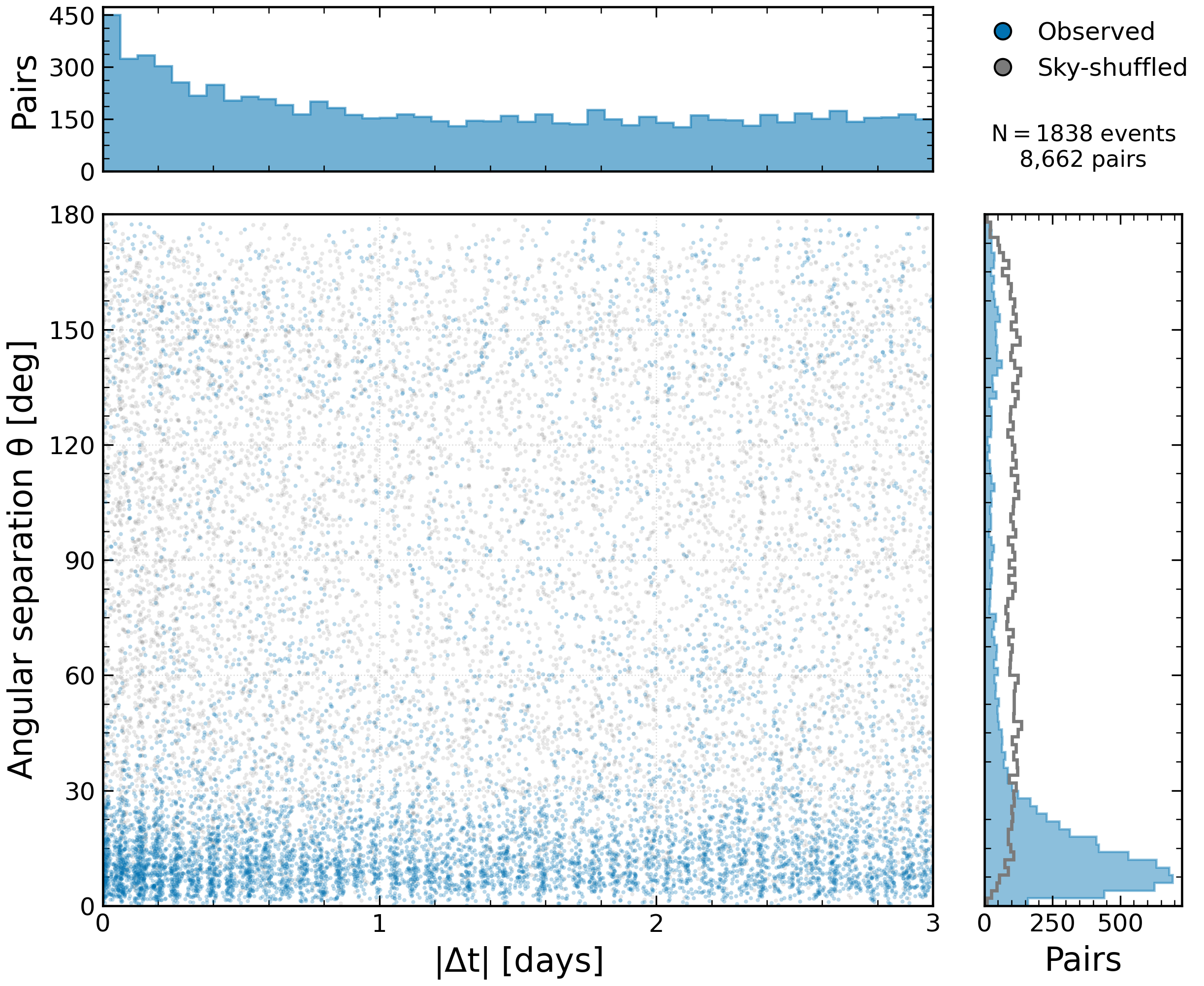}
            \caption{\textbf{Temporal and angular pair distribution excluding triggers from 2020-2021.} The distribution is constructed in the same manner as that shown in Fig.~\ref{fig: metric}, but with triggers from 2020-2021 excluded. The prominent feature near an angular separation of $90^{\circ}$ seen in Fig.~\ref{fig: metric} is absent.}
            \label{fig: metric excluded}
        \end{figure}

        As discussed in Sec.~\ref{sec: clustering}, the foreground distribution in Fig.~\ref{fig: metric} exhibits a prominent feature at an angular separation of $\sim90^{\circ}$. This feature is associated with activity during 2020--2021. Figure \ref{fig: metric excluded} shows the same distribution as Fig.~\ref{fig: metric}, but with triggers from 2020-2021 excluded. The feature at $\sim90^{\circ}$ is no longer present, indicating that it is associated with a temporally localized period of enhanced source activity rather than a general property of the trigger population.

\begin{figure}
            \centering
            \includegraphics[width=0.65\linewidth]{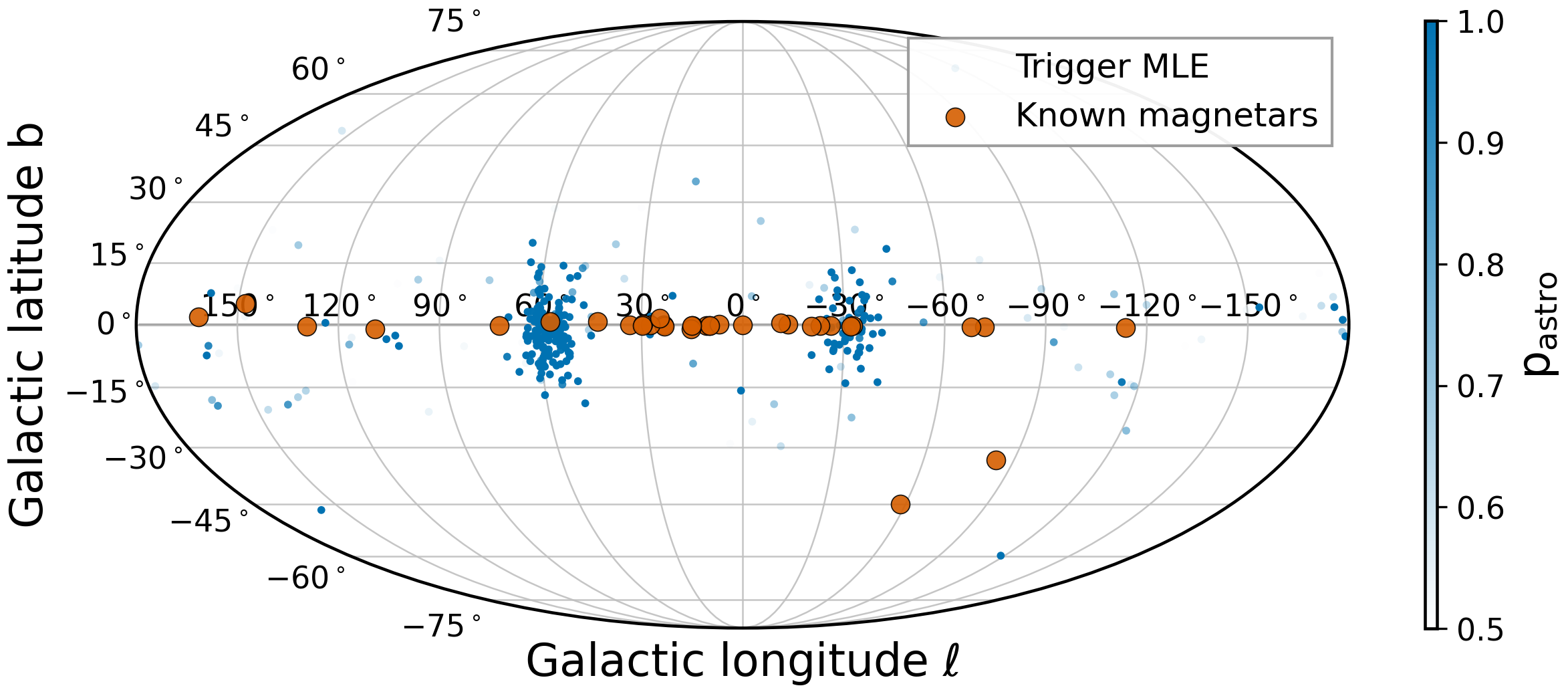}
            \caption{\textbf{Sky positions of triggers contributing to the $90^{\circ}$ feature.} The MLE sky positions of triggers belonging to pairs with angular separations of $90^{\circ}\pm1^{\circ}$ are shown in Galactic coordinates. The opacity of each point represents its $p_{\rm astro}$ value, with fainter points corresponding to lower $p_{\rm astro}$. The concentration of triggers around two distinct sky locations indicates that two active sources dominate the excess of pairs near $90^{\circ}$ in Fig.~\ref{fig: metric}.}
            \label{fig: 90 sky positions}
        \end{figure}

        The $90^{\circ}$ feature results from two magnetars, SGR 1935+2154 and Swift J1555.2-5402, separated by approximately $90^{\circ}$ on the sky, undergoing periods of substantial bursting activity at approximately the same time. To illustrate the origin of this feature, we select the trigger pairs contributing to Fig.~\ref{fig: metric} with angular separations of $90^{\circ}\pm1^{\circ}$ and plot the MLE sky position of each trigger. Fig.~\ref{fig: 90 sky positions} shows these positions in Galactic coordinates. The concentration of events around two distinct regions of the sky demonstrates that the excess of pairs near $90^{\circ}$ is dominated by the contemporaneous activity of these two sources.

\begin{figure}
            \centering
            \includegraphics[width=1.0\linewidth]{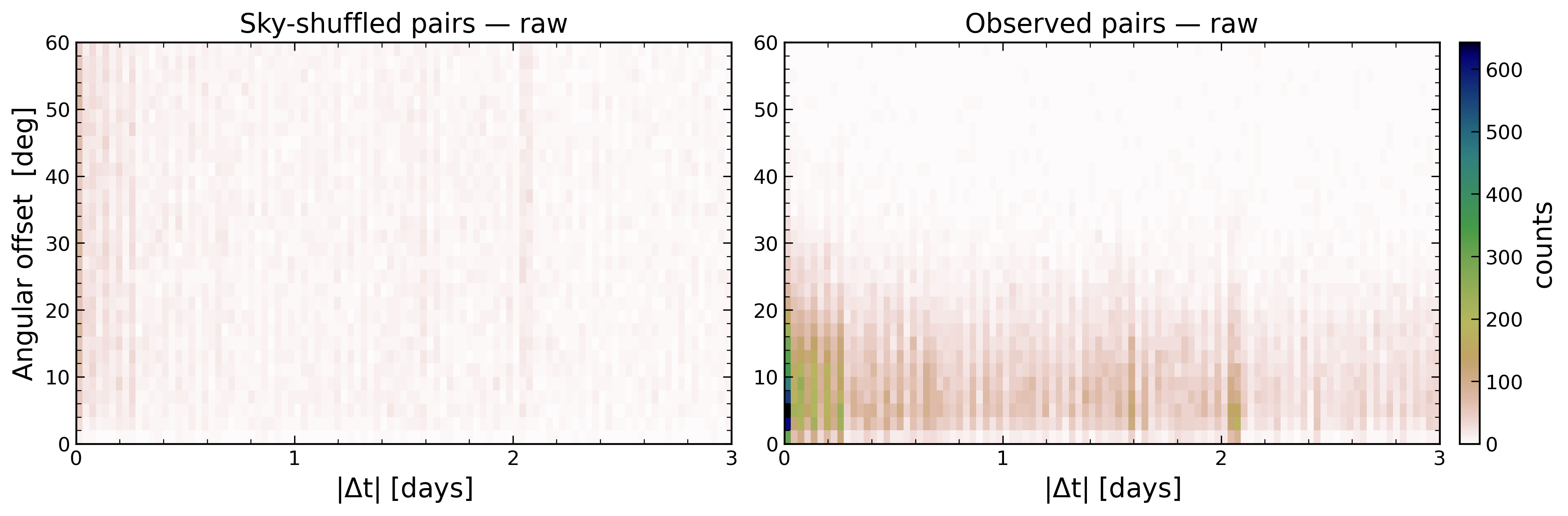}
            \includegraphics[width=1.0\linewidth]{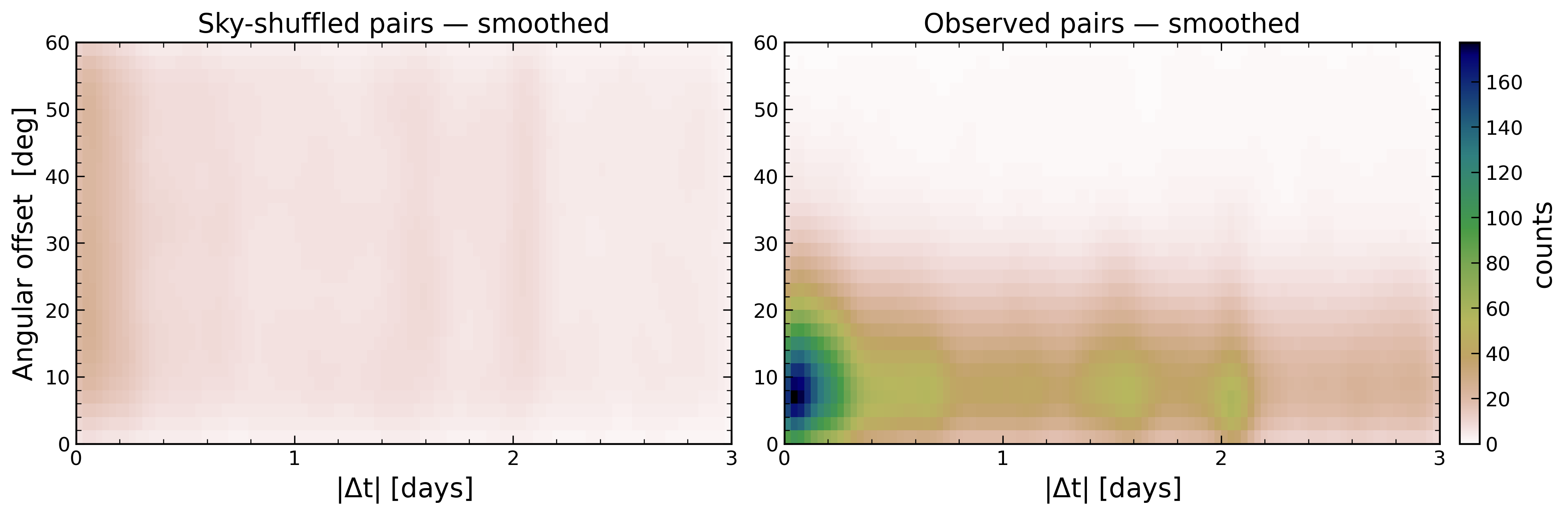}
            \caption{\textbf{Histograms used to calculate $p(T|\Delta t,\Delta\theta)$.} The upper panels show the raw two-dimensional histograms of the observed trigger pairs and the corresponding sky-shuffled pairs. The lower panels show the smoothed distributions used to numerically evaluate Eq.~\ref{eq: within image prob} and construct the association probability shown in Fig.~\ref{fig: metric}.}
            \label{fig: metric histograms}
        \end{figure}

        Finally, Fig.~\ref{fig: metric histograms} shows the two-dimensional histograms used to construct the association probability $p(T|\Delta t,\Delta\theta)$. The upper panels show the raw foreground and sky-shuffled background distributions corresponding to the trigger pairs presented in Fig.~\ref{fig: metric}. The lower panels show the corresponding smoothed distributions used to numerically evaluate Eq.~\ref{eq: within image prob}. Smoothing suppresses statistical fluctuations arising from the finite number of trigger pairs while preserving the broader structure of the foreground and background distributions.

    \section{All clustered bursts} \label{appendix: all subclusters}
    
\startlongtable
\begin{deluxetable*}{ccccccclclc}
\tabletypesize{\scriptsize}
\tablecaption{GBM subclusters and the two nearest known magnetars to each combined-posterior MLE. Boldface rows have a clear named-magnetar association. An italic association entry marks a subcluster later associated with SGR 1935+2154 or Swift J1555.2-5402, or later classified as non-magnetar (NM).\label{tab:subcluster_nearest_magnetars}}
\tablehead{
\colhead{ID} & \colhead{$N$} & \colhead{$\bar t$} & \colhead{$\Delta t$} & \colhead{$\alpha_{\mathrm{MLE}}$} & \colhead{$\delta_{\mathrm{MLE}}$} & \colhead{$\Delta\theta_1$} & \colhead{Nearest} & \colhead{$\Delta\theta_2$} & \colhead{Second} & \colhead{Assoc.} \\
\colhead{} & \colhead{} & \colhead{(UTC)} & \colhead{(h)} & \colhead{(deg)} & \colhead{(deg)} & \colhead{(deg)} & \colhead{magnetar} & \colhead{(deg)} & \colhead{magnetar} & \colhead{}
}
\startdata
0 & 3 & 2013-02-27 & 25.4 & 274.92 & -23.32 & 3.8 & SGR 1808 & 3.9 & SGR 1806 & \nodata \\
1 & 3 & 2014-02-12 & 20.6 & 264.38 & -30.69 & 2.4 & SGR J1745 & 8.9 & CXOU J171405.7 & \textit{NM} \\
2 & 4 & 2014-02-27 & 17.6 & 267.19 & -30.69 & 1.8 & SGR J1745 & 8.2 & SGR 1801 & \textit{NM} \\
3 & 3 & 2014-03-02 & 14.1 & 274.22 & -29.31 & 6.8 & SGR J1745 & 7.3 & SGR 1801 & \textit{NM} \\
\textbf{4} & \textbf{3} & \textbf{2014-07-05} & \textbf{0.1} & \textbf{298.12} & \textbf{26.61} & \textbf{6.2} & \textbf{SGR 1935} & \textbf{9.0} & \textbf{SGR 2013} & \textbf{SGR 1935} \\
\textbf{5} & \textbf{21} & \textbf{2015-02-23} & \textbf{75.6} & \textbf{296.02} & \textbf{22.02} & \textbf{2.1} & \textbf{SGR 1935} & \textbf{13.9} & \textbf{SGR 2013} & \textbf{SGR 1935} \\
\textbf{6} & \textbf{3} & \textbf{2015-02-28} & \textbf{33.4} & \textbf{15.97} & \textbf{67.19} & \textbf{7.1} & \textbf{4U 0142} & \textbf{16.0} & \textbf{1E 2259} & \textbf{4U 0142} \\
7 & 3 & 2015-06-24 & 22.6 & 309.38 & 32.09 & 5.4 & SGR 2013 & 17.2 & SGR 1935 & \textit{NM} \\
8 & 11 & 2015-06-26 & 54.3 & 303.05 & 32.80 & 1.6 & SGR 2013 & 13.7 & SGR 1935 & \textit{NM} \\
9 & 5 & 2015-08-27 & 0.1 & 246.09 & -36.42 & 9.5 & 1RXS J170849.0 & 10.0 & CXOU J171405.7 & \textit{NM} \\
10 & 3 & 2016-05-14 & 14.1 & 291.09 & 22.67 & 2.6 & SGR 1935 & 14.0 & SGR 1900 & \textit{SGR 1935} \\
\textbf{11} & \textbf{28} & \textbf{2016-05-19} & \textbf{91.4} & \textbf{293.20} & \textbf{20.74} & \textbf{1.3} & \textbf{SGR 1935} & \textbf{13.0} & \textbf{SGR 1900} & \textbf{SGR 1935} \\
\textbf{12} & \textbf{43} & \textbf{2016-06-24} & \textbf{116.0} & \textbf{291.80} & \textbf{22.02} & \textbf{1.8} & \textbf{SGR 1935} & \textbf{13.6} & \textbf{SGR 1900} & \textbf{SGR 1935} \\
\textbf{13} & \textbf{15} & \textbf{2016-07-27} & \textbf{37.5} & \textbf{178.75} & \textbf{-63.45} & \textbf{4.6} & \textbf{PSR J1119} & \textbf{8.4} & \textbf{1E 1048.1} & \textbf{PSR J1119} \\
\textbf{14} & \textbf{8} & \textbf{2017-07-13} & \textbf{0.0} & \textbf{21.09} & \textbf{66.44} & \textbf{5.3} & \textbf{4U 0142} & \textbf{17.8} & \textbf{1E 2259} & \textbf{4U 0142} \\
15 & 3 & 2017-10-19 & 4.3 & 253.12 & -48.14 & 2.4 & CXOU J164710.2 & 2.8 & SGR 1627 & \nodata \\
16 & 11 & 2017-11-02 & 105.6 & 35.27 & 62.70 & 4.1 & 4U 0142 & 15.4 & SGR 0418 & \textit{NM} \\
17 & 3 & 2017-11-07 & 82.2 & 43.93 & 58.92 & 9.0 & 4U 0142 & 10.9 & SGR 0418 & \textit{NM} \\
18 & 5 & 2017-11-14 & 41.2 & 38.72 & 58.16 & 7.0 & 4U 0142 & 13.7 & SGR 0418 & \textit{NM} \\
19 & 4 & 2017-11-17 & 31.0 & 40.61 & 59.68 & 7.1 & 4U 0142 & 12.6 & SGR 0418 & \textit{NM} \\
\textbf{20} & \textbf{6} & \textbf{2018-02-06} & \textbf{17.1} & \textbf{247.88} & \textbf{-45.78} & \textbf{2.0} & \textbf{SGR 1627} & \textbf{2.7} & \textbf{CXOU J164710.2} & \textbf{CXOU J164710.2} \\
21 & 6 & 2018-03-29 & 46.3 & 273.52 & 2.39 & 9.4 & AX J1845.0 & 9.7 & PSR J1846 & \nodata \\
22 & 9 & 2018-04-04 & 114.3 & 277.03 & 5.38 & 7.8 & 3XMM J185246.6 & 9.3 & AX J1845.0 & \nodata \\
23 & 3 & 2018-04-04 & 18.9 & 136.12 & -60.43 & 13.1 & 1E 1048.1 & 16.2 & PSR J1119 & \nodata \\
\textbf{24} & \textbf{22} & \textbf{2019-11-04} & \textbf{29.9} & \textbf{292.50} & \textbf{21.38} & \textbf{1.3} & \textbf{SGR 1935} & \textbf{13.2} & \textbf{SGR 1900} & \textbf{SGR 1935} \\
25 & 6 & 2020-02-03 & 33.1 & 257.46 & -44.20 & 4.1 & 1RXS J170849.0 & 4.3 & CXOU J164710.2 & \nodata \\
\textbf{26} & \textbf{119} & \textbf{2020-04-27} & \textbf{44.4} & \textbf{293.20} & \textbf{22.02} & \textbf{0.5} & \textbf{SGR 1935} & \textbf{14.1} & \textbf{SGR 1900} & \textbf{SGR 1935} \\
27 & 4 & 2020-05-20 & 27.2 & 291.80 & 22.02 & 1.8 & SGR 1935 & 13.6 & SGR 1900 & \textit{SGR 1935} \\
\textbf{28} & \textbf{37} & \textbf{2020-07-18} & \textbf{17.8} & \textbf{279.14} & \textbf{-4.78} & \textbf{1.2} & \textbf{1E 1841} & \textbf{2.5} & \textbf{SGR 1830} & \textbf{1E 1841} \\
29 & 80 & 2020-11-18 & 188.2 & 87.19 & 27.95 & 19.7 & SGR 0501 & 33.6 & SGR 0418 & \textit{NM} \\
30 & 98 & 2020-11-30 & 407.5 & 84.37 & 27.95 & 18.8 & SGR 0501 & 32.7 & SGR 0418 & \textit{NM} \\
31 & 6 & 2020-12-02 & 393.2 & 274.92 & -15.71 & 0.3 & AX J1818.8 & 0.6 & Swift J1818.0 & \nodata \\
32 & 3 & 2020-12-13 & 14.6 & 272.11 & -16.96 & 2.5 & Swift J1818.0 & 2.7 & AX J1818.8 & \nodata \\
33 & 3 & 2021-01-06 & 5.0 & 282.66 & -6.58 & 2.8 & 1E 1841 & 3.8 & PSR J1846 & \nodata \\
\textbf{34} & \textbf{4} & \textbf{2021-01-24} & \textbf{36.8} & \textbf{298.12} & \textbf{21.38} & \textbf{4.1} & \textbf{SGR 1935} & \textbf{13.8} & \textbf{SGR 2013} & \textbf{SGR 1935} \\
\textbf{35} & \textbf{22} & \textbf{2021-01-30} & \textbf{93.2} & \textbf{293.91} & \textbf{18.84} & \textbf{3.1} & \textbf{SGR 1935} & \textbf{11.7} & \textbf{SGR 1900} & \textbf{SGR 1935} \\
\textbf{36} & \textbf{4} & \textbf{2021-02-11} & \textbf{23.5} & \textbf{291.80} & \textbf{24.62} & \textbf{3.3} & \textbf{SGR 1935} & \textbf{14.0} & \textbf{SGR 2013} & \textbf{SGR 1935} \\
37 & 4 & 2021-04-25 & 23.8 & 216.15 & -44.20 & 17.2 & 1E 1547.0 & 17.7 & Swift J1555.2 & \nodata \\
38 & 4 & 2021-06-07 & 40.5 & 243.90 & -52.80 & 3.2 & PSR J1622 & 3.3 & Swift J1555.2 & \textit{Swift J1555.2} \\
39 & 3 & 2021-06-12 & 29.5 & 247.50 & -52.80 & 3.2 & PSR J1622 & 5.3 & SGR 1627 & \textit{Swift J1555.2} \\
40 & 3 & 2021-06-15 & 38.5 & 246.43 & -42.61 & 5.0 & CXOU J164710.2 & 5.3 & SGR 1627 & \textit{Swift J1555.2} \\
41 & 27 & 2021-06-22 & 225.2 & 238.50 & -52.80 & 1.3 & Swift J1555.2 & 1.6 & 1E 1547.0 & \textit{Swift J1555.2} \\
42 & 4 & 2021-07-01 & 55.2 & 238.19 & -46.57 & 6.0 & PSR J1622 & 7.4 & SGR 1627 & \textit{Swift J1555.2} \\
43 & 15 & 2021-07-13 & 310.7 & 241.53 & -53.57 & 1.7 & Swift J1555.2 & 2.4 & 1E 1547.0 & \textit{Swift J1555.2} \\
\textbf{44} & \textbf{25} & \textbf{2021-07-11} & \textbf{296.3} & \textbf{291.09} & \textbf{22.67} & \textbf{2.6} & \textbf{SGR 1935} & \textbf{14.0} & \textbf{SGR 1900} & \textbf{SGR 1935} \\
45 & 3 & 2021-07-22 & 38.3 & 251.85 & -43.41 & 2.5 & CXOU J164710.2 & 4.6 & SGR 1627 & \nodata \\
46 & 3 & 2021-07-28 & 53.6 & 267.13 & -55.11 & 13.4 & CXOU J164710.2 & 13.5 & SGR 1627 & \nodata \\
47 & 8 & 2021-08-10 & 76.8 & 238.58 & -50.48 & 3.6 & Swift J1555.2 & 3.9 & 1E 1547.0 & \textit{Swift J1555.2} \\
48 & 3 & 2021-08-19 & 29.6 & 230.92 & -61.94 & 8.4 & 1E 1547.0 & 8.9 & Swift J1555.2 & \textit{Swift J1555.2} \\
49 & 7 & 2021-08-23 & 76.3 & 238.93 & -58.92 & 4.7 & 1E 1547.0 & 4.9 & Swift J1555.2 & \textit{Swift J1555.2} \\
50 & 3 & 2021-08-26 & 34.7 & 236.25 & -57.40 & 3.2 & 1E 1547.0 & 3.6 & Swift J1555.2 & \textit{Swift J1555.2} \\
51 & 7 & 2021-08-31 & 35.6 & 236.49 & -55.11 & 1.1 & 1E 1547.0 & 1.7 & Swift J1555.2 & \textit{Swift J1555.2} \\
52 & 27 & 2021-09-16 & 345.9 & 242.10 & -52.80 & 2.3 & Swift J1555.2 & 3.0 & 1E 1547.0 & \textit{Swift J1555.2} \\
\textbf{53} & \textbf{153} & \textbf{2021-09-12} & \textbf{351.7} & \textbf{297.42} & \textbf{19.47} & \textbf{4.2} & \textbf{SGR 1935} & \textbf{14.4} & \textbf{SGR 1900} & \textbf{SGR 1935} \\
\textbf{54} & \textbf{11} & \textbf{2021-09-30} & \textbf{108.2} & \textbf{298.83} & \textbf{20.74} & \textbf{4.9} & \textbf{SGR 1935} & \textbf{14.2} & \textbf{SGR 2013} & \textbf{SGR 1935} \\
55 & 27 & 2021-10-02 & 203.8 & 238.50 & -52.80 & 1.3 & Swift J1555.2 & 1.6 & 1E 1547.0 & \textit{Swift J1555.2} \\
\textbf{56} & \textbf{3} & \textbf{2021-10-12} & \textbf{43.6} & \textbf{237.35} & \textbf{-52.03} & \textbf{2.2} & \textbf{Swift J1555.2} & \textbf{2.3} & \textbf{1E 1547.0} & \textbf{Swift J1555.2} \\
57 & 7 & 2021-10-19 & 74.5 & 239.69 & -53.57 & 0.7 & Swift J1555.2 & 1.4 & 1E 1547.0 & \textit{Swift J1555.2} \\
\textbf{58} & \textbf{4} & \textbf{2021-10-28} & \textbf{30.1} & \textbf{235.76} & \textbf{-55.87} & \textbf{1.9} & \textbf{1E 1547.0} & \textbf{2.5} & \textbf{Swift J1555.2} & \textbf{Swift J1555.2} \\
59 & 8 & 2021-11-03 & 107.8 & 237.50 & -49.70 & 4.4 & Swift J1555.2 & 4.6 & 1E 1547.0 & \textit{Swift J1555.2} \\
\textbf{60} & \textbf{7} & \textbf{2021-11-18} & \textbf{69.6} & \textbf{239.69} & \textbf{-53.57} & \textbf{0.7} & \textbf{Swift J1555.2} & \textbf{1.4} & \textbf{1E 1547.0} & \textbf{Swift J1555.2} \\
61 & 6 & 2021-11-22 & 44.4 & 237.00 & -56.64 & 2.4 & 1E 1547.0 & 2.8 & Swift J1555.2 & \textit{Swift J1555.2} \\
62 & 3 & 2021-11-25 & 18.3 & 241.88 & -48.14 & 3.0 & PSR J1622 & 4.8 & SGR 1627 & \textit{Swift J1555.2} \\
63 & 3 & 2021-12-05 & 11.8 & 227.00 & -56.64 & 6.5 & 1E 1547.0 & 7.2 & Swift J1555.2 & \textit{Swift J1555.2} \\
64 & 3 & 2021-12-11 & 31.7 & 235.76 & -55.87 & 1.9 & 1E 1547.0 & 2.5 & Swift J1555.2 & \textit{Swift J1555.2} \\
\textbf{65} & \textbf{3} & \textbf{2021-12-19} & \textbf{8.1} & \textbf{244.15} & \textbf{-55.11} & \textbf{3.3} & \textbf{Swift J1555.2} & \textbf{3.8} & \textbf{1E 1547.0} & \textbf{Swift J1555.2} \\
\textbf{66} & \textbf{4} & \textbf{2021-12-25} & \textbf{27.4} & \textbf{291.80} & \textbf{20.74} & \textbf{2.1} & \textbf{SGR 1935} & \textbf{12.4} & \textbf{SGR 1900} & \textbf{SGR 1935} \\
67 & 3 & 2022-01-04 & 83.2 & 244.69 & -54.34 & 3.5 & Swift J1555.2 & 4.1 & 1E 1547.0 & \textit{Swift J1555.2} \\
\textbf{68} & \textbf{7} & \textbf{2022-01-04} & \textbf{49.9} & \textbf{292.50} & \textbf{21.38} & \textbf{1.3} & \textbf{SGR 1935} & \textbf{13.2} & \textbf{SGR 1900} & \textbf{SGR 1935} \\
\textbf{69} & \textbf{11} & \textbf{2022-01-13} & \textbf{237.1} & \textbf{243.90} & \textbf{-52.80} & \textbf{3.2} & \textbf{PSR J1622} & \textbf{3.3} & \textbf{Swift J1555.2} & \textbf{Swift J1555.2} \\
\textbf{70} & \textbf{152} & \textbf{2022-01-14} & \textbf{231.1} & \textbf{294.61} & \textbf{20.74} & \textbf{1.4} & \textbf{SGR 1935} & \textbf{13.7} & \textbf{SGR 1900} & \textbf{SGR 1935} \\
71 & 5 & 2022-01-20 & 13.9 & 293.91 & 23.97 & 2.1 & SGR 1935 & 13.3 & SGR 2013 & \textit{SGR 1935} \\
72 & 5 & 2022-01-26 & 143.0 & 291.80 & 18.21 & 4.1 & SGR 1935 & 10.1 & SGR 1900 & \textit{SGR 1935} \\
73 & 4 & 2022-01-25 & 138.5 & 238.50 & -52.80 & 1.3 & Swift J1555.2 & 1.6 & 1E 1547.0 & \textit{Swift J1555.2} \\
74 & 3 & 2022-02-05 & 30.5 & 239.69 & -53.57 & 0.7 & Swift J1555.2 & 1.4 & 1E 1547.0 & \textit{Swift J1555.2} \\
75 & 4 & 2022-02-13 & 23.5 & 242.65 & -52.03 & 2.9 & PSR J1622 & 3.1 & Swift J1555.2 & \nodata \\
76 & 4 & 2022-03-05 & 55.1 & 232.35 & -53.57 & 3.2 & 1E 1547.0 & 3.8 & Swift J1555.2 & \nodata \\
\textbf{77} & \textbf{3} & \textbf{2022-04-23} & \textbf{15.8} & \textbf{238.50} & \textbf{-52.80} & \textbf{1.3} & \textbf{Swift J1555.2} & \textbf{1.6} & \textbf{1E 1547.0} & \textbf{Swift J1555.2} \\
78 & 3 & 2022-05-26 & 47.6 & 293.91 & 27.95 & 6.1 & SGR 1935 & 10.4 & SGR 2013 & \nodata \\
79 & 3 & 2022-05-26 & 12.2 & 242.03 & -62.70 & 8.7 & 1E 1547.0 & 8.8 & Swift J1555.2 & \nodata \\
80 & 4 & 2022-10-05 & 15.4 & 296.72 & 22.67 & 2.9 & SGR 1935 & 13.1 & SGR 2013 & \textit{SGR 1935} \\
81 & 161 & 2022-10-13 & 122.2 & 295.31 & 22.67 & 1.7 & SGR 1935 & 13.7 & SGR 2013 & \textit{SGR 1935} \\
\textbf{82} & \textbf{11} & \textbf{2022-10-18} & \textbf{59.2} & \textbf{295.31} & \textbf{21.38} & \textbf{1.6} & \textbf{SGR 1935} & \textbf{14.6} & \textbf{SGR 1900} & \textbf{SGR 1935} \\
83 & 3 & 2022-10-23 & 65.1 & 269.30 & -24.62 & 1.9 & SGR 1801 & 4.7 & SGR 1808 & \nodata \\
84 & 22 & 2022-11-09 & 51.9 & 291.09 & 20.11 & 3.0 & SGR 1935 & 11.5 & SGR 1900 & \textit{SGR 1935} \\
85 & 5 & 2022-11-15 & 82.9 & 292.50 & 23.97 & 2.4 & SGR 1935 & 14.1 & SGR 2013 & \textit{SGR 1935} \\
\textbf{86} & \textbf{4} & \textbf{2022-11-19} & \textbf{7.4} & \textbf{291.80} & \textbf{23.32} & \textbf{2.3} & \textbf{SGR 1935} & \textbf{14.8} & \textbf{SGR 1900} & \textbf{SGR 1935} \\
87 & 5 & 2022-12-16 & 0.5 & 273.52 & -19.47 & 1.0 & XTE J1810 & 1.6 & SGR 1806 & \nodata \\
88 & 3 & 2022-12-18 & 39.7 & 271.41 & -22.67 & 1.1 & SGR 1801 & 2.1 & SGR 1808 & \nodata \\
89 & 147 & 2023-02-07 & 498.5 & 71.85 & 43.41 & 3.1 & SGR 0501 & 14.8 & SGR 0418 & \textit{NM} \\
90 & 3 & 2023-01-30 & 95.1 & 255.51 & -45.78 & 2.6 & CXOU J164710.2 & 4.8 & SGR 1627 & \nodata \\
91 & 9 & 2023-02-19 & 47.2 & 74.51 & 44.20 & 1.2 & SGR 0501 & 14.7 & SGR 0418 & \textit{NM} \\
92 & 15 & 2023-02-23 & 100.2 & 75.00 & 47.36 & 2.1 & SGR 0501 & 12.0 & SGR 0418 & \textit{NM} \\
93 & 5 & 2023-02-24 & 19.0 & 277.73 & -16.96 & 2.3 & Swift J1822.3 & 3.1 & AX J1818.8 & \nodata \\
94 & 5 & 2023-02-28 & 36.3 & 275.62 & -23.97 & 4.7 & SGR 1808 & 4.8 & SGR 1806 & \nodata \\
\textbf{95} & \textbf{13} & \textbf{2024-08-21} & \textbf{65.6} & \textbf{279.84} & \textbf{-6.58} & \textbf{1.7} & \textbf{1E 1841} & \textbf{2.2} & \textbf{SGR 1830} & \textbf{1E 1841} \\
96 & 7 & 2024-09-15 & 42.0 & 274.22 & -25.28 & 4.3 & SGR 1801 & 5.0 & SGR 1808 & \nodata \\
97 & 3 & 2024-09-25 & 14.7 & 284.77 & -8.39 & 5.6 & 1E 1841 & 6.0 & Swift J1834.9 & \nodata \\
98 & 4 & 2024-11-19 & 3.5 & 273.52 & -19.47 & 1.0 & XTE J1810 & 1.6 & SGR 1806 & \nodata \\
99 & 5 & 2024-11-29 & 36.3 & 275.62 & -12.64 & 3.4 & Swift J1822.3 & 3.5 & AX J1818.8 & \nodata \\
\textbf{100} & \textbf{6} & \textbf{2024-12-06} & \textbf{71.5} & \textbf{284.77} & \textbf{-4.78} & \textbf{3.6} & \textbf{PSR J1846} & \textbf{4.0} & \textbf{AX J1845.0} & \textbf{1E 1841} \\
101 & 4 & 2025-04-11 & 45.5 & 270.70 & -16.96 & 3.2 & XTE J1810 & 3.7 & SGR 1806 & \nodata \\
102 & 4 & 2025-05-04 & 23.7 & 243.98 & -41.81 & 6.8 & SGR 1627 & 6.9 & CXOU J164710.2 & \nodata \\
\enddata
\tablecomments{Boldface marks an association with a known magnetar. An italic Assoc.\ entry marks a subcluster later associated with SGR 1935+2154 or Swift J1555.2-5402, or later classified as non-magnetar (NM). Magnetar names are truncated before the Declination sign. $N$ is the number of GBM triggers; $\bar t$ is the mean trigger date; $\Delta t$ is the time between the first and last GBM trigger in the subcluster; $\alpha_{\mathrm{MLE}}$ and $\delta_{\mathrm{MLE}}$ are the combined-posterior maximum-likelihood position; $\Delta\theta_1$ and $\Delta\theta_2$ are the offsets to the nearest and second-nearest catalog magnetars.}
\end{deluxetable*}

\subsection{Combining Localization Posteriors}
\label{appendix:combined_posteriors}

    For events belonging to the same subcluster, we estimate a common source
    position by combining the posteriors of the individual events and marginalizing over the spectral parameters to obtain the posterior of the sky position.
    Let $\Omega$ denote the sky position, $\xi$ the spectral parameters and
    $\{d_i\}_{i=1}^{n}$ the data associated with the $n$ events in a given
    subcluster. From Bayes' theorem, the posterior for the common source
    position is
    \begin{equation}
    p(\Omega | d_1,\ldots,d_n)
    =
    \frac{
    p(d_1,\ldots,d_n | \Omega)
    p(\Omega)
    }{
    p(d_1,\ldots,d_n)
    }.
    \label{eq:joint_posterior}
    \end{equation}
    Assuming that the noise realizations of the individual events are
    independent, conditioned on
    $\Omega$, the joint likelihood factorizes as
    \begin{equation}
    p(d_1,\ldots,d_n | \Omega)
    =
    \prod_{i=1}^{n} p(d_i | \Omega).
    \label{eq:factorized_likelihood}
    \end{equation}
    Since the evidence $p(d_1,\ldots,d_n)$ is independent of
    $\Omega$, Eq.~\ref{eq:joint_posterior} can therefore be written as
    \begin{align}
    p(\Omega | d_1,\ldots,d_n)
    &\propto
    p(\Omega)
    \prod_{i=1}^{n} p(d_i | \Omega) \nonumber \\
    &\propto p(\Omega)^{1-n}\prod_{i=1}^{n}p(\Omega|d_i),
    \label{eq:joint_posterior_proportional}
    \end{align}
    where in the second proportionality we used the Bayes theorem for individual bursts and $p(\Omega|d_i)=\int d\xi_i\,p(\Omega,\xi_i|d_i)$ is the marginalized single event posterior. 
    For the uniform priors we use in our parameter estimation framework \citep{perera_expanding_2026}
    $p(\Omega)$ is constant over the sky. The prior term can therefore also be absorbed into the normalization, such that
    \begin{equation}
    p(\Omega | d_1,\ldots,d_n)
    =
    \frac{1}{\mathcal{Z}}
    \prod_{i=1}^{n}
    p(\Omega | d_i),
    \label{eq:combined_posterior_uniform}
    \end{equation}
    where $\mathcal{Z}$ is the normalization constant,
    \begin{equation}
    \mathcal{Z}
    =
    \int
    \prod_{i=1}^{n}
    p(\Omega | d_i)
    \,d\Omega.
    \label{eq:combined_posterior_norm}
    \end{equation}
    The best-estimated position of the subcluster is then
    obtained from the maximum of the marginalized combined posterior,
    \begin{equation}
    \hat{\Omega}
    =
    \underset{\Omega}{\operatorname*{arg\,max}}\,
    p(\Omega | d_1,\ldots,d_n).
    \label{eq:combined_position}
    \end{equation}

\section{Model Selection Complementary Figures}
\label{appendix: bics}

Here we present the complementary $\Delta\mathrm{BIC}$ distributions to
Fig.~\ref{fig:dbic_ottb}, showing the $\Delta\mathrm{BIC}$ of the alternative spectral models relative to the preferred model in each case. The panels show the results for BB, COMPT, PL, BB+BB, and Band, respectively.
    
\begin{figure}
    \centering
    \includegraphics[width=0.45\linewidth]{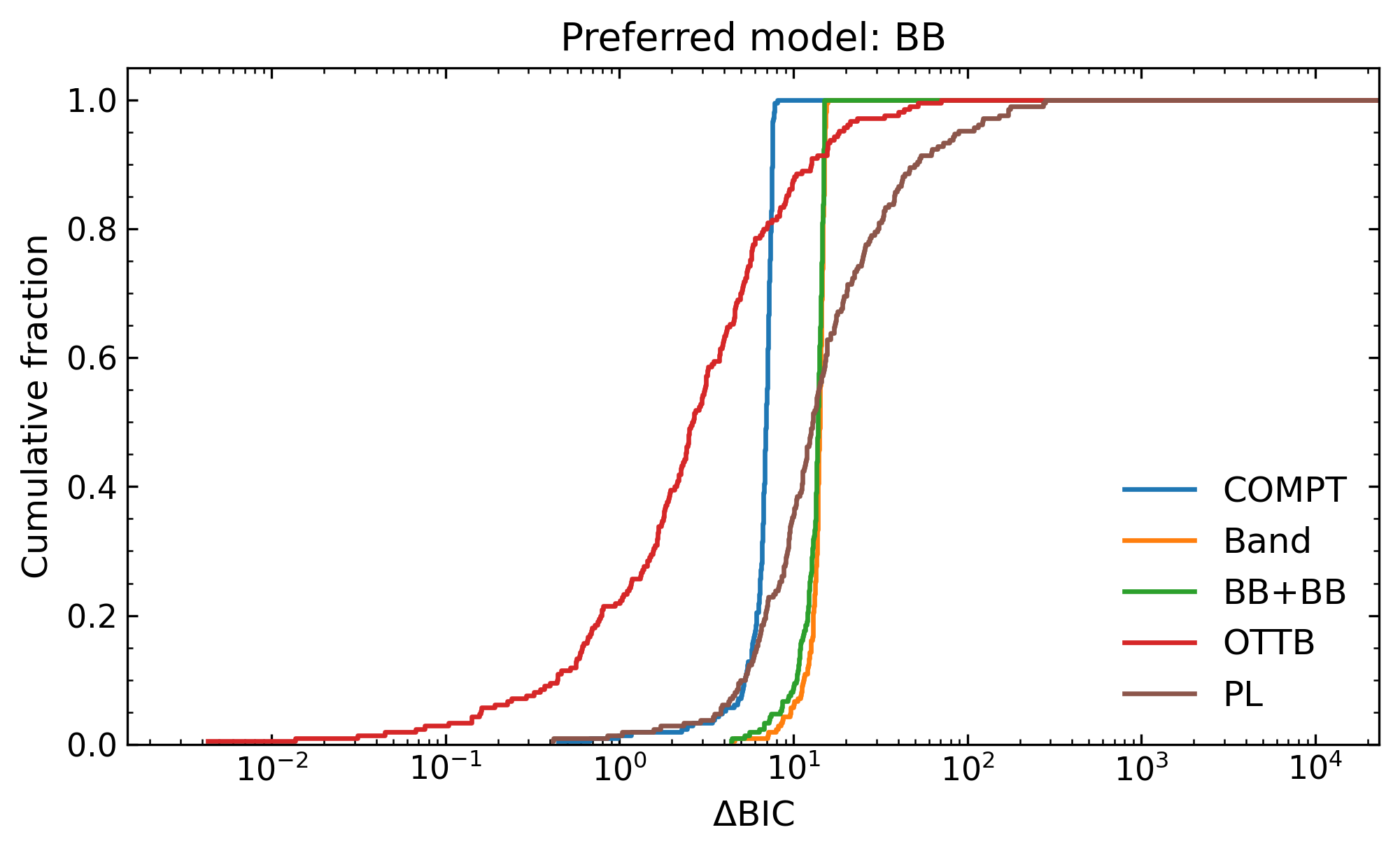}
    \includegraphics[width=0.45\linewidth]{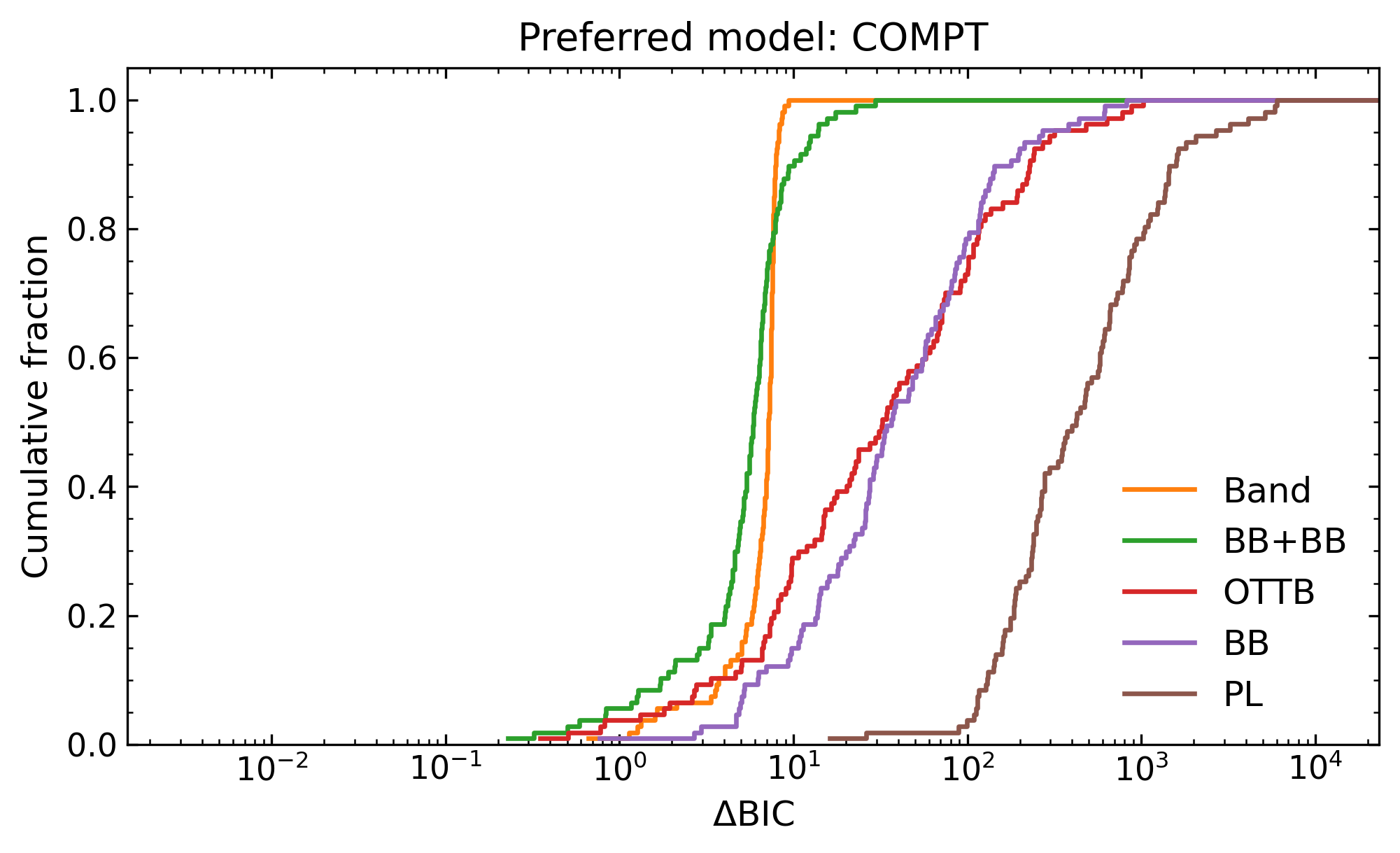}
    \includegraphics[width=0.45\linewidth]{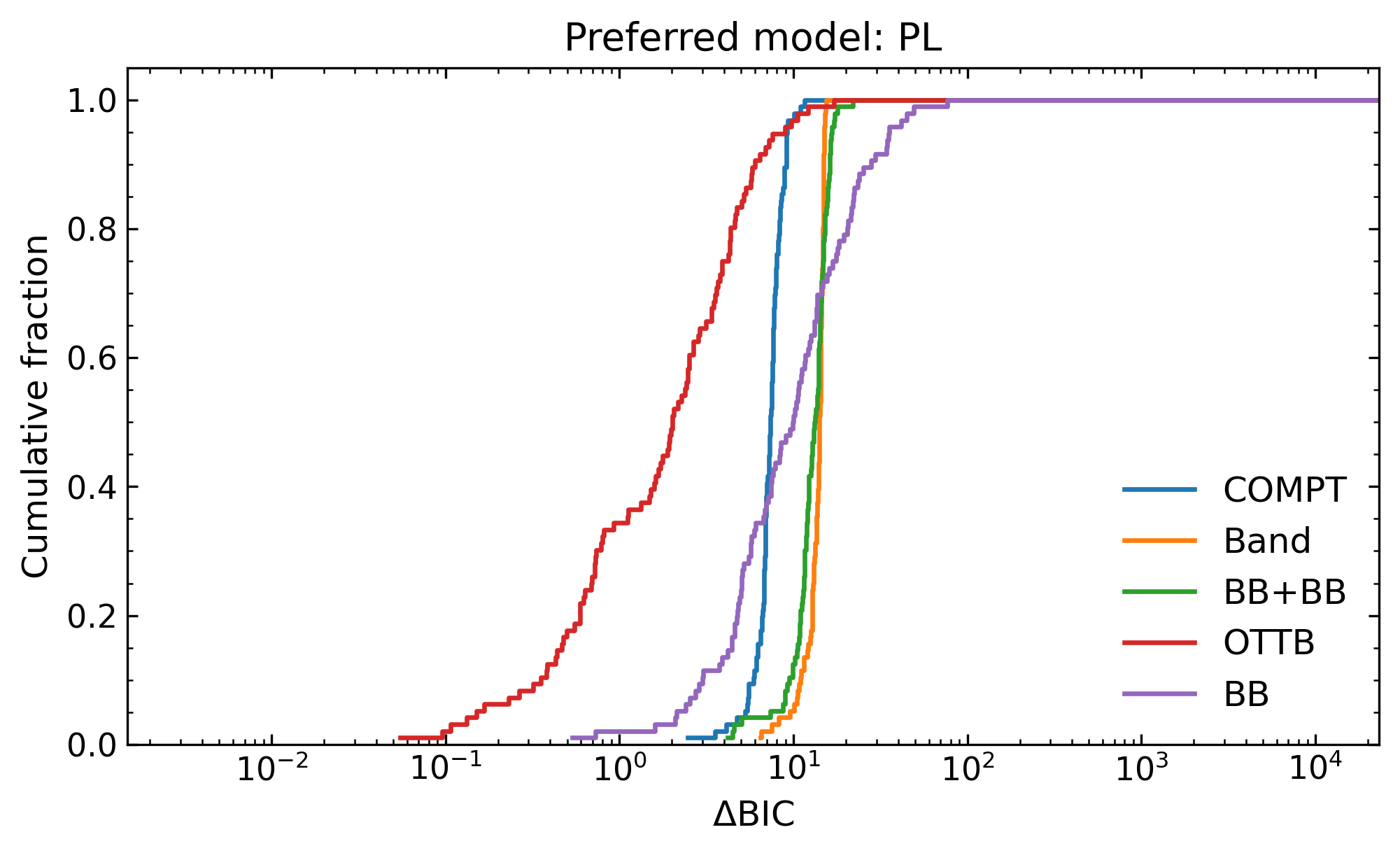}
    \includegraphics[width=0.45\linewidth]{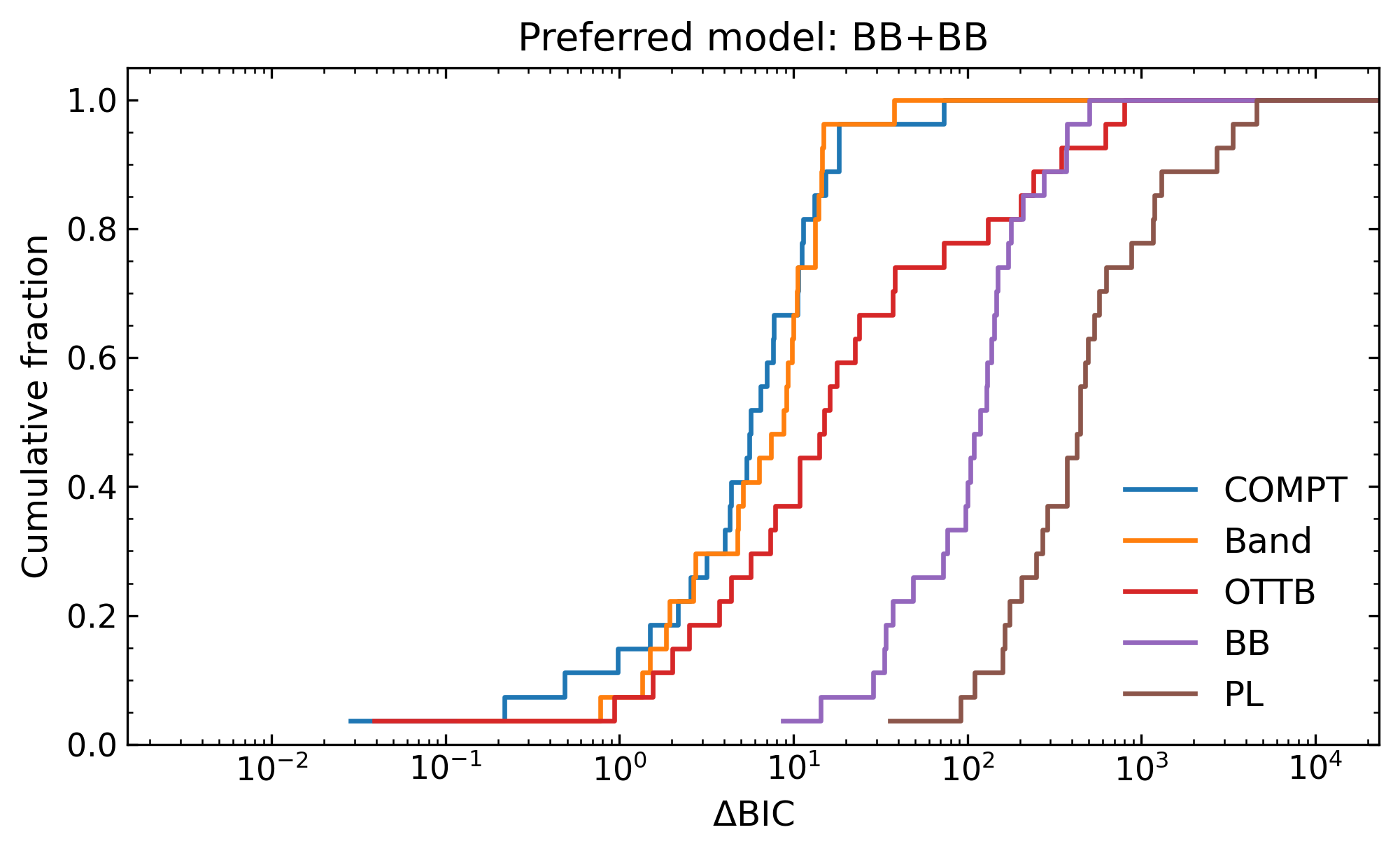}
    \includegraphics[width=0.45\linewidth]{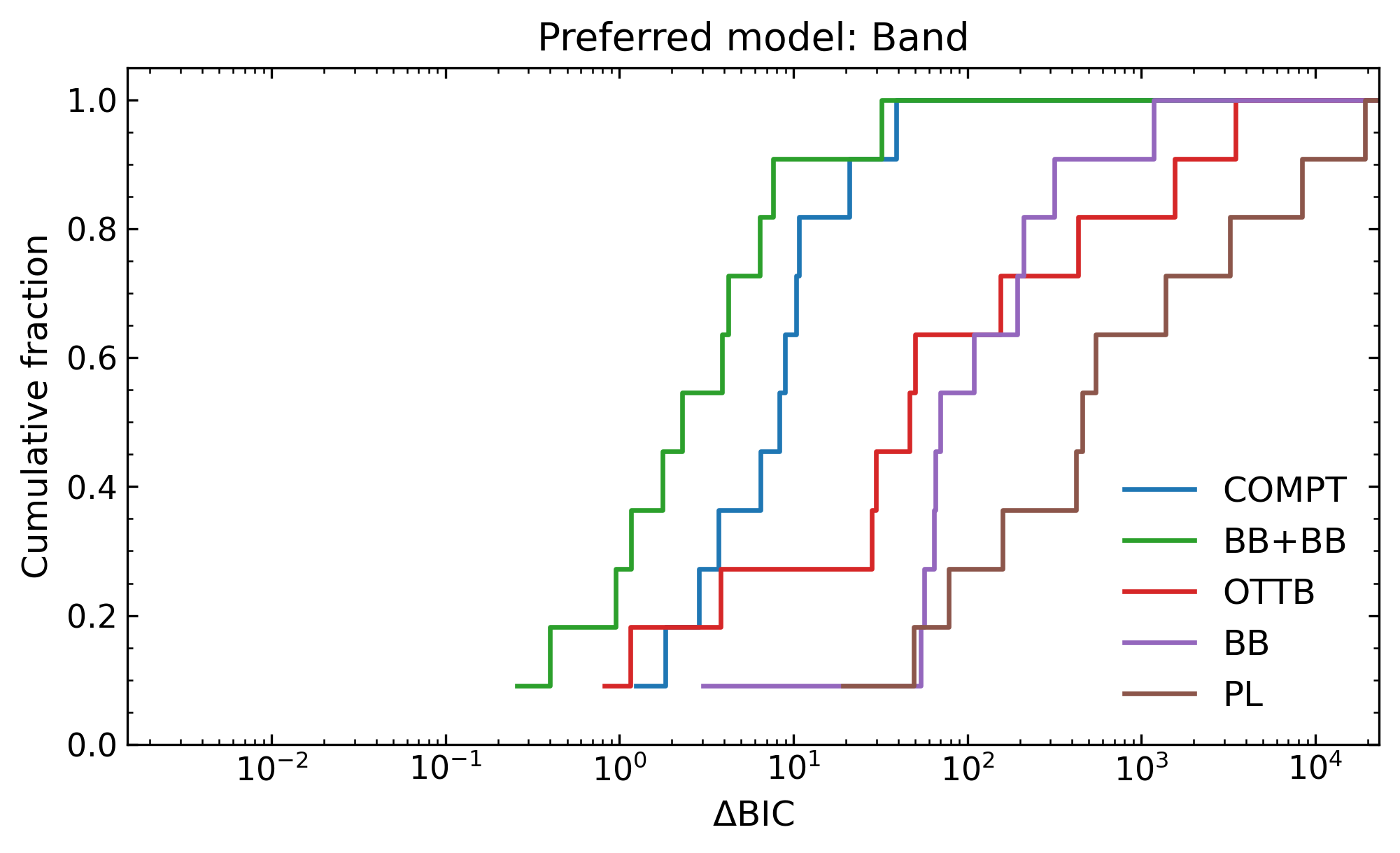}
    \caption{Cumulative distributions of $\Delta\mathrm{BIC}$ for the alternative spectral models relative to the preferred model, for bursts whose preferred models are BB (top left), COMPT (top right), PL (middle left), BB+BB (middle right), and Band (bottom). These distributions complement Fig.~\ref{fig:dbic_ottb} and show that the spectral model degeneracy is present regardless of which model is preferred.}
    \label{fig:rest_of_bics}
\end{figure}

\section{Burst Catalog} \label{sec:appendix_catalog}

The complete magnetar burst catalog and the full sample defined in Sec. \ref{sec:data_selection}, together with the posterior samples described below, is available in a Zenodo repository. The burst catalog is also available in a machine-readable format, and a portion is shown in Table~\ref{tab:catalog_stub} to illustrate its form and content; the printed columns are a subset and are not in the same order as the machine-readable table.

Descriptions of the catalog columns are provided here.

The table contains two distinct sets of inferred parameters. Columns inherited from the coherent search and localization pipeline (\texttt{duration}, \texttt{pe\_duration}, \texttt{snr}, \texttt{pastro}, \texttt{ra\_*}, \texttt{dec\_*}, the Earth and Sun Bayes factors, and the Swift/BAT rates follow-up) are obtained exactly as in the GRB catalog of \citet{perera_expanding_2026}: the sky position and a Band-function spectrum are sampled jointly. The spectral-fit columns (\texttt{best\_model}, \texttt{t90}, \texttt{fluence}, and all quantities suffixed by \texttt{\_compt}, \texttt{\_band}, \texttt{\_bb2}, \texttt{\_ottb}, \texttt{\_bb}, or \texttt{\_pl}) are obtained from a separate, time-integrated analysis performed only for bursts associated with a known source. In that analysis the sky position is fixed to the associated magnetar, event-specific detector responses are used, and six spectral models are fitted (see Sec.~\ref{sec:burst_catalog} and Table~\ref{tab:spec_models}). These catalog-fit parameters should not be confused with the Band-function parameters of the search-stage localization.

The posterior samples from the search-stage localization, including the Band-function parameters used for the sky position, are provided in the same Zenodo repository under the same catalog index as the tabulated burst. Per-model samples from the catalog spectral fits are released in the same way. Uncertainties on the localization and spectral parameters are not tabulated and can be obtained from these samples. Bootstrap uncertainties are reported for the preferred-model $8$-$200\,\mathrm{keV}$ fluence and $T_{90}$.

\begin{deluxetable*}{lccccccccc}
\tablecaption{Example of the magnetar burst catalog.
\label{tab:catalog_stub}}
\tablehead{
\colhead{\texttt{trigtime}} &
\colhead{\texttt{association}} &
\colhead{\texttt{duration}} &
\colhead{\texttt{snr}} &
\colhead{\texttt{pastro}} &
\colhead{\texttt{ra\_max}} &
\colhead{\texttt{dec\_max}} &
\colhead{\texttt{best\_model}} &
\colhead{\texttt{t90}} &
\colhead{\texttt{fluence}} \\
\colhead{(UTC)} &
\colhead{} &
\colhead{(s)} &
\colhead{($\sigma$)} &
\colhead{} &
\colhead{(deg)} &
\colhead{(deg)} &
\colhead{} &
\colhead{(s)} &
\colhead{($\mathrm{erg}\,\mathrm{cm}^{-2}$)}
}
\startdata
2013-06-17 09:52:30.361 & 1E 1547.0$-$5408 & 0.179 & 32.22 & 1.000 & 237.82 & $-$56.66 & bb2   & 0.161 & $1.21\times10^{-7}$ \\
2014-04-29 13:13:28.494 & SGR 1806$-$20    & 0.054 & 29.62 & 1.000 & 270.68 & $-$13.63 & bb2   & 0.276 & $1.09\times10^{-7}$ \\
2014-07-05 09:32:48.688 & SGR 1935+2154    & 0.098 & 21.30 & 1.000 & 299.82 & 24.57  & compt & 0.088 & $6.79\times10^{-8}$ \\
2014-07-05 09:37:34.497 & SGR 1935+2154    & 0.016 & 22.84 & 1.000 & 296.25 & 28.53  & bb2   & 0.027 & $3.45\times10^{-8}$ \\
2014-07-05 09:41:06.150 & SGR 1935+2154    & 0.040 & 8.24  & 1.000 & 291.65 & 42.05  & compt & \nodata & \nodata \\
$\cdots$ & $\cdots$ & $\cdots$ & $\cdots$ & $\cdots$ & $\cdots$ & $\cdots$ & $\cdots$ & $\cdots$ & $\cdots$ \\
\enddata
\tablecomments{Only a subset of columns is shown, and not in the machine-readable format table order.
\texttt{duration} is the search boxcar; \texttt{ra\_max} and \texttt{dec\_max}
are the search-stage localization.
\texttt{t90} and \texttt{fluence} ($8$--$200\,\mathrm{keV}$) are for
\texttt{best\_model} only; the full table reports these quantities for every fitted model.
This table is published in its entirety in machine-readable form.}
\end{deluxetable*}

\begin{itemize}
    \item \textbf{\texttt{catalog\_index}}: Burst index linking the row to its sample files.
    \item \textbf{\texttt{trigtime}}: The trigger time of the transient in UTC.
    \item \textbf{\texttt{association}}: The source assigned to the burst by the localization and clustering analysis (a magnetar name, \texttt{unassociated}, or \texttt{Non magnetar}).
    \item \textbf{\texttt{subcluster}}: Integer identifier of the second-stage spatial-temporal subcluster. A value of $-1$ indicates that the burst is not a member of a subcluster.
    \item \textbf{\texttt{duration}}: The duration of the search boxcar template (in seconds) that maximized the detection test statistic.
    \item \textbf{\texttt{search\_binning}}: The time resolution of the TTE data segment used during the initial search pipeline (either 1 ms or 10 ms).
    \item \textbf{\texttt{slc\_index}}: The internal index of the data slice analyzed by the pipeline. This information was added for the reproducibility of the results. Relevant only for \texttt{search\_binning==0.001} sec.
    \item \textbf{\texttt{pe\_duration}}: The refined peak-flux duration (in seconds) derived during the (search) parameter estimation stage.
    \item \textbf{\texttt{snr}}: The calibrated, drift-corrected Poisson matched-filter signal-to-noise ratio from the detection pipeline. Reported in units of standard normal distribution $\sigma$.
    \item \textbf{\texttt{pastro}}: The probability of astrophysical origin ($p_{\rm astro}$).
    \item \textbf{\texttt{bat\_snr}}: The test statistic ($\rho_{\rm bat}$) resulting from the Swift/BAT targeted follow-up search. Reported in units of standard normal distribution $\sigma$.
    \item \textbf{\texttt{pbat}}: The probability that a temporally coincident signal in Swift/BAT is a genuine joint detection ($p_{\rm bat}$).
    \item \textbf{\texttt{classification}}: The source class assigned to the trigger by our decision tree framework (e.g., SGR3).
    \item \textbf{\texttt{gbm\_catalog}}: The identifier of a temporally coincident event in the GBM catalog.
    \item \textbf{\texttt{loc\_source}}: Origin of the precise localization adopted for the burst. Set to \texttt{BAT} when our targeted BAT imaging search yields $p_{\rm src}>0.5$, an angular offset $<30^{\circ}$, and an image exposure $<10\,\mathrm{s}$; otherwise the localization source reported in the GBM catalog (e.g.\ Swift/BAT, Swift/XRT) if it exists, otherwise set to None.
    \item \textbf{\texttt{ra\_median}, \texttt{dec\_median}}: The median values of the marginalized posterior distribution for the Right Ascension and Declination (in degrees), obtained from the search-stage free-sky localization.
    \item \textbf{\texttt{ra\_max}, \texttt{dec\_max}}: The Maximum Likelihood Estimates (MLE) for the sky position from the same search-stage localization.
    \item \textbf{\texttt{earth\_bayes\_factor}}: A statistical measure quantifying the probability of the trigger originating from the Earth.
    \item \textbf{\texttt{sun\_bayes\_factor}}: A statistical measure quantifying the probability of the trigger originating from the Sun.
    \item \textbf{\texttt{bat\_ra}, \texttt{bat\_dec}}: Right Ascension and Declination (in degrees) of the counterpart identified in the targeted Swift/BAT imaging search.
    \item \textbf{\texttt{bat\_image\_snr}}: The signal-to-noise ratio of that BAT image detection.
    \item \textbf{\texttt{bat\_ang\_offset}}: Angular separation (in degrees) between the GBM search localization and the BAT image position.
    \item \textbf{\texttt{bat\_exposure}}: Exposure duration of the BAT image (in seconds).
    \item \textbf{\texttt{psource}}: The probability that the BAT image source is a genuine counterpart to the GBM trigger ($p_{\rm src}$).
    \item \textbf{\texttt{best\_model}}: The preferred spectral model from the catalog spectral analysis, defined as the model with the lowest BIC. One of \texttt{compt}, \texttt{band}, \texttt{bb2}, \texttt{ottb}, \texttt{bb}, or \texttt{pl}. Blank if no catalog spectral fit is available.
    \item \textbf{\texttt{t90}}: The $T_{90}$ duration (in seconds) of the preferred catalog spectral model (\texttt{best\_model}), measured on the deconvolved $8$--$200\,\mathrm{keV}$ energy fluence. $T_{90}$ is the interval accumulating $5\%$--$95\%$ of the fluence.
    \item \textbf{\texttt{t90\_err\_plus}, \texttt{t90\_err\_minus}}: The positive and negative $1\sigma$ bootstrap uncertainties on \texttt{t90}.
    \item \textbf{\texttt{fluence}}: The $8$--$200\,\mathrm{keV}$ energy fluence (in $\mathrm{erg}\,\mathrm{cm}^{-2}$) of the preferred catalog spectral model, accumulated over the catalog spectral-analysis window.
    \item \textbf{\texttt{fluence\_err\_plus}, \texttt{fluence\_err\_minus}}: The positive and negative $1\sigma$ bootstrap uncertainties on \texttt{fluence}.
    \item \textbf{\texttt{bic\_\{model\}}, \texttt{dbic\_\{model\}}}: The Bayesian information criterion of the catalog fit for that spectral model, and $\Delta\mathrm{BIC}$ relative to the preferred model ($\Delta\mathrm{BIC}=0$ for \texttt{best\_model}). Higher $\Delta\mathrm{BIC}$ indicates a less preferred model.
    \item \textbf{\texttt{alpha\_median\_\{model\}}, \texttt{epeak\_median\_\{model\}}, \texttt{beta\_median\_\{model\}}, \texttt{kt\_median\_\{model\}}, \texttt{kt\_cool\_median\_\{model\}}, \texttt{kt\_hot\_median\_\{model\}}, \texttt{frac\_cool\_median\_\{model\}}}: Median values of the marginalized posterior for the catalog-fit spectral parameters of that model ($E_{\rm peak}$ and $kT$ in keV). Only the parameters belonging to the given model are populated (COMPT: $\alpha$, $E_{\rm peak}$; Band: $\alpha$, $\beta$, $E_{\rm peak}$; BB and OTTB: $kT$; BB+BB: $kT_{\rm c}$, $kT_{\rm h}$, $f$; PL: $\alpha$).
    \item \textbf{\texttt{alpha\_max\_\{model\}}}, etc.: The corresponding Maximum Likelihood Estimates (MLE) for the catalog-fit spectral parameters.
\end{itemize}

\bibliography{references}{}
\bibliographystyle{aasjournalv7.1}

\end{document}